\documentclass[sort&compress]{elsarticle}
\usepackage{graphicx}
\usepackage{amsmath}
\usepackage{amsfonts}
\usepackage{amssymb}
\usepackage{epsfig}
\usepackage{bm}
\usepackage{color}
\usepackage[usenames,dvipsnames]{xcolor}

\providecommand*{\iu}{\ensuremath{\mathrm{i}}}

\newcommand \beq{\begin{eqnarray}}
\newcommand \eeq{\end{eqnarray}}

\def\j{{\boldsymbol j}}
\def\p{{\boldsymbol p}}
\def\q{{\boldsymbol q}}

\def\x{{\boldsymbol x}}
\def\y{{\boldsymbol y}}
\def\X{{\boldsymbol X}}
\def\F{{\boldsymbol F}}
\def\P{{\boldsymbol P}}

\def\Y{{\boldsymbol Y}}

\def\R{{\boldsymbol R}}
\def\r{{\boldsymbol r}}

\def\v{{\boldsymbol v}}

\def\b{{\boldsymbol b}}

\def\a{{\boldsymbol a}}
\def\s{{\boldsymbol s}}

\def\bmt{{\boldsymbol t}}

\def\bmxi{\mbox{\boldmath$\xi$}}

\def\bmnabla{\mbox{\boldmath$\nabla$}}

\def\bmgamma{\mbox{\boldmath$\gamma$}}
\def\bmeta{\mbox{\boldmath$\eta$}}

\def\bmcalR{\mbox{\boldmath${\cal R}$}}

\def\bra#1{\langle#1\vert}
\def\ket#1{\vert#1\rangle}

\newcommand{\rmd}{{\rm d}}
\newcommand{\rme}{{\rm e}}
\newcommand{\del}{\partial}
\newcommand{\nn}{\nonumber\\ }

\begin{document}

\begin{frontmatter}

\title{Quantum and Classical Dynamics of Heavy Quarks in a Quark-Gluon Plasma}
\author{Jean-Paul Blaizot}

\address
{
	Institut de Physique Th{\'e}orique, Universit\'e Paris Saclay, 
        CEA, CNRS, 
	F-91191 Gif-sur-Yvette, France\\
}

\author{Miguel Angel Escobedo }
\address
{
Department of Physics, P.O. Box 35, FI-40014 University of Jyv\"{a}skyl\"{a}, Finland
}
\begin{abstract}
We derive equations for the time evolution of the reduced density matrix of a collection of heavy quarks and antiquarks immersed in a quark gluon plasma. These equations, in their original form, rely on two approximations: the weak coupling between the heavy quarks and the plasma, the fast response of the plasma to the perturbation caused by the heavy quarks. An additional semi-classical approximation is performed. This allows us to recover  results previously obtained for the abelian plasma using the influence functional formalism. In the case of QCD, specific features of the color dynamics make the implementation of the semi-classical approximation more involved. We explore two approximate strategies to solve numerically the resulting equations in the case of a quark-antiquark pair. One involves Langevin equations with additional random color forces, the other treats the transition between the singlet and octet color configurations as collisions in a Boltzmann equation which can be solved with Monte Carlo techniques. 

 \end{abstract}
 
\end{frontmatter}

 \section{Introduction}
 
 Heavy quarkonia, bound states of charm or bottom quarks, constitue a prominent probe of the quark-gluon plasma produced in ultra-relativistic heavy ion collisions, and are the object of many  investigations, both theoretically and experimentally. Recent data from the LHC provide evidence for a sequential suppresion, with the most fragile (less bound) states being more strongly suppressed \cite{Khachatryan:2016xxp}, while there are  indications that charm quarks are sufficiently numerous to recombine, an effect that is seen to counterbalance the expected suppression  \cite{Adam:2015isa}.  These findings are in line with general expectations. 
  The dissociation of quarkonium was suggested long  ago \cite{Matsui:1986dk} as resulting from the screening of the binding forces by the  quark gluon plasma. Recombination is a natural phenomenon to expect \cite{BraunMunzinger:2000px,Thews:2000rj} whenever the number of heavy quarks is sufficiently large, which seems to be the case of charm quarks at the LHC. However, in order to go beyond these qualitative remarks, and extract precise information about the dynamics,  we have to address  a rather complicated many-body problem.

Even leaving aside the production mechanisms of heavy quarks in hadronic collisions, which is not fully understood yet (see e.g. \cite{Brambilla:2010cs}), the description of the interactions of these heavy quarks with an expanding quark-gluon plasma is indeed complicated for a number of reasons. Many effects  can contribute, among which: screening affecting the binding potential, collisions with the plasma constituents, absorption of gluons of the plasma by the bound states. It should be added to this that the bound states do not exist as  objects ``deposited'' in the plasma: it takes time before a newly created quark-antiquark pair can be considered as a bound state, and during this time it is interacting with the plasma.  This is a feature that is often forgotten in many models that attempts to describe the data (for representatives of recent phenomenological analyses, see e.g. \cite{Strickland:2011aa,Nendzig:2012cu,Du:2017qkv} and references therein). Thus, most models  emphasize static or stationary aspects (even when the expansion of the plasma is take into account): this is the case of potential models \cite{Mocsy:2013syh}, spectral function calculations \cite{Petreczky:2012rq}, or kinetic approaches based  on rate equations \cite{Du:2017qkv,Yao:2017fuc}. Clearly a fully time-dependent, out of equilibrium treatment is called for. Such a treatment should also establish contact between the dynamics of heavy quark-antiquark pairs and their bound state, and that of isolated heavy quarks  in a quark-gluon plasma (see e.g. \cite{Alberico:2011zy,Aarts:2016hap}).  In short,  there is a need for a general, simple, and robust formalism, where all the relevant effects  can be treated within the same framework. In this respect, the observation that the collisions could be taken into account by an imaginary potential is a significant  one \cite{Laine:2006ns,Beraudo:2007ky,Brambilla:2008cx,Escobedo:2008sy,Rothkopf:2011db}.

In recent years  it has been recognized that techniques from the theory of open quantum systems (see e.g. \cite{OQS,Weiss}) could offer a fruitful perspective on this problem. A system of heavy quarks in a quark gluon plasma falls indeed in the category of typical problems addressed by this theory: a small system, weakly interacting with a large ``reservoir'', the quark-gluon plasma.  This point of view has emerged explicitly or implicitly in a number of recent works: derivation of a master equation, and corresponding rate equations \cite{Borghini:2011ms}, use of the influence functional method \cite{Akamatsu:2014qsa,Blaizot:2015hya}, 
 solution of a stochastic Schr\"{o}dinger equation \cite{Katz:2015qja,Kajimoto:2017rel}, or direct computation of the evolution of the density matrix \cite{Brambilla:2016wgg,DeBoni:2017ocl}. 

The present work follows similar lines. Its initial  motivation  was to generalize the results of \cite{Blaizot:2015hya} to the non-Abelian case (QCD). Part of that generalization is straightforward, and relies on the same approximations as those used in the case of the abelian plasma. To some extent, this program has already been considered in the recent work by Akamatsu \cite{Akamatsu:2012vt}, albeit using a formalism slightly different from that used in \cite{Blaizot:2015hya} and in the present paper.  However, color degrees of freedom modifies the picture in a very substantial way. The reason is that, in a collision involving one gluon exchange for instance,  color  changes in a discrete way, in contrast to  position or momentum which vary  continuously. Thus, while we can treat the motion of the heavy quarks within a semi-classical approximation,  there is no such semi-classical limit for the color dynamics (except perhaps in the large $N_c$  limit). It follows that the derivation of Fokker-Planck or Langevin equations made in the abelian case needs to be reconsidered, which we do in this paper. We shall see that the complete dynamics, including the color degrees of freedom, can still be described by Fokker-Plack and Langevin equations, but only in very specific circumstances.    

  This paper focusses on conceptual issues. It  is organized as follows.  In Sect.~\ref{sec:quantum} we derive the quantum master equation for the reduced density matrix of a system of heavy quarks and antiquarks immersed in a quark-gluon plasma, in thermal equilibrium. This equation, whose structure is close to that of a Lindblad equation, is used as a starting point of all later developments. In Sect.~\ref{sec:qed} we  rederive from it  the results that we had previously obtained for the abelian plasma \cite{Blaizot:2015hya} using a path integral formalism. In particular we recover, after performing a semi-classical approximation, the Fokker-Planck and Langevin equations that describe the random walks of center of mass and relative coordinates of a quark-antiquark pair. This section on the abelian plasma paves the way for the treatment of the non abelian case discussed  in Sect.~\ref{sec:qcd}. The equations that we present there, before we do the semi-classical approximation, are fully quantum equations. But they are difficult to solve in general. Thus,  in Sect.~\ref{numerics}  we look for additional approximations that allow us to obtain solutions in some particular regimes, in order to start getting insight into the general solution. In particular,  we explore two ways of implementing the semi-classical approximation. In the first case, we restrict the dynamics to stay close to a maximum entropy color state, where the colors of the heavy quarks are random. In this case the dynamics is described by a Langevin equation with a new random color force. The method used in this case is easily extended to the case of an arbitrary number of quark-antiquark pairs, and allows us to address the question of recombination.  However, it is based on a perturbative approach that breaks down for some values of the parameters. Another strategy focuses on the case of a single quark-antiquark pair. The transition between singlets and octets are treated as ``collisions'' in a kinetic equation that we solve using  Monte Carlo techniques. The last section summarizes our main results, and presents a brief outlook.  
Several appendices at the end gather various technical material.

\section{Equation for the density matrix of heavy quarks in a quark-gluon plasma}
\label{sec:quantum}

Our description of the heavy quark dynamics in a quark-gluon plasma is based on the assumption that the interaction between the heavy quarks and the quark-gluon plasma is weak, and can be treated in perturbation theory (with appropriate resummations).  The generic hamiltonian for such a system reads
\beq
H=H_Q+H_1+H_{\rm pl},
\eeq
where $H_Q$ describes the dynamics of the heavy quarks in the absence of the plasma, $H_{\rm pl}$ is the hamiltonian of the plasma in the absence of the heavy quarks, and $H_1$ is the interaction between the heavy quarks and the plasma constituents. The heavy quarks are treated as non relativistic particles, and the spin degree of freedom is ignored: the state of a heavy quark is then entirely specified by its position and color. As we have mentioned already, we shall consider $H_1$ to be small and treat it as a perturbation. In Coulomb gauge, and neglecting magnetic interactions, this interaction takes the form
\beq\label{H1}
H_1=-g\int_\r A_0^a(\r) n^a(\r),
\eeq
where $n^a$ denotes the color charge density of the heavy particles. For a quark-antiquark pair, for instance, this is given by\footnote{We denote here the position operator by $\hat r$, but most often  the symbol  \^{ } will be omitted, the context being enough to recognize the operators.} 
\beq\label{colordensity}
n^a(\x)=\delta(\x-\hat \r)\, t^a\otimes \mathbb{I} -\mathbb{I}\otimes\delta(\x-\hat\r) \, \tilde t^a,
\eeq
where we use the first quantization to describe the heavy quark and antiquark, and 
 the two components of the tensor product refer respectively to the Hilbert spaces of the heavy quark (for the first component) and the heavy antiquark (for the second component). 
In Eq.~(\ref{colordensity}), $t^a$ is a color matrix in the fundamental representation of SU(3) and describes the coupling between the heavy quark and the gluon. The coupling of the heavy antiquark and the gluon is described by $-\tilde t^a$, with $\tilde t^a$ the transpose of $t^a$.

We are looking for an effective theory for the heavy quark dynamics, obtained by eliminating the plasma degrees of freedom. In  previous works,  this was achieved explicitly by constructing the Feynman-Vernon influence functional \cite{FeynmanVernon}, using the path integral formalism  (see e.g. \cite{Akamatsu:2012vt,Blaizot:2015hya}). In the present paper, we shall proceed differently, by writing directly the equations of motion for the reduced density matrix of the heavy quarks. Although the derivations presented here are self-contained, we emphasize that the main approximations that are implemented in the present section are quite common in various fields, and belong to what is commonly referred to as the theory of open quantum systems (see e.g. \cite{OQS}).

We assume that the system contains a fixed number, $N_Q$, of heavy quarks (and, in general, an equal number of antiquarks). We call ${\cal D}$ the density matrix of the full system, and  ${\cal D}_Q$ the reduced density matrix for the heavy quarks. The latter is defined as the partial trace of the full density matrix over the plasma degrees of freedom, that is
\beq
{\cal D}_Q={\rm Tr}_{\rm pl} {\cal D}.
\eeq
 In order to make contact with the work of Ref.~\cite{Blaizot:2015hya}, we recall that a typical question addressed there was the following: Given a set of heavy quarks at position $\X_i$ at time $t_i$, where $\X$ denotes collectively the set of coordinates of the quarks and antiquarks (temporarily ignoring color),  what is the probability $P(\X_f, t_f|\X_i, t_i)$ to find them as position $\X_f$ at time $t_f$?  This probability  is given by 
\beq\label{probability0}
P(\X_f, t_f|\X_i, t_i) =\left| \langle\X_f,t_f|\X_i,t_i\rangle\right|^2=\bra{\X_f}{\cal D}_Q\ket{\X_i},
\eeq
that is, it can be obtained as a specific element of the reduced density matrix.
In \cite{Blaizot:2015hya} a representation of this quantity was obtained in terms of a path integral which is remains difficult to evaluate in general.\footnote{The analogous path integral for a single heavy quark in an abelian plasma has been evaluated in \cite{Beraudo:2010tw}. However, this evaluation was performed in Euclidean space. An analytic continuation is needed to recover the real time information, and procedures to do so numerically are not without ambiguities.} However, in the regime where a semi-classical approximation is valid,  the dynamics that it describes is equivalent to that of a Fokker-Planck equation which can be easily solved numerically, in particular by solving the associated Langevin equation. Two approximations are involved in the construction of the influence functional such as presented in \cite{Akamatsu:2012vt,Blaizot:2015hya}. The first one is the weak coupling approximation for the interaction of the heavy quarks with the plasma, the second assumes that  the response of the plasma to the perturbation caused by the heavy quarks is fast compared to the characteristic time scales of the heavy quark motion. An additional approximation, to which we refer to as a semi-classical approximation, leads, as we have just mentioned, to Fokker Planck and Langevin equations.

 The last two approximations exploit the fact that the mass of  the heavy quark is large, i.e., $M\gg T$.  Thus, when the heavy quark is not too far from thermal equilibrium, its thermal wavelength $\lambda_{\rm th}\sim 1/\sqrt{MT}$ is small compared to the typical microscopic length scale $\sim 1/T$. Under such condition, the density matrix can be considered as nearly diagonal (in position space), motivating a semi-classical approximation: indeed the off-diagonal matrix elements $\bra{\X}{\cal D_Q}\ket{\X'}$ die off when $|\X-\X'|\gtrsim \lambda_{\rm th}$.  The typical heavy quark velocity is of the order of the thermal velocity $\sim\sqrt{T/M}\ll 1$,  and the dynamics of the heavy fermions is much slower than that of the plasma.  The typical frequency for the plasma dynamics is the plasma frequency which, for ultra-relativistic plasmas, is of the order of the Debye screening mass $m_D$.  During a time $t\sim m_D^{-1}$,  the heavy quark moves a  distance which is small compared to the size of the screening cloud, $\sim m_D^{-1}$. Thus, over a time scale characteristic of the plasma collective dynamics,  the heavy quark positions are almost frozen (they are completely frozen in the limit $M\to\infty$). One can also recognize that the collisions of the heavy particles with the light constituents of the plasma involve the exchange of soft gluons, with typical momenta $q\lesssim m_D\ll M$. The corresponding energy transfer $\sim q^2/M\sim m_D^2/M$ is small on the scale of the plasma frequency, $m_D^2/M\ll m_D$.
 
 \subsection{Equation for the density matrix}
 
\begin{figure}[!hbt]
\begin{center}
\includegraphics[width=1.1\textwidth]{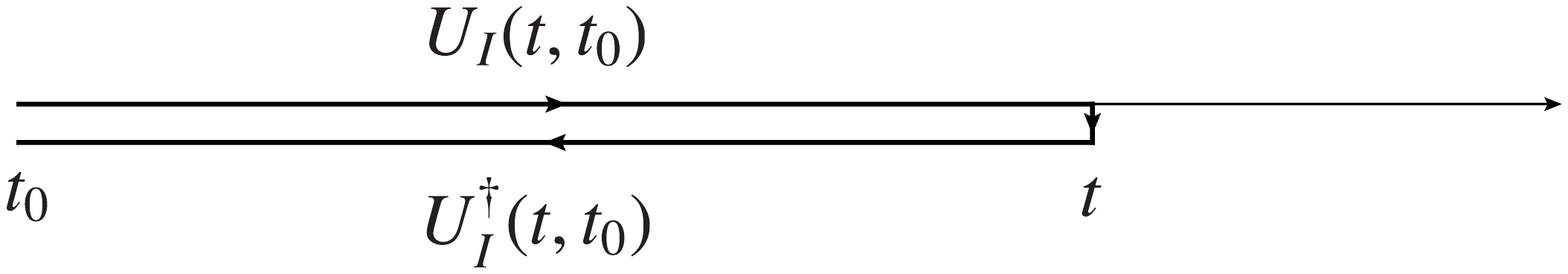}
\vspace{-7cm}
\caption{The Schwinger-Keldysh contour} \vspace{-0.25in}
\label{fig:SKcontour}
\end{center}
\end{figure}

 The density matrix obeys the general equation of motion
 \beq
 i\frac{\rmd {\cal D}}{\rmd  t}=[H,{\cal D}].
 \eeq
 I order to treat the interaction between the plasma and the heavy particle using perturbation theory, we  move to the interaction representation. We set $H=H_0+H_1$, with $H_0=H_Q+H_{\rm pl}$ and define 
 \beq
 {\cal D}(t)=U_0(t,t_0) {\cal D}^I(t) U_0^\dagger(t,t_0),
 \eeq
 where ${\cal D}^I(t) $, the interaction representation of the density matrix, satisfies the equation
 \beq\label{exacteqintrepres0}
 \frac{\rmd {\cal D}^I}{\rmd  t}&=&-i [H_1(t),{\cal D}^I(t)],\qquad  H_1(t)=U_0(t,t_0)^\dagger H_1 U_0(t,t_0).
 \eeq
 Here, $H_1(t)$ denotes the interaction representation of $H_1$. The evolution  operator in the interaction representation, $U_I(t_0,t)=U_0^\dagger(t,t_0) U(t,t_0)$, can be expanded in powers of $H_1(t)$ in the usual way
 \beq
 U_I(t,t_0)={\rm T}\exp\{ -i\int_{t_0}^t \rmd t' H_1(t')  \},
 \eeq
 where the symbol ${\rm T}$ denotes time ordering. 
 Similarly, Eq.~(\ref{exacteqintrepres0}) can be integrated formally using the Schwinger-Keldysh contour \cite{Schwinger61,Keldysh64}: \beq\label{exprhoSK}
{\cal D}^I(t)&=& U_I(t,t_0)   {\cal D}(t_0) U_I^\dagger(t,t_0)\nn
&=&{\rm T}_{\cal C}\left[  \exp\left\{ -i\int_{\cal C} \rmd t_{_{\cal C}} H_1(t_{_{\cal C}})  \right\}{\cal D}(t_0) \right],
\eeq
where the operator ${\rm T}_{\cal C}$ orders the operators $ H_1(t_{_{\cal C}})  $ along the contour parameterized by the contour time $t_{_{\cal C}}$, with the operators carrying the largest $t_{_{\cal C}}$ coming before those with smaller $t_{_{\cal C}}$  (see Fig.~\ref{fig:SKcontour}). 
The upper branch of the contour, with time running from $t_0$ to $t$,  represents the evolution operator  $U_I(t,t_0)$, the lower branch of the contour, with time running from $t$ to $t_0$, represents the operator $U_I^\dagger (t,t_0)$.  As can be seen in Eq.~(\ref{exprhoSK}), in the expansion of ${\cal D}^I(t)$ in powers of $H_1(t)$, the operators $H_1(t)$ that sit on the left of ${\cal D}(t_0)$ live on the upper branch of the contour, while those that appear on the right of ${\cal D}(t_0)$ live on the lower branch (they come later along the contour).  
 \vspace{-0.0cm}

 To proceed further, we  assume that, at the initial time $t_0$, the density matrix factorizes
 \beq\label{factorizedD}
 {\cal D}^I(t_0)= {\cal D}_Q^I(t_0)\otimes  {\cal D}_{\rm pl}^I(t_0).
 \eeq
We also assume that at time $t_0$,  the plasma is  in a state of thermal equilibrium, so that its density matrix ${\cal D}^I_{\rm pl}(t_0)={\cal D}_{\rm pl}(t_0)$ is a canonical density matrix, 
 \beq
 {\cal D}_{\rm pl}(t_0)=\frac{1}{Z_{\rm pl}} \sum_m \rme^{-\beta E_m},
 \eeq
 where $\beta=1/T$, with $T$ the equilibrium temperature.  This factorization of the density matrix allows for a simple calculation of the trace over the plasma degrees of freedom. 
 
 Let us then examine perturbation theory at second 
 order in $H_1$, with $H_1$ given by Eq.~(\ref{H1}). Performing the trace over the plasma degrees of freedom is immediate, thanks to the specific structure of $H_1$ and the factorization of the density matrix at $t=t_0$. One obtains
\beq\label{exprhoSK2b}
&& {\cal D}_Q^I(t)= {\cal D}_Q(t_0) -i\int_{t_0}^t \rmd t' \int_\x \langle A_0^a(\x)\rangle [n^a(\x,t), {\cal D}_Q(t_0)]\nn
&&-\frac{1}{2}\int_{t_0}^t \rmd t_1 \int_{t_0}^{t} \rmd t'_1\int_{\x\x'}\,{\rm T}[n^a(t_1,\x) n^b(t_1',\x')] {\cal D}_Q(t_0) \,\langle {\rm T}[A_0^a(t_1,\x) A_0^b(t_1',\x')] \rangle_0\nn
&&-\frac{1}{2}\int_{t_0}^t \rmd t_2 \int_{t_0}^{t} \rmd t_2'\int_{\x\x'}\, {\cal D}_Q(t_0)\tilde{\rm T}[n^a(t_2,\x) n^b(t'_2,\x')]  \,\langle \tilde{\rm T}[A_0^a(t_2,\x) A_0^b(t'_2,\x')] \rangle_0\nn
&&+\int_{t_0}^t \rmd t_1 \int_{t_0}^{t} \rmd t_2\int_{\x\x'}\, [n^a(t_1,\x){\cal D}_Q(t_0) n^b(t_2,\x')]  \,\langle  A_0^a(t_2,\x')A_0^b(t_1,\x)  \rangle_0 ,\nn
\eeq
where, in the last three lines,  we have used the convention that $t_1, t'_1$ run on the upper part of the contour, while $t_2,t'_2$ run on the lower branch.   Note that the linear term vanishes since the plasma is color neutral (so that  $ \langle A_0^a(\x)\rangle_0=0$). Here the notation  $\langle\cdots \rangle_0$ stands for the average with the plasma equilibrium density matrix, that is
\beq
\langle \cdots\rangle_0={\rm Tr}_{\rm pl}\left[ \frac{1}{Z_{\rm pl}} \rme^{-\beta H_{\rm pl}} \cdots\right]. 
\eeq Similarly the correlators of the gauge fields are diagonal in color, i.e. they are proportional to $\delta^{ab}$. These correlators are the exact correlators in the plasma (the fields are in the interaction representation, which corresponds to the Heisenberg representation when considering the plasma alone).  They are written as
\beq\label{correlators}
\langle {\rm T}[A_0^a(t_1,\x) A_0^b(t_1', \x')] \rangle_0&=&-i\delta^{ab}\Delta(t_1-t_1',\x-\x')\nn
 \langle\tilde {\rm T}[A_0^a(t_2,\x) A_0^b(t_2' ,\x')] \rangle_0&=&-i\delta^{ab}\tilde\Delta(t_2-t_2',\x-\x')\nn
 \langle {\rm T}_C  A_0^a(t_2,\x')A_0^b(t_1,\x)  \rangle_0&=&\delta^{ab}\Delta^>(t_2-t_1,\x'-\x)\nn
 &=&\delta^{ab}\Delta^<(t_1-t_2,\x-\x').
 \eeq
The apparent inversion of the order of times in the last correlator results from the relation  ${\rm Tr}_{\rm pl}A_0^b(t_1,\x) {\cal D}_{\rm pl}(t_0)A_0^a(t_2,\x')= \langle   A_0^a(t_2,\x')A_0^b(t_1,\x)  \rangle_0$ which follows from  the cyclic invariance of the trace. 

\begin{figure}[!hbt]
\begin{center}
\includegraphics[width=0.8\textwidth]{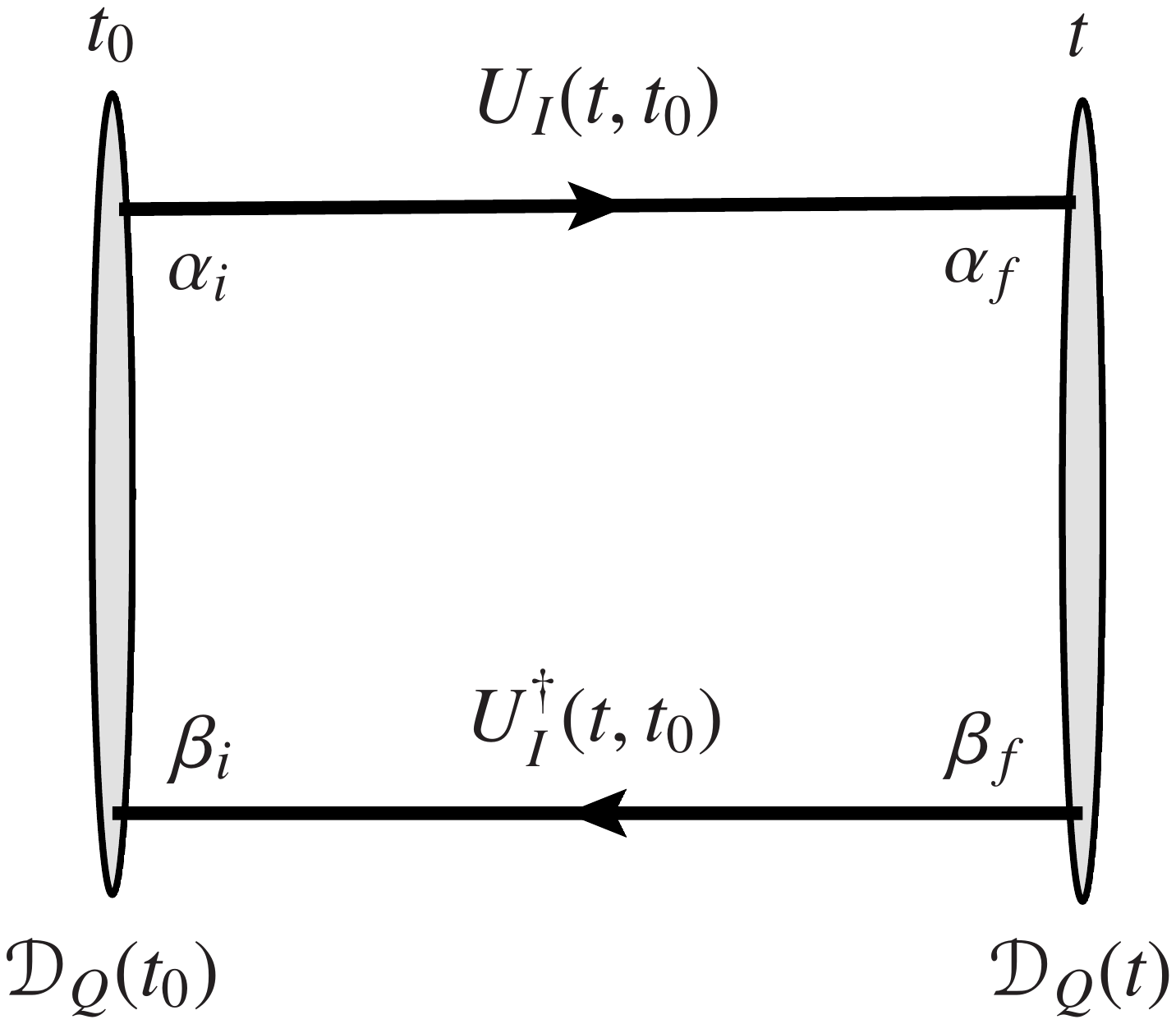}
\vspace{-2.5cm}
\caption{Graphical representation of the evolution of the density matrix from $t_0$ to $t$. The horizontal lines represent the evolution operators $U_I(t,t_0)$ (upper branch) or $U_I^\dagger(t,t_0)$ (lower branch). When ${\cal D}_Q $ is the density matrix of a single heavy quark, these horizontal lines may be interpreted as the associated propagators of the heavy particle. When ${\cal D}_Q $ is the density matrix of a heavy quark-antiquark pair,  a single horizontal line is replaced by a pair of lines, associated with the propagator of the pair (see Fig.~\ref{fig:densitymatrix2} below).}  \vspace{-0.25in}
\label{fig:densitymatrix}
\end{center}
\end{figure}

\begin{figure}[!hbt]
\begin{center}
\includegraphics[width=1\textwidth]{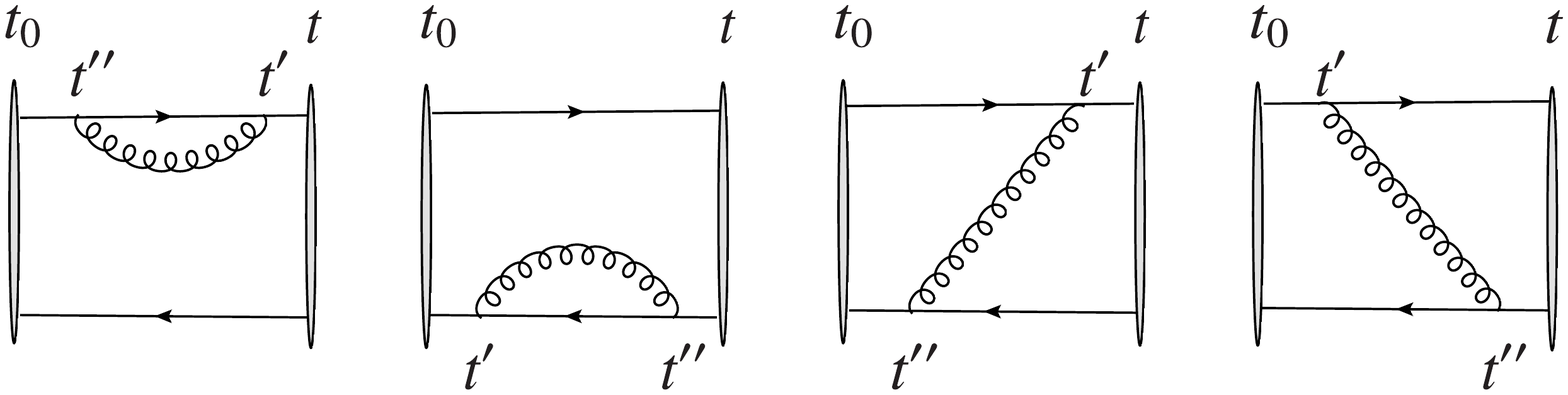}
\vspace{-5cm}
\caption{These diagrams are in one-to-one correspondence with the terms in the last three lines of the  right-hand side of Eq.~(\ref{exprhoSK2b}) for the single particle density matrix ${\cal D}_Q^I(t)$. } \vspace{-0.25in}
\label{fig:prederiv}
\end{center}
\end{figure}

It is convenient to represent the evolution of the density matrix by a diagram such as that in Fig.~\ref{fig:densitymatrix}, where the upper and lower parts of the diagram may be associated to the corresponding upper and lower parts of the Schwinger-Keldysh contour. The explicit expression that this diagram represents is
\beq
\bra{\alpha_f} {\cal D}_Q(t)\ket{\beta_f}=\sum_{\alpha_i\beta_i}\bra{\alpha_f}U_I(t,t_0)\ket{\alpha_i}\bra{\alpha_i} {\cal D}_Q(t_0) \ket{\beta_i}\bra{\beta_i} U_I^\dagger(t,t_0)\ket{\beta_f},
\eeq 
where $\alpha$ of $\beta$ represent the set of quantum numbers that are necessary to specify the state of the heavy particles (essentially the position and color). 
\vspace{-0cm}
The diagrammatic interpretation of Eq.~(\ref{exprhoSK2b}) is then given in Fig.~\ref{fig:prederiv} (for the case of a single particle density matrix). 
\vspace{-0cm}
 \\

 In order to implement our further approximations, it is convenient  to consider the time derivative of the density matrix. This can be obtained by taking the derivative of Eq.~(\ref{exprhoSK2b}) above (see Eq.~(\ref{exprhoSK2bder}) in \ref{App:alternative}). But it is more instructive to return to Eq.~(\ref{exacteqintrepres0}), and rewrite it as
  \beq\label{exacteqintrepres1}
 \frac{\rmd {\cal D}^I}{\rmd  t}=-i [H_1(t),{\cal D}^I(t_0)]-\int_{t_0}^t\rmd t' [H_1(t),[H_1(t'),{\cal D}^I(t')]].
 \eeq
 This exact equation is obtained by formally integrating Eq.~(\ref{exacteqintrepres0}) and inserting  the solution  back into the equation. Perturbation theory at second order in $H_1$ is recovered by replacing ${\cal D}^I(t')$ by ${\cal D}^I(t_0)$ in the double commutator in the right hand side.  
 One may then proceed to the average over the plasma degrees of freedom, as we did before, and get the following equation for the reduced density matrix ${\cal D}_Q$
\beq\label{eqrhoAt0b2b}
\frac{\rmd {\cal D}^I_Q(t)}{\rmd t}
&=&-\int_{t_0}^t \rmd t'\int_{\x\x'}\left( [n^a(t,\x), n^a(t',\x'){\cal D}^I_Q(t_0)]\Delta^>(t-t',\x-\x')\right.\nn
&&\qquad\qquad\quad \left.  +[{\cal D}^I_Q(t_0) n^a(t',\x'),n^a(t,\x)]\Delta^<(t-t',\x-\x')\right).\nn
\eeq
We have used the fact that the linear term vanishes in  a neutral plasma, and the sum over the color index $a$ is implicit. 

At this point, we can improve on strict perturbation theory. To do so we notice that the integral over $t'$ in Eq.~(\ref{exacteqintrepres1}) is in fact limited to a region near $t'\lesssim t$: this is because $\Delta(t-t')$ dies out when $t-t'\gtrsim m_D^{-1}$, and we are interested on the evolution of the density matrix over time scales that are much larger than $m_D^{-1}$. Thus, noticing also that the difference ${\cal D}^I(t)-{\cal D}^I(t')$ involves in any case an extra power of $H_1$, we replace ${\cal D}^I(t')$ by ${\cal D}^I(t)$ in the right hand side of Eq.~(\ref{exacteqintrepres1})\footnote{An alternative procedure, which leads to slightly different equations, is presented in \ref{App:alternative}}, turning the equation into an equation for ${\cal D}(t)$ which is now local in time. We shall furthermore exploits the fact  that the density matrix  approximately factorizes at all times, as does the density matrix at the initial time $t_0$. Again, this is consistent with the weak coupling approximation since the correction to the factorized from necessarily involves additonal powers of $H_1$. The latter approximation allows us to perform the trace over the plasma degree of freedom, in the same way as we did earlier for strict perturbation theory. The resulting equation is in fact identical to Eq.~(\ref{eqrhoAt0b2b}) in which we replace in the right hand side ${\cal D}^I_Q(t_0)$by ${\cal D}^I_Q(t)$. It is convenient for the following to write this equation in the Schr\"{o}dinger picture. A simple calculation yields 
\beq
&&\!\!\!\!\!\!\!\!\!\! \frac{\rmd {\cal D}_Q}{\rmd t}+i[H_Q,{\cal D}_Q(t)]=\nn
&&-\int_{\x\x'}
 \int_{0}^{t-t_0} \rmd \tau \,[n^a_\x ,U_Q(\tau)n^a_{\x'}U_Q^\dagger(\tau){\cal D}_Q(t)]  \, \Delta^>(\tau;\x-\x'))\nn
&&-\int_{\x\x'}\int_{0}^{t-t_0} \rmd \tau \, [ {\cal D}_Q(t)U_Q(\tau)n^a_{\x'}U_Q^\dagger(\tau), n^a_\x]  \,\Delta^<(\tau;\x-\x').\nn
\eeq
where we have set $t-t'=\tau$. This equation has the same physical content as the influence functional derived in \cite{Blaizot:2015hya}, and it is based on analogous approximations. It relies on a weak coupling approximation, but goes beyond strict second order perturbation theory, in particular by resumming secular terms. 

This equation still contains a memory integral that we shall simplify thanks to our last approximation: In line with the fact that only small values of $\tau $ are relevant, it consists in replacing $\rme^{-iH_Q\tau}\simeq 1-iH_Q\tau$, and   keep terms up to linear order in $\tau$ in the integrals. More precisely, we write
\beq
U_Q(\tau)n_{\x'}^a U_Q^\dagger(\tau)=U_Q^\dagger(-\tau)n_{\x'}^a U_Q(-\tau)=n_{\x'}^a(-\tau)
\eeq
and 
\beq
n_{\x'}^a(-\tau)=n_{\x'}^a-\tau \dot n_{\x'}^a,\qquad \dot n_{\x'}^a=i\left[H_Q,n_{\x'}^a\right],
\eeq
 the time-dependence of $n_{\x'}(t)$ being given by the Heisenberg representation, $n_{\x'}^a(t)=\rme^{iH_Qt} n_{\x'}^a\rme^{-iH_Qt}$.
We get
\beq\label{premain}
\frac{\rmd {\cal D}_Q}{\rmd t}+i[H_Q,{\cal D}_Q(t)]&\approx& -\int_{\x\x'} [n^a_\x,n^a_{\x'} {\cal D}_Q ] \,  \int_{0}^{\infty} \rmd \tau \Delta^>(\tau;\x-\x'))\nn
&-&\int_{\x\x'} [ {\cal D}_Q n^a_{\x'}, n^a_\x]  \,\int_{0}^{\infty} \rmd \tau \,\Delta^<(\tau;\x-\x')\nn
&+&\int_{\x\x'}[n^a_\x, \dot n^a_{\x'}{\cal D}_Q]  \,  \int_{0}^{\infty} \rmd \tau\,\tau\, \Delta^>(\tau;\x-\x'))\nn
&+&\int_{\x\x'}[ {\cal D}_Q \dot n^a_{\x'}, n^a_\x] \,\int_{0}^{\infty} \rmd \tau \,\tau\,\Delta^<(\tau;\x-\x').\nn
\eeq

At this point we use the values of the time integrals given in \ref{App:correlators}. These involve the zero frequency part of the time-order propagator $\Delta(\omega=0)=\Delta^R(\omega=0,\r)+i\Delta^<(\omega=0,\r)$, which we identify with the real and imaginary part of a ``potential''. More precisely, we set
\beq
V(\r)=-\Delta^R(\omega=0,\r), \qquad W(\r)=-\Delta^<(\omega=0,\r).
\eeq
This terminology stems from the fact that $V(\r)+iW(\r)$ plays the role of a complex potential in a Schr\"{o}dinger equation describing the relative motion of a quark-antiquark pair: the real part represents the screening corrections, and adds to the  familiar interaction arising in leading order from one-gluon exchange, the imaginary part accounts effectively for the collisions between the heavy quarks and the plasma constituents \cite{Laine:2006ns,Beraudo:2007ky}.

 After a simple calculation that uses the properties $V(\x-\x')=V(\x'-\x)$ and $W(\x-\x')=W(\x'-\x)$, we get 
 \beq\label{main}
\frac{\rmd {\cal D}_Q}{\rmd t}+i[H_Q,{\cal D}_Q(t)]&\approx&-\frac{i}{2}\int_{\x\x'} V(\x-\x')   [n_\x^a n^a_{\x'},{\cal D}_Q ],\nn
&+&\frac{1}{2}\int_{\x\x'} W(\x-\x')    \left( \{n^a_\x n^a_{\x'}, {\cal D}_Q\}-2 n^a_{\x} {\cal D}_Q n^a_{\x'}\right)\nn
&+&\frac{i}{4T}\int_{\x\x'}\, W(\x-\x')  \left(  [n^a_\x , \dot n^a_{\x'}{\cal D}_Q]  +[n^a_\x ,{\cal D}_Q\dot n^a_{\x'} ]   \right) \nn\eeq
Note that first line of the right hand side of this equation describes a hamiltonian evolution, that is, it can be written as the commutator in the left hand side, with $H_Q $ replaced by $\frac{1}{2}\int_{\x\x'} V(x-\x') n_\x^an_{\x'}^a$. It follows that we can shift the ``direct'', one-gluon exchange potential initially contained in $H_Q$ into $V$,  and keep in $H_Q$ only the kinetic energy of the heavy quarks. This is the strategy that was followed in 
\cite{Blaizot:2015hya} and that we shall adopt in this paper. In this way the potential $V(\r)$ becomes the screened Coulomb potential, and  $H_Q$ represents only the kinetic energy of the heavy particles (see also the discussion after Eq.~(\ref{phaserealVqed}) below).

The equation  (\ref{main}) is our main equation.  It is a fully quantum mechanical equation.  It is a Markovian equation for the reduced density matrix ${\cal D}_Q(t)$.    We shall write this equation in the following way
\beq\label{main2}
\frac{\rmd {\cal D}_Q}{\rmd t}={\cal L}\,{\cal D}_Q,
\eeq
with ${\cal L}={\cal L}_0+{\cal L}_1+{\cal L}_2+{\cal L}_3$, and
\beq\label{main3}
{\cal L}_0\,{\cal D}_Q&\equiv &-i[H_Q,{\cal D}_Q],\nn
{\cal L}_1\,{\cal D}_Q&\equiv& -\frac{i}{2}\int_{\x\x'} V(\x-\x')   [n^a_\x n^a_{\x'},{\cal D}_Q ],
\nn
{\cal L}_2\,{\cal D}_Q&\equiv&\frac{1}{2}\int_{\x\x'} W(\x-\x')    \left( \{n^a_\x n^a_{\x'}, {\cal D}_Q\}-2 n^a_{\x} {\cal D}_Q n^a_{\x'}\right),
\nn
{\cal L}_3\,{\cal D}_Q&\equiv&\frac{i}{4T}\int_{\x\x'}\, W(\x-\x')  \left(  [n^a_\x , \dot n^a_{\x'}{\cal D}_Q]  +[n^a_\x ,{\cal D}_Q\dot n^a_{\x'} ]   \right) .
\eeq
The structure of Eq.~(\ref{main2}) is close to that of a Lindblad equation \cite{Lindblad}, but Eq.~(\ref{main2})  is not quite a Lindblad equation: while the operator ${\cal L}_2$ can be put in the Lindblad form, this is not so of the operator ${\cal L}_3$, unless one adds extra, subleading terms (see the discussion in \ref{App:alternative}). For a recent discussion of the Lindblad equation for an abelian plasma, in a formalism not too different from the present one, see \cite{DeBoni:2017ocl}). 
 The notation is, at this stage, symbolic and just expresses the fact that the right hand side of Eq.~(\ref{main}) is a linear functional of the density matrix. It will acquire a more precise meaning as we proceed.  We may however make the following observation. When taking matrix elements between localized states, specified by the coordinates of the heavy particles, the density operators $n_\x$ play the role of projection operators, and are diagonal in the coordinate representation. Thus the same matrix elements, as far as the coordinates are concerned, will appear on the left and the right. The operator ${\cal L}$ will then appear as a differential operator acting on this matrix element (in fact ${\cal L}_1$ and ${\cal L}_2$ are simply multiplicative, as we shall see).

\begin{figure}[!hbt]
\begin{center}
\includegraphics[width=0.9\textwidth]{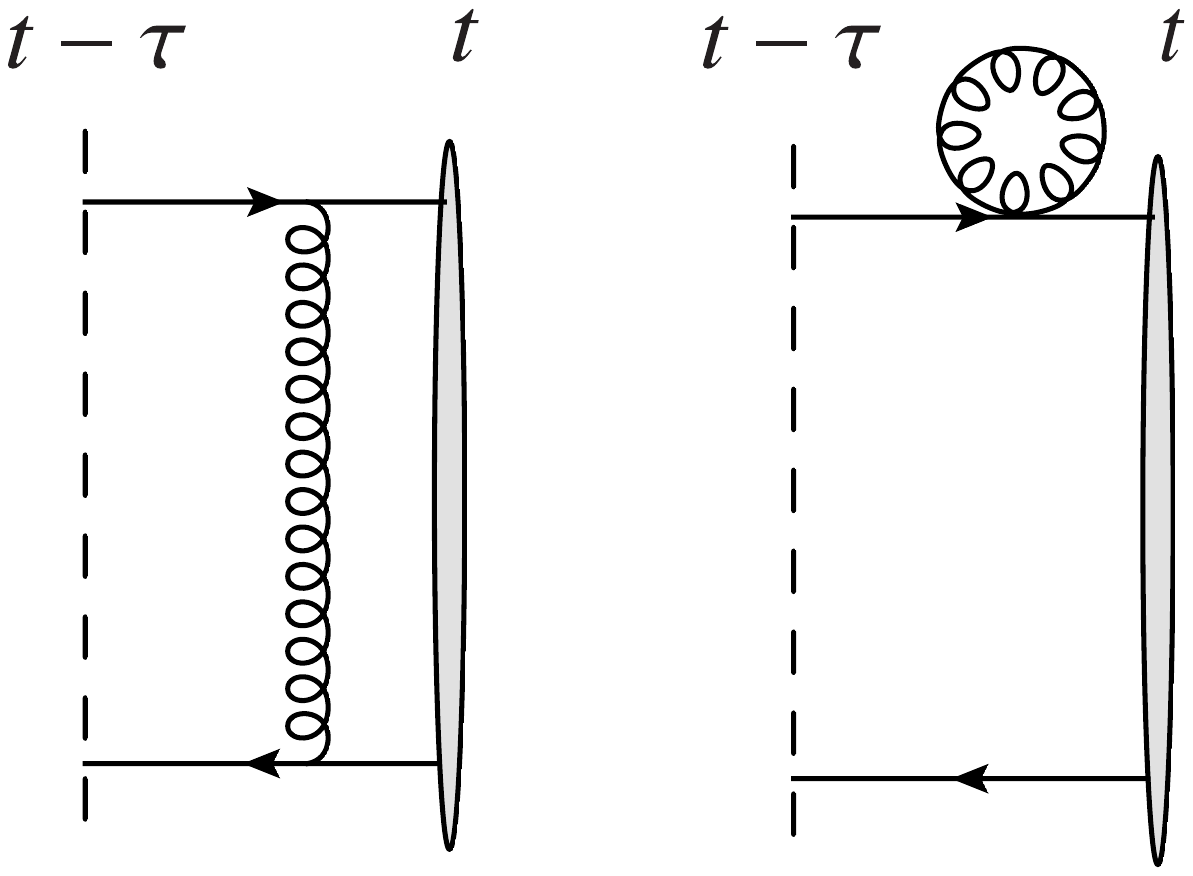}
\vspace{-3.5cm}
\caption{These diagrams illustrate  generic processes taken into account in Eq.~(\ref{main3}), in the case of the single particle density matrix. Note that there is another diagram with a tadpole insertion,  on the lower line, not drawn. Depending on the operator considered, the propagator of the gluon corresponds to $V$, $W$ or involves spatial derivatives of $W$. Note that since we treat the heavy quarks and antiquarks as non relativistic particles, the direction of the arrows in such a diagram does not refer to the nature (quark or antiquark) of the heavy particle: rather, it is correlated to the SK contour time, forward (to the right) in the upper branch, backward (to the left)  in the lower branch.} \vspace{-0.25in}
\label{fig:deriv0b}
\end{center}
\end{figure}

 It is convenient to associate a diagrammatic representation of the various contributions that we shall calculate. The relevant diagrams will preserve the topological structures of those already introduced, but because of the various approximations that we have performed, they cannot be calculated with standard rules. As an illustration, we display in Fig.~\ref{fig:deriv0b}  diagrams corresponding to the time derivative of the single particle density matrix (diagrams corresponding to the two particle density matrix are displayed in Fig.~\ref{fig:densitymatrix2} below). All interactions in Eq.~(\ref{main}) have become instantaneous. For this reason, we  draw these  as vertical gluon lines, or as tadpole insertions, located anywhere between $t-\tau$ and $t$.  Note that terms where the two densities sit close together in Eq.~(\ref{main2}), like in $[n^a_\x n^a_{\x'}, {\cal D}_Q]$, correspond to diagrams where the two ends of the gluon is hooked on the upper (or lower) part of the diagram, while a term such as $n^a_{\x} {\cal D}_Q n^a_{\x'}$ corresponds to a gluon joining the upper and lower parts of the diagram. Since, as we shall see, in QCD these two types of terms correspond also to different color structures,  we shall find convenient to split the operators ${\cal L}_i$ into two components, ${\cal L}_i={\cal L}_{ia}+{\cal L}_{ib}$, with for instance ${\cal L}_{2a}\propto \{n^a_\x n^a_{\x'}, {\cal D}_Q\}$ and ${\cal L}_{2b}\propto n^a_{\x} {\cal D}_Q n^a_{\x'}$. Note that ${\cal L}_1$ has only contributions of type $a$, i.e., ${\cal L}_1={\cal L}_{1a}$.\\
 
\vspace{-0.0cm}

In the rest of this paper, we shall deal only with the heavy quark reduced density matrix. We shall then drop the subscript $Q$ in order to simplify the notation and write simply ${\cal D}$ in place of ${\cal D}_Q$.

\section{Semi-classical approximation for abelian plasmas}
\label{sec:qed}
The equation (\ref{main}) is quite general. It holds for any system of heavy quarks and antiquarks. Depending on the system considered, the color density $n^a(\x)$ and the density matrix ${\cal D}$ take different forms. In this section, we study the specific form of Eq.~(\ref{main}), and the associated operators ${\cal L}_i$ in Eq.~(\ref{main3}), for the single particle and the two particle density matrices, in the case of an electromagnetic (abelian) plasma. For simplicity, we shall continue to refer to the charged particles as quarks (positive charge) and antiquarks (negative charge). The interaction hamiltonian reads as in Eq.~(\ref{H1}), with $n^a(\x)$ replaced by the density of charged particles.

Our goal here is twofold: i) This section is a preparation for the more complicated case of non abelian plasmas presented in the next section. Some of the results obtained here will indeed extend trivially to  QCD, to within multiplicative color factors.  ii) We wish to establish the relation with results obtained previously for the influence functional  obtained in the path integral formalism.  In particular, we shall show that we obtain, after a semi-classical approximation,  the same Fokker-Planck equations, and  the associated Langevin equations, as derived earlier in the path integral formalism in Ref.~\cite{Blaizot:2015hya}.

\subsection{Single particle density matrix}

In first quantization, the charge density $n(\x)$ of  a single heavy quark  is the  operator $n(\x)=\delta(\x-\hat\r)$,  with matrix elements 
\beq
\bra{\r}n(\x)\ket{\r'}=\delta(\x-\r)\delta(\r-\r').
\eeq
We also need the matrix elements of the time derivative of the density. These can be easily obtained from the continuity equation, $\dot n(\x)=-\bmnabla_\x \j(\x)$, where the matrix elements of the current $\j(\x)$ are given by 
\beq
\bra{\r}\j(\x)\ket{\r'}=\frac{1}{2iM} [\bmnabla_\r \delta(\r-\r')][\delta(\x-\r')+\delta(\x-\r)].
\eeq
One then gets
 \beq\label{divjx}
\bra{\r}\dot n(\x)\ket{\r'}=-\frac{1}{2iM}\left\{\left[\bmnabla_\r \delta(\r-\r')\right] \cdot\bmnabla_\x [\delta(\x-\r')+\delta(\x-\r)]\right\}.
  \eeq
\\

We can then proceed to the evaluation of the various contributions ${\cal L}_i$ in Eq.~(\ref{main3}), first in the case of the single particle density matrix. 
It is  easy to show that the first line of   Eq.~(\ref{main3}) yields
\beq
\bra{\r} {\cal L}_1{\cal D}\ket{\r'}=-\frac{i}{2} \int_{\x,\x'} V(\x-\x')\bra{\r}\left[n(\x)n(\x'), {\cal D}\right]\ket{\r'}=0.
\eeq
Thus, the real part of the potential does not contribute to the evolution of the single particle density matrix. In terms of diagrams, this results from the cancellation of the tadpole insertions in the upper and lower branches (see the second diagram in Fig.~\ref{fig:deriv0b}), which represent here (unphysical) self-interactions. \\

Taking the matrix element of the second line of Eq.~(\ref{main3}), one obtains
\beq
\bra{\r} {\cal L}_2{\cal D}\ket{\r'}=[W(0)-W(\r-\r') ]\bra{\r}{\cal D}\ket{\r'}=-\Gamma(\r-\r') \bra{\r}{\cal D}\ket{\r'},
\eeq
where we have set
\beq\label{eq:deGamma}
\Gamma(\r)\equiv W(\r)-W(0).
\eeq
This equation illustrates the role of the collisions, captured here by the imaginary part of the potential, in the phenomenon of decoherence (the damping of the off-diagonal matrix elements of the density matrix).  In contrast to what happens with the real part of the potential that we have just discussed,  in the present case the two tadpole contributions add up, instead of cancelling. They are in fact needed to properly define the damping rate $\Gamma$, and insure in particular that it cancels when $\r\to 0$, so that the density (the diagonal part of the density matrix) is not affected by the collisions. 

A useful estimate of $\Gamma(\r)$ is obtained in the Hard Thermal Loop approximation \cite{Pisarski:1988vd,Braaten:1989mz,Frenkel:1989br} which gives 
\beq
\Gamma(\r)=\alpha T \phi(m_D r),
\eeq  
where $T$ is the temperature, and $\phi(x)$ a monotonously increasing function such that $\phi(x=0)=0$ and $\phi(x\to \infty)=1$ \cite{Laine:2006ns}. The same formula holds in the case of QCD, with $\alpha$ replaced by $\alpha_s$, the strong coupling constant, and the multiplication by appropriate color factors (see Sect.~\ref{numerics}). In the limit of a large separation, $\Gamma(\r)\simeq 2\gamma_Q$, where $\gamma_Q=\alpha_sT/2$ is the so-called damping factor of a heavy quark (or antiquark) \cite{Braaten:1992gd}. At small separation, interferences cancel the effect of collisions: the heavy quark pair is seen then as a small electric dipole, i.e., an electrically neutral object on the scale of the  wavelengths of the typical modes of the plasma. Note that at large separation, the imaginary part of the potential itself vanishes, so that $W(0)=-2\gamma_Q$.  \\

Considering finally the third line of Eq.~(\ref{main3}) one gets\footnote{Here, and throughout this paper, we use the shorthand $\nabla W(0)$ for $\left.\nabla_x W(x)\right|_{x=0}$, and similarly for $\nabla^2 W(0)$.}
\beq\label{eqforrhoQEDM1}
\bra{\r} {\cal L}_3{\cal D}\ket{\r'}&=&\frac{1}{4MT}\left[ \nabla^2 W(0) -\nabla^2W(\r-\r') \right]\bra{\r}{\cal D}\ket{\r'}\nn
&-&\frac{1}{4MT} \nabla_\r W(\r-\r')\cdot (\nabla_\r -\nabla_{\r'}) \bra{\r}{\cal D}\ket{\r'}.
\eeq
The spatial derivatives originate from the time derivatives of the density (see Eq.~(\ref{divjx})), which involve the velocity of the heavy quark (hence the factor $1/M$). In fact, there is a close correspondence between ${\cal L}_3$ and ${\cal L}_2$. Observe indeed that ${\cal L}_3$ can be obtained from ${\cal L}_2$ by multiplying the latter by the overall factor $1/(4MT)$, and performing the following substitutions: $W(0)\to \nabla^2 W(0)$, $W(\r-\r')\to 
\nabla_\r W(\r-\r')\cdot (\nabla_\r -\nabla_{\r'})$. We shall see that analogous correspondences also exist in the more complicated case of the 2 particle density matrix. \\

At this point, we make the following change of variables
\beq \label{newvariables0}
{\bmcalR}=\frac{\r+\r'}{2},\quad \y=\r-\r',
\eeq
and set 
\beq
\bra{\r}{\cal D}(t)\ket{\r'}={\cal D}({\bmcalR},\y,t).
\eeq
The equation (\ref{main})  becomes then $\frac{\rmd }{\rmd t}{\cal D} ({\bmcalR},\y,t)={\cal L} {\cal D} ({\bmcalR},\y,t)$, with ${\cal L}$ appearing now explicitly as a  differential operator acting on the function ${\cal D}({\bmcalR},\y,t)$:
\beq\label{eqforrhoQEDM2}
{\cal L}=  \frac{i}{M}\nabla_{\bmcalR}\cdot\nabla_\y  -\Gamma(\y)+\frac{1}{4MT}\left[ \nabla^2 W(0) -\nabla^2_\y W(\y)-2\nabla_\y W(\y)\cdot\nabla_\y \right].
\eeq
The first term arises from the kinetic energy, i.e.,  it represents ${\cal L}_0$.
Note that the other terms, which represent the effect of the collisions, vanish for $y=0$, in particular thanks to the property $\nabla W(0)=0$. As already mentioned, this reflects the fact that the collisions do not change the local density of heavy quarks. 
\\

The equation (\ref{eqforrhoQEDM2}) above is   the explicit form of the operators ${\cal L}_i$ in Eq.~(\ref{main3}) for the density matrix of  a single heavy quark (in the abelian case). It has been obtained without any additional approximation beyond those leading to Eq.~(\ref{main3}).  We may proceed further and simplify Eq.~(\ref{eqforrhoQEDM2}) by performing a small $\y$ expansion. The variable $\y$ measures by how much the density matrix deviates from a diagonal matrix, a situation which is reached in the classical limit. Thus, the small $\y$ expansion may be viewed as a semi-classical expansion.  We have
\beq
W(\y)=W(0)+\frac{1}{2} \y\cdot {\cal H}(0) \cdot \y +\cdots
\eeq
where ${\cal H}(0)$ is the (positive definite) Hessian matrix of $W$,
\beq
{\cal H}_{ij}(\y)\equiv \frac{\del^2 W(\y)}{\del y_i\del y_j},
\eeq evaluated at $\y=0$, and we have used $\del_\y W(y)|_{\y=0}=0$. Note that if we stop the expansion of $W(\y)$ at quadratic order, $\nabla^2 W(0) -\nabla^2_\y W(\y)=0$.  The differential operator (\ref{eqforrhoQEDM2}) reads then
\beq
\label{eqforrhoQEDM4}
{\cal L}=  \frac{i}{M} \nabla_{\cal R}\cdot\nabla_\y -\frac{1}{2} \y\cdot {\cal H}(0) \cdot \y 
-\frac{1}{2MT}\y \cdot {\cal H}(0) \cdot\nabla_\y.
\eeq

At this point some comments on the order of magnitude of the various terms are appropriate. It is convenient to measure the time in terms of the inverse temperature, setting for instance $\tau=T\, t$. Dividing both members of the equation by $T$, on gets on the left hand side  $\del_\tau$, and  the operator ${\cal L}/T$ on the right hand side is dimensionless. We shall assume in this paper that the heavy particles are initially close to rest. In interacting with the medium they ultimately thermalize,  their velocity reaching values of order $\sqrt{T/M}$,  so that  $\nabla_{\cal R}\lesssim \sqrt{MT}$. The variable $\y$ measures the non locality of the density matrix. When the heavy quark is not too far from equilibrium, this non locality is of the order of the thermal wavelength, that is ${\cal D}({\cal R},\y,t)$ dies out when $y\gtrsim \lambda_{\rm th}\sim 1/\sqrt{MT}$. Thus in the first term, typically $\nabla_y\sim \sqrt{MT}$, so that  $\nabla_{\cal R}\cdot\nabla_\y\sim MT$. It follows that the term ${\cal L}_0/T$, where ${\cal L}_0$ represents the kinetic energy of the heavy quark,  is of order unity,  while the other two terms are both of order $\Gamma(\y)$. The range of variation of  $\Gamma(\y)$ is controlled by the Debye mass, i.e., it varies little on the scale of the thermal wavelength of the heavy particles. More precisely, using the HTL estimate  $\Gamma(\r)\approx \alpha T (m_D r)^2$, we get  $\Gamma(\y)/T\approx \alpha m_D^2/(MT)\ll 1$, the inequality following from our assumption $M\gg T$, and the fact that $m_D\lesssim T$ (in strict weak coupling $m_D^2\approx \alpha T^2$).   In summary, the ratio of the last two terms in Eq.~(\ref{eqforrhoQEDM4}) to the kinetic term is of order $\alpha m_D^2/(MT)\ll 1$, which justifies the semi-classical expansion. \\

To see better the physical content of Eq.~(\ref{eqforrhoQEDM4}), we  take its Wigner transform with respect to $\y$. We define, with a slight abuse of notation, 
\beq
{\cal D} ({\bmcalR},\p,t)=\int \rmd^3\y \,\rme^{-i\p\cdot\y}\, {\cal D} ({\bmcalR},\y,t),
\eeq
and obtain
\beq
\label{eqforrhoQEDM4W}
{\cal L}=  -\frac{\p}{M} \cdot \bmnabla_{\bmcalR} +\frac{1}{2} \bmnabla_\p\cdot {\cal H}(0) \cdot \bmnabla_\p 
+\frac{1}{2MT}\bmnabla_\p \cdot {\cal H}(0) \cdot\p .
\eeq
The corresponding equation for ${\cal D} ({\cal R},\p,t)$ may be interpreted as a Fokker-Planck equation. The second term in Eq.~(\ref{eqforrhoQEDM4W}), proportional to $\bmnabla_\p^2$ can be viewed as a diffusion term (in momentum space), and is associated with a noise term  in the corresponding Langevin equation (see below). It originates from the contribution ${\cal L}_2$. The last term, steming from the opeartor ${\cal L}_3$, can be associated with friction.  This can be made more transparent by introducing the following definitions
\beq\label{noisedef}
{\cal H}_{ij}(0)=\frac{1}{3} \nabla^2 W(0)\,\delta_{ij}\equiv \eta\,\delta_{ij},\qquad \eta=2\gamma T.
\eeq
Then we operator above yields the followwing Fokker-Planck equation 
\beq
\left( \frac{\del}{\del t}+\v\cdot\bmnabla_{\bmcalR}    \right) {\cal D} ({\bmcalR},\p,t)=\frac{1}{2}\eta \,\nabla_\p^2{\cal D} ({\bmcalR},\p,t)+\gamma \,\bmnabla_\p\cdot \left(\v \, {\cal D} ({\bmcalR},\p,t) \right), 
\eeq
where $\v\equiv\p/M$  is the velocity of the particle. 
It is easily shown that this equation can be obtained  from the following Langevin equation
\beq
M\ddot {\bmcalR} =-\gamma \dot {\bmcalR} +\bmxi(t), \qquad \langle \xi_i(t)\xi_j(t')\rangle=\eta\, \delta_{ij}\delta(t-t').
\eeq
The relation $\eta=2\gamma T$ between the diffusion constant $\eta$ and the fiction coefficient $\gamma$ can be viewed as an Einstein equation and expresses the fact that both noise and friction have the same origin, as can be made obvious by rewriting Eq.~(\ref{eqforrhoQEDM4W}) as follows
\beq
{\cal L}=  -\v \cdot \bmnabla_{\cal R} +\frac{1}{2} \bmnabla_\p\cdot {\cal H}(0) \cdot \left(  \bmnabla_\p +\frac{\v}{T}\right).
\eeq

\subsection{The two particle density matrix}

We consider now a heavy quark-antiquark pair. The charge density operator is   written as
\beq\label{rhoadef0qed2}
n(\x)=\delta(\x-\hat\r)\otimes \mathbb{I} -\mathbb{I}\otimes\delta(\x-\hat\r),
\eeq
where the first  term refers to the quark  and the second to the antiquark, the minus sign reflecting the fact that the antiquark has a charge opposite to that of the quark. 
The matrix elements of $n(\x)$ are given by
\beq
\bra{\r_1\r_2} n(\x)\ket{\r_1'\r_2'}=\delta(\r_1-\r_1')\delta(\r_2-\r_2')\left[\delta(\x-\r_1)-\delta(\x-\r_2)\right].
\eeq
Similarly, the matrix elements of the time derivative of the density are given by 
\beq
 &&\bra{\r_1,\r_2}\dot n(\x)\ket{\r_3,\r_4}\nn
 &&\qquad=-\frac{1}{2iM} [\bmnabla_{\r_1}\delta(\r_{13})]\cdot \bmnabla_\x[\delta(\x-\r_3)+\delta(\x-\r_1)] \delta(\r_{24})\nn
  &&\qquad\quad+\frac{1}{2iM} [\bmnabla_{\r_2}\delta(\r_{24})]\cdot \bmnabla_\x[\delta(\x-\r_4)+\delta(\x-\r_2)]\delta(\r_{13}),
 \eeq
which is easily obtained from   Eq.~(\ref{divjx}). Note that we have introduced here a short notation, $\r_{ij}\equiv \r_i-\r_j$, that will  be used often in the following. We shall also occasionally write $\nabla_1$ for $\nabla_{\r_1}$, and introduce similar other shorthands,  in order to reduce the size of some formulae. 

\vspace{-0.cm}
\begin{figure}[!hbt]
\begin{center}
\includegraphics[width=0.8\textwidth]{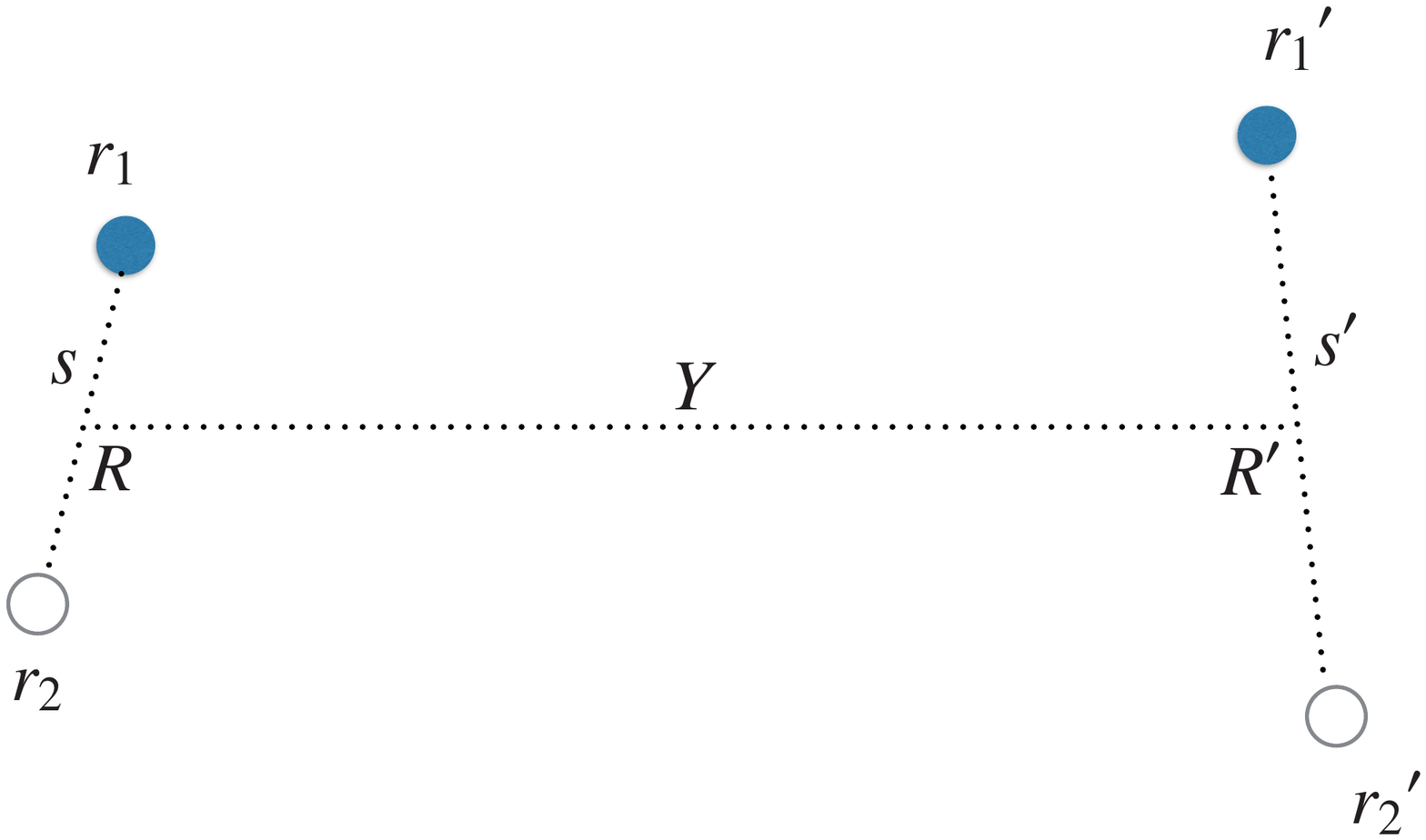}
\vspace{-2.5cm}
\caption{The various coordinates that are used in the evaluation of the two particle density matrix elements $\bra{\r_1\r_2} {\cal D}_Q\ket{\r_1'\r_2'}$.} \vspace{-0.25in}
\label{fig:coordinates}
\end{center}
\end{figure}
\vspace{+0.5cm}

It will be also convenient at a later stage to change variables. Thus, we define the center of mass and relative coordinates, 
\beq\label{changeofvar1}
\R=\frac{\r_1+\r_2}{2},\quad \s=\r_1-\r_2,\quad \R'=\frac{\r_1'+\r_2'}{2},\quad \s'=\r_1'-\r_2',
\eeq
as well as the set of coordinates that generalize those introduced in Eq.~(\ref{newvariables0}) for the single particle case, 
\beq\label{changeofvar5}
{\bmcalR}=\frac{\R+\R'}{2},\qquad \Y=\R-\R',\qquad \y=\s-\s',\qquad \r=\frac{\s+\s'}{2}.
\eeq
The latter are useful to derive the semi-classical approximation. In this approximation,  $Y\to 0, y\to 0$, and ${\bmcalR}$ and $\r$ become respectively  the center of mass and the relative coordinates. These various coordinates are illustrated in Fig.~\ref{fig:coordinates}.\\

We now turn to the specific evaluation of the matrix elements of Eq.~(\ref{main}) in the case of a quark-antiquark pair. Consider first the matrix element of the free evolution, governed by the hamiltonian 
\beq
H_Q=\frac{p_1^2}{2M}+\frac{p_2^2}{2M}.
\eeq
We have 
\beq
-i\bra{\r_1 \r_2} [H_Q,{\cal D}]\ket{\r_1'\r_2'}=i\left(\frac{\nabla_{\cal R}\cdot \nabla_\Y}{2M}+\frac{2\nabla_\r\cdot\nabla_\y}{M}  \right) \bra{\r_1 \r_2} {\cal D}\ket{\r_1'\r_2'},
\eeq
that is
\beq
{\cal L}_0=i\left(\frac{\nabla_{\cal R}\cdot \nabla_\Y}{2M}+\frac{2\nabla_\r\cdot\nabla_\y}{M}  \right) .
\eeq

Turning now to the operator ${\cal L}_1$, a simple calculation yields 
\beq\label{phaserealVqed}
\bra{\r_1\r_2}{\cal L}_1\,{\cal D}\ket{\r_1'\r_2'}&=&i [V(\r_{12})-V(\r_{1'2'})]\bra{\r_1\r_2}{\cal D}\ket{\r_1'\r_2'}.
\eeq
Note the cancellation of the self-interaction terms, as was the case for the single particle density matrix. 
The real part of the potential produces just a phase in the evolution of the density matrix. This can be understood as a hamiltonian evolution, ${\cal L}\,{\cal D}=-i  [H,{\cal D}]$, with here $H\to -V$, the minus sign resulting from the fact that the two interacting heavy particles have opposite charges. As we have mentioned earlier (see the discussion after Eq.~(\ref{main})), the structure of the equation makes it possible to include in the potential $V$ both the ``direct'' interaction between the heavy quarks, by which we mean the interaction that exists in the absence of the plasma, as well as the ``induced'' interaction that results from the interaction of the heavy quarks with the plasma constituents. The latter is responsible for the screening phenomenon. In the HTL approximation, we have 
\beq
V(r)=\alpha m_D +\alpha \frac{e^{-m_Dr}}{r},
\eeq
where the first term cancels the constant contribution hidden in the screened Coulomb potential (the second term) at short distance. Thus as  $r\to 0$, $V(r)$ reduces to the Coulomb potential, $V(r)\sim \alpha/r$. Note that the function $V(\r)$ thus defined corresponds to the interaction potential of two particles with identical charges. 
 \\
\vspace{-0.0cm}
\begin{figure}[!hbt]
\begin{center}
\includegraphics[width=1.0\textwidth]{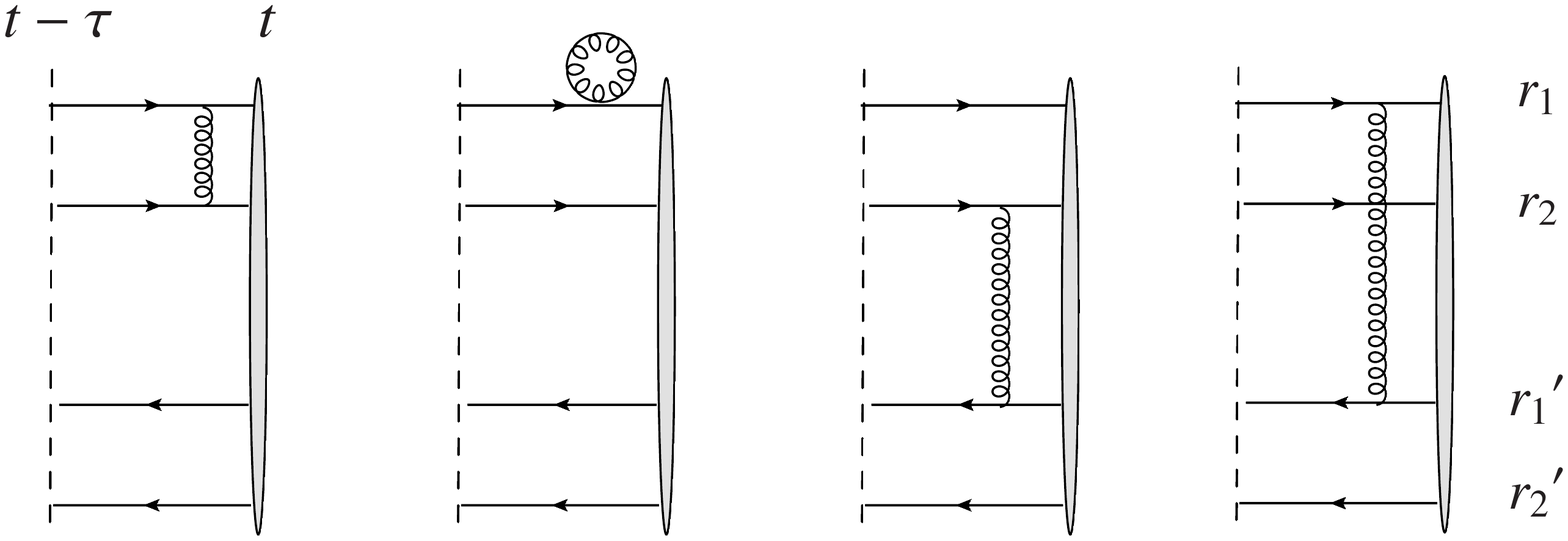}
\vspace{-3cm}
\caption{Graphical illustrations for typical contributions to the operators ${\cal L}_i$ for the two-particle density matrix. In the first two diagrams, the gluon line represents either $V$ or $W$, while in the last two, only $W$ and its spatial derivatives are involved (the hamiltonian evolution, involving the real part of the potential, does not connect the upper and lower parts of the diagrams). In the last diagram, the gluon line connect two particles with the same charge, and contribute to the quantity called $W_a$. In the third diagram, the gluon line connect two particles with opposite charges, and contributes to the quantity called $W_a$. When $W$ is involved in the first diagram, it represents a contribution to $W_c$, and finally the tadpole insertion in the second diagram is associated to $V(0)$, or to $W(0)$ and its spatial drivatives. The defiitions of $W_{a,b,c}$ are given in Eq.~(\ref{Wabcdef}). }  \vspace{-0.0cm}
\label{fig:densitymatrix2}
\end{center}
\end{figure}

By taking the matrix element of the second line of Eq.~(\ref{main3}), we obtain 
\beq\label{eqWlinear}
\bra{\r_1\r_2}{\cal L}_2\,{\cal D}\ket{\r_1'\r_2'}
= [2W(0)-W_{12}-W_{1'2'}- W^-]\bra{\r_1\r_2}{\cal D}\ket{\r_1'\r_2'}.
 \eeq
 The various terms in this expression correspond to the various ways the exchanged gluon can be hooked on the upper and lower lines. To simplify the bookkeeping of the various contributions, and the writing of the equations, we define the following quantities, which will often appear in forthcoming formulae:
\beq\label{Wabcdef}
&&W_a\equiv W_{11'}+W_{22'}, \qquad W_b\equiv W_{21'}+W_{12'}\nn
&&W_c\equiv W_{12}+W_{1'2'},\qquad W^\pm \equiv W_a\pm W_b
\eeq
  These quantities correspond to the diagrams in Fig.~\ref{fig:densitymatrix2}, where $W$ plays the role of the propagator: $W_a$ connects particles with the same charge in the bra and the ket, while $W_b$ connect particles with opposite charges; $W_c$ connects particles within the bra, or within the ket. In the infinite mass limit, $\r_1=\r_1'$, $\r_2=\r_2'$ and $W_a=2W(0)$, $W_b=W_c=2W(\r)$, and $W_-(\r)=-2\Gamma(\r)$.
  
Note that $2W(0)-W_{12}-W_{1'2'}=-\Gamma_{12}-\Gamma_{1'2'}$, where $\Gamma(\r)$ is defined in Eq.~(\ref{eq:deGamma}). As was  the case for the single particle density matrix, the collisions tend to equalize the coordinates (here the relative coordinates) in the ket and in the bra, bringing the density matrix to the diagonal form. The structure of the entire damping factor takes actually a form more complicated than in the case of the single particle density matrix.   The combination of terms in the right hand side of Eq.~(\ref{eqWlinear}) can indeed be written
\beq
2W(0)-W_{12}-W_{1'2'}- W^-= -\left( \Gamma_{12}+ \Gamma_{1'2'}+ \Gamma_{11'}+ \Gamma_{22'}- \Gamma_{12'}- \Gamma_{21'}  \right).
\eeq
Note that the entire contribution vanishes when $\r_1'=\r_1$ and $\r_2'=\r_2$: $\Gamma_{11'}\to \Gamma(0)=0$, and similarly for $\Gamma_{22'}$ while the other terms mutually cancel. This is of course related to  the fact that the collisions do not change the local density of heavy particles, as we have already discussed.  
For future reference, we write ${\cal L}_2$ as a sum of two contributions (as explained at the end of Sect.~\ref{sec:quantum}), and write 
\beq\label{L2aQED}
 {\cal L}_{2a}^{\rm QED}=2W(0)-W_c, 
 \qquad {\cal L}_{2b}^{\rm QED}=-W^- .
 \eeq
The diagonal elements ($\r_1'=\r_1,\r_2'=\r_2$) of  ${\cal L}_{2a}$ and ${\cal L}_{2b}$ mutually cancel, as we have seen. 
 \\

Finally,  we turn to the $1/M$ corrections, which are more involved.  
The calculations are straightforward, but  lengthy. However, as we shall see, the results are simply related to those obtained for ${\cal L}_2$. 
Again, we split ${\cal L}_3$ into two contributions, ${\cal L}_3={\cal L}_{3a}+{\cal L}_{3b}$. We obtain
\beq\label{QEDanticom}
{\cal L}_{3a}^{\rm QED}&=&-\frac{i}{8T}\int_{\x\x'}\, W(\x-\x')  \left(  2{\cal D}\dot n_{\x'} n_\x   -2n_\x \dot n_{\x'}{\cal D}   \right)\nn
&=&\frac{1}{4MT}\,[2\bmnabla^2 W(0)-\nabla^2 W_c -\bmnabla W_c\cdot \bmnabla_c]  ,
\eeq
where we have used $\nabla W(0)=0$, and introduced the following shorthand notation
\beq
\bmnabla W_c\cdot \bmnabla_c\equiv \bmnabla_{1} W_{12}\cdot \bmnabla_{12}+ \bmnabla_{1'} W_{1'2'}\cdot \bmnabla_{1'2'}.
\eeq
with analogous definitions for $W_a$, $W_b$, $W_\pm$ (to be used later). In this formula, 
\beq
 \bmnabla_{12}\equiv \bmnabla_1-\bmnabla_{2},
 \eeq
 and recall that $\bmnabla_1$ stands for $\bmnabla_{\r_1}$ and $W_{11'}$ for $W(\r_1-\r_{1'})$.

The second contribution to ${\cal L}_3$ reads
\beq\label{QEDndotn1}
{\cal L}_{3b}^{\rm QED}&=& -\frac{i}{4T}\int_{\x,\x'}  W(\x-\x')\,\left[   \dot n(\x) {\cal D} n(\x')-n(\x){\cal D}\dot n(\x') \right]\nn
&=&-\frac{1}{4MT}\left\{ \nabla^2 W_a  +\bmnabla W_a\cdot \bmnabla_a-\nabla^2 W_b  -\bmnabla W_b\cdot \bmnabla_b\right\}\nn
&=&-\frac{1}{4MT}\left\{ \nabla^2 W^-  +\bmnabla W^-\cdot \bmnabla^-\right\}.
\eeq
Note the analogy between Eqs.~(\ref{L2aQED}) and Eqs.~(\ref{QEDanticom}) and (\ref{QEDndotn1}). The latter follow from the former after the replacements $W(0)\to \nabla^2 W(0)$, $W_c\to \nabla^2 W_c +\bmnabla W_c\cdot \bmnabla_c$, $W_-\to  \nabla^2 W_- +\bmnabla W_-\cdot \bmnabla_-$, and the multiplication by the overall factor $1/(4MT)$. This correspondence is analogous to that already observed after Eq.~(\ref{eqforrhoQEDM1}), and its origin is the same. 
\\

Until now, the expressions that we have obtained are an exact transcription of the operators ${\cal L}_i$ in Eq.~(\ref{main3}) to the case of a (abelian) quark-antiquark pair.. 
At this point it is useful to go to the coordinates (\ref{changeofvar5}), i.e., $\bra{\r_1\r_2}{\cal D}\ket{\r_1'\r_2'}\to {\cal D}({\cal R},\Y,\r,\y)$,  and perform a semi-classical expansion similar to that which leads to Eq.~(\ref{eqforrhoQEDM4}) for the single particle density matrix. We obtain then 
\beq
\label{eqforrhoQEDM4b}
&&\frac{\rmd }{\rmd t}{\cal D}({\cal R},\Y,\r,\y)=[{\cal L}_0+{\cal L}_1+{\cal L}_2+{\cal L}_3]\,{\cal D}({\cal R},\Y,\r,\y),
\eeq
where 
 \beq\label{L123QEDa}
&&{\cal L}_{1}\approx i \y\cdot\bmnabla V(\r),\nn
&&{\cal L}_{2}\approx \Y\cdot({\cal H}(\r)-{\cal H}(0))\cdot \Y-\frac{1}{4} \y\left( {\cal H}(\r)+{\cal H}(0)\right)\y,\nn
&&{\cal L}_{3}\approx-\frac{1}{2MT}\left\{   \Y\cdot\left({\cal H}(0)-{\cal H}(\r)  \right)\cdot \nabla_\Y
+ \y\cdot \left( {\cal H}(0)+  {\cal H}(\r)\right) \cdot\nabla_\y     \right\}. \nn
   \eeq
 After performing the  Wigner transform with respect to the variables $\Y$ and $\y$, 
\beq
{\cal D}({\cal R},\P, \r,\p)=\int_\Y\int_\y \,\rme^{-i\P\cdot\Y}  \rme^{-i\p\cdot\y}\, {\cal D}({\cal R},\Y,\r,\y),
\eeq
we obtain
\beq
&&{\cal L}_0=-\left(\frac{\P\cdot\bmnabla_{\cal R}}{4M}+\frac{2\p\cdot\bmnabla_\r}{M}  \right),\nn
&& {\cal L}_1=-\bmnabla V(\r)\cdot \bmnabla_\p ,\nn
&&{\cal L}_2=\left[\bmnabla_\P\cdot ({\cal H}(0)-{\cal H}(\r))\cdot\bmnabla_\P+\frac{1}{4} \nabla_\p\cdot \left( {\cal H}(\r)+{\cal H}(0)\right)\cdot \nabla_\p\right] ,\nn
&&{\cal L}_3=\frac{1}{2MT} \bmnabla_\P\cdot ({\cal H}(0)-{\cal H}(\r))\cdot \P+\frac{1}{2MT}\bmnabla_\p\cdot\left[ {\cal H}(\r)+{\cal H}(0)\right] \cdot \p. \nn
\eeq 

 We note that the operators for the relative coordinates are independent of the center of mass coordinates. It is  then easy to identify  the operators for the relative coordinates, and determine the elements of the corresponding Langevin equation. The relative velocity is given by $2\p/M=\p/(M/2)$. Thus 
\beq\label{FPrel0}
{\cal L}_0^{\rm rel}=-\frac{2\p\cdot\bmnabla_\r}{M}=-\v\cdot\bmnabla_\r.
\eeq
Similarly,
\beq\label{FPrel1}
{\cal L}_1^{\rm rel}=-\bmnabla V(\r)\cdot \bmnabla_\p=-\frac{2}{M}\bmnabla V(\r)\cdot \bmnabla_\v.
\eeq
The term ${\cal L}_2$ corresponds to the noise term. We write it as 
\beq\label{FPrel2}
{\cal L}_2^{\rm rel}= \frac{1}{4} \nabla_\p\cdot \left( {\cal H}(\r)+{\cal H}(0)\right)\cdot \nabla_\p =\frac{2}{M^2}\eta_{ij}(\r) \nabla_\v^i \nabla_\v^j,
\eeq
with 
\beq
 \eta_{ij}(\r)=\frac{1}{2} \left( {\cal H}_{ij}(\r)+{\cal H}_{ij}(0)\right).
 \eeq
Finally ${\cal L}_3$ corresponds to the friction, and we write it as
\beq\label{FPrel3}
{\cal L}_{3}^{\rm rel}=\frac{1}{2MT}({\cal H}_{ij}(0)+{\cal H}_{ij}(\r)) \nabla^i_\p \p^j=\frac{2}{M}\gamma_{ij}(\r)\nabla_\v^i v^j.
\eeq
Friction and noise are related by an Einstein relation
\beq
\gamma_{ij}(\r)=\frac{1}{2T} \eta_{ij}(\r).
\eeq
The Langevin equation associated with the relative motion is then of the form
\beq\label{langevinrelative}
\frac{M}{2} \ddot \r^i=-\gamma_{ij} \v^j-\bmnabla^i V(\r) +\xi^i(\r,t),\quad \langle\xi^i(\r,t)\xi^i(\r,t')\rangle=\eta_{ij}(\r)\delta(t-t').
\eeq
Note that for an isotropic plasma, we have (cf. Eq.~(\ref{noisedef}))
\beq
\eta_{ij}(\r)=\delta_{ij}\eta(\r),\qquad \eta(\r)=\frac{1}{6}\left(\nabla^2 W(0)+\nabla^2 W(\r)  \right).
\eeq

One can repeat the same for the center of mass coordinates. Here we set $\v=\P/(2M)$. We get
\beq
&&{\cal L}_0^{\rm CM}=-\frac{\P\cdot\bmnabla_{\cal R}}{2M} = -\v\cdot\bmnabla_{\cal R}\nn
&&{\cal L}_1^{\rm CM}=0\nn
&&{\cal L}_2^{\rm CM}= \nabla_\P\cdot ({\cal H}(0)-{\cal H}(\r))\cdot\nabla_\P=\frac{1}{8 M^2} \eta_{ij}(\r)\nabla_\v^i \nabla_\v^j \nn
&&{\cal L}_{3}^{\rm CM}=\frac{1}{2MT}({\cal H}_{ij}(0)-{\cal H}_{ij}(\r))\nabla_\P^i P^j=\frac{1}{2M}\gamma_{ij}(\r)\nabla_\v^i \v^j, \nn
\eeq
with 
\beq
\eta_{ij}(\r)=2 \left( {\cal H}_{ij}(0)-{\cal H}_{ij}(\r)\right)
\eeq
and the Einstein relation
\beq
\gamma_{ij}(\r)=\frac{1}{2T}\eta_{ij}(\r).
\eeq
The Langevin equation associated with the center of mass motion is then
\beq\label{langevincm}
2M\ddot {\cal R}^i=-\gamma_{ij} \v^j+\xi^i(\r,t)
\eeq

These two equations (\ref{langevinrelative}) and (\ref{langevincm}) are identical to  those obtained in \cite{Blaizot:2015hya} (see Eqs.~(4.69) there). Note that  while the Langevin equation for the relative motion does not depend on the center of mass, this is not so for the Langevin equation describing the center of mass motion, which depends on the relative coordinate: this is because, as we have already emphasized, the effect of the collisions on the system depends on its size, with small size dipoles being little affected by the typical plasma field fluctuations.

\section{QCD}
\label{sec:qcd}

We turn now to QCD. Much of the calculations are similar to those of the QED case, with however the obvious additional complications related to the color algebra. As we did in QED, we shall consider successively the case of the single particle density matrix, and that of a quark-antiquark pair.

\subsection{Single quark density matrix}

For a single quark, the color charge density can be written as (see Eq.~(\ref{colordensity}))
\beq\label{rhoadef0a1} 
n^a(\x)=\delta(\x-\hat\r)\, t^a,
\eeq
with matrix elements
\beq
\bra{\r,\alpha}n^a(\x)\ket{\r',\alpha'}=\delta(\r-\r')n(\r)\bra{\alpha}t^a\ket{\alpha'},
\eeq
where $n(\r)$ is the density of heavy quarks, that is, the number of heavy quarks located at point $\r$ irrespective of their color state. 

The reduced density matrix of a single quark can be written as follows (see ~\ref{sect:densitymatrix})
\beq\label{Dcolor1}
{\cal D}=D_0\, \mathbb{I}+ D_1\cdot\bmt .
\eeq
 It depends on 9 real parameters, and contains a scalar as well as a vector (octet) contributions. In fact, since we assume the plasma to be color neutral, we need consider only the scalar part of the density matrix (see however ~\ref{ap:singlequark}), that is the quantity $\bra{\r}D_0\ket{\r'}$ . 

With ${\cal D}$ having only a scalar component, i.e., ${\cal D}=D_0\, \mathbb{I}$, one can perform immediately the sum over the color indices in 
Eq.~(\ref{main}), using 
\beq\label{CasimirCF}
t^a t^a=C_F=\frac{N_c^2-1}{2N_c}.
\eeq
The result is then identical to that obtained in QED, to within the multiplicative factor $C_F$: there is no specific effect of the color degree of freedom on the color singlet component of the density matrix, aside from this multiplicative color factor. The resulting Fokker-Planck equation is then essentially identical to that first derived by Svetitsky long ago \cite{Svetitsky:1987gq}, which has been used in numerous phenomenological applications. 

\subsection{Density matrix of a quark-antiquark pair}

The color density of a quark-antiquark pair is given by Eq.~(\ref{colordensity}). The color structure of the reduced density matrix ${\cal D}$ is discussed in ~\ref{sect:densitymatrix}. We shall use two convenient representations. In the first one, to which we refer as the $(D_0, D_8)$ basis, the density matrix takes the form 
\beq
{\cal D}=D_0 \,\mathbb{I}\otimes \mathbb{I} +D_8 \, t^a\otimes \tilde t^a,
\label{eq:qaqdcomp}
\eeq
where $D_0$ and $D_8$ are matrices in coordinate space (product of coordinate spaces of the quark and the antiquark), e.g. $\bra{\r_1,\r_2}D_0\ket{\r_1',\r_2'}$, and similarly for $D_8$. The second representation is  in terms of components  $D_{\rm s}$ and $D_{\rm o}$ defined by  (see ~\ref{sect:densitymatrix})
\beq
{\cal D}=D_{\rm s} \ket{\rm s}\bra{\rm s}+D_{\rm o} \sum_c\ket{{\rm o}^c}\bra{{\rm o}^c}
\eeq
where $\ket{\rm s}\bra{\rm s}$ denotes a projector on a color singlet state, and $\ket{{\rm o}^c}\bra{{\rm o}^c}$ a projector on a color octet state with given projection $c$. We shall refer to this basis as the ($D_{\rm s},D_{\rm o}$) basis, or as the  singlet-octet basis. The relation between the two basis is given by the following equations
\beq
&&D_{\rm s}=D_0+C_F D_8,\qquad D_{\rm o}=D_0-\frac{1}{2N_c} D_8\nn
&&D_0=\frac{1}{N_c^2}(D_{\rm s}+(N_c^2-1) D_{\rm o}),\qquad D_8=\frac{2}{N_c}(D_{\rm s}-D_{\rm o}).
\eeq
The advantage of the singlet-octet basis is that it involves states of the quark-antiquark pair with well defined color (singlet or octet), which is not the case in the $(D_0,D_8)$ basis. The latter will play a role when we address the issue of equilibration of color. Then the matrix $D_0$, which represents a completely unpolarized color state, or a maximum (color) entropy state, plays an essential role. 

Because of the existence of two independent functions of the coordinates to describe the density matrix of the quark-antiquark pair, the equation of motion for ${\cal D}$ takes the form of a matrix equation
\beq
\frac{\del}{\del t}{\cal D}={\cal L}\,{\cal D},
\eeq
where  ${\cal D} $ can be viewed as a two dimensional vector, e.g. 
\beq
{\cal D}=\left(
\begin{array}{c}
 D_0   \\
  D_8 
\end{array}
\right),\qquad {\rm or}\qquad {\cal D}=\left(
\begin{array}{c}
 D_{\rm s}  \\
  D_{\rm o} 
\end{array}
\right),
\eeq
and  ${\cal L}$ is a $2\times 2$ matrix in the corresponding  space. 

The derivation of the equations for the various components of the reduced density matrix in the two basis is straightforward, but lengthy. The results are listed in ~\ref{ap:equations}. In this section we shall give brief indications on how to obtain the equations in the singlet-octet basis: the color algebra is then transparent, and most of the equations can be related to the corresponding ones of the abelian case. 

 \subsection{The equations in the singlet-octet basis}\label{sec:qcd-pair}
 
\noindent{\bf The contribution to ${\cal L}_1$.  }

When written in terms of $D_{\rm s}$ and $D_{\rm o}$ the two equations decouple. This is because the product $n^a_\x n^a_{\x'} $ is a scalar in color space, and hence does not change the color state of the pair. In other terms, the one-gluon exchange does not change the color state (singlet or octet) of the heavy quark-antiquark pair. The color algebra is then trivial, and yields
\beq\label{DsDoForce}
&&\frac{\rm d D_{\rm s}}{\rmd t}=iC_F  [V_{12}-V_{1'2'}]\,D_{\rm s}\nn
&&\frac{\rmd D_{\rm o}}{\rmd t}=-\frac{i}{2N_c}[V_{12}-V_{1'2'}]D_{\rm o} .
\eeq
The diagrams contributing here are the first two diagrams in Fig.~\ref{fig:densitymatrix2}, and the equations above are analog to that obtained in QED. In fact we have 
\beq
{\cal L}_1^{\rm ss}=C_F {\cal L}_1^{\rm QED},\qquad {\cal L}_1^{\rm oo}=-\frac{1}{2N_c} {\cal L}_1^{\rm QED},
\eeq
where ${\cal L}_1^{\rm QED}$ is given by Eq.~(\ref{phaserealVqed}).  Note the well known property that the interaction is attractive (and proportional to $C_F$) in the singlet channel, and repulsive (and proportional to $1/(2N_c)$) in the octet channel. \\

\noindent{\bf The contribution to ${\cal L}_2$.}
 
 In this case we have two types of contributions. The first ones involve products of the color charges, making up a color scalar. These contribute to  ${\cal L}_{2a}$, which is diagonal in color. The second type of contribution involves  transitions from singlet to octet or from octet to octet. We denote these contributions by ${\cal L}_{2b}$.
 
 Consider first ${\cal L}_{2a}$. We have, for the singlet
\beq\label{eqforrhoA4css}
{\cal L}_{2a}^{\rm ss} =C_F [2W(0)-W_c]D_{\rm s}=C_F{\cal L}_{2a}^{\rm QED} .
\eeq
with  $ {\cal L}_{2a}^{\rm QED}$ given in Eqs.~(\ref{L2aQED}). For the octet
\beq\label{eqforrhoA4coo}
{\cal L}^{\rm oo}_{2a}= 2C_FW(0)+\frac{1}{2N_c}W_c.
\eeq

Consider next ${\cal L}_{2b}$. The corresponding contributions involve transitions from singlet to octet intermediate states, or, when considering the octet density matrix, transitions from octet to singlet  and also octet to octet transitions that generate an additional diagonal contribution.  We have
\beq\label{L2bsoQCD}
{\cal L}_{2b}^{\rm so}=-C_F W^- =C_F\, {\cal L}_{2b}^{\rm QED} ,\qquad {\cal L}_{2b}^{\rm os}=-\frac{1}{2N_c} W^-=\frac{1}{2N_c} {\cal L}_2^{\rm QED} ,
\eeq
where $ {\cal L}_{2b}^{\rm QED}$ is given in Eqs.~(\ref{L2aQED}).
Similarly, for the octet to octet transition, we have
\beq \label{L2booQCD}
{\cal L}_{2b}^{\rm oo}=-\left(\frac{N_c^2-4}{4N_c}W^- +\frac{N_c}{4} W^+ \right)=-\left( \frac{N_c^2-2}{2N_c}W_a+\frac{1}{N_c} W_b  \right),
\eeq
which has no counterpart in QED.  \\

\noindent{\bf The contribution to ${\cal L}_3$.}

The calculation of these contributions is more involved, but we can relate simply the results to those obtained earlier for the operators ${\cal L}_2$. Let us first list the results. We have
\beq
&&{\cal L}_{3a}^{\rm ss}=\frac{C_F}{4MT}\,[2\bmnabla^2 W(0)-\nabla^2 W_c- \bmnabla W_c\cdot \bmnabla_c ]=C_F {\cal L}_{3a}^{\rm QED},\\
&&{\cal L}_{3a}^{\rm oo}=\frac{C_F}{2MT}\,[\bmnabla^2 W(0)] +\frac{1}{4MT}\frac{1}{2N_c}\left(\nabla^2 W_c+ \bmnabla W_c\cdot \bmnabla_c \right),\\
&&
{\cal L}_{3b}^{\rm so}=C_F\,{\cal L}_{3b}^{\rm QED},\qquad {\cal L}_{3b}^{\rm os}=\frac{1}{2N_c}\,{\cal L}_{3b}^{\rm QED},
\eeq
with 
\beq\label{L3bQED2}
{\cal L}_{3b}^{\rm QED}= -\frac{1}{4MT}\left\{\bmnabla^2 W^- +\bmnabla W^-\cdot \bmnabla^- \right\}, 
\eeq
and 
\beq
{\cal L}_{3b}^{\rm oo} =-\frac{1}{4MT}\left\{\frac{N_c^2-4}{4N_c}\left[\nabla^2 W^- +\bmnabla W^-\cdot \nabla^-  \right] + \frac{N_c}{4}  \left[\nabla^2 W^+ +\bmnabla W^+\cdot \nabla^+  \right] \right\}\nn
\eeq
It is easy to verify that  the equations giving ${\cal L}_2$ and the corresponding equations for ${\cal L}_3$  are related via the same substitutions that are discussed after Eq.~(\ref{QEDndotn1}).

\subsubsection{Semiclassical approximation}

The formulae listed in the previous subsection are an exact transcription of the main equation, Eq.~(\ref{main}) for a quark-antiquark pair. Analogous formulae can be written for a pair of quarks. They are given in ~\ref{sec:HQpair}. We shall be mostly concerned in this paper, in particular for the numerical studies presented in the next section, by the semi-classical limits of these equations. These can be obtained easily by using the formulae given in  ~\ref{ap:semiclassical}. In this subsection,  we just reproduce the few equations that will be used in the next section. 

We consider first the equation for the component $D_{\rm s}$, which we write as 
\beq
\frac{\rmd D_{\rm s}}{\rmd  t}=(D_{\rm  s}|{\cal L}|{\cal D}). 
\eeq
We obtain
\beq\label{eqfinDs}
&&(D_{\rm s}|{\cal L}|{\cal D})= \left( 2i \frac{\nabla_\r\cdot \nabla_\y}{M}+ i \frac{\nabla_{\cal R}\cdot \nabla_\Y}{2M} +iC_F \y\cdot \bmnabla V(\r)\right) D_{\rm s}\nn
&&\qquad\qquad-2C_F \Gamma(\r) (D_{\rm s}-D_{\rm o})\nn
&&\qquad\qquad-\frac{C_F}{4}\left(\y\cdot {\cal H}(\r)\cdot\y\, D_{\rm s}+\y\cdot {\cal H}(0)\cdot\y \,D_{\rm o} \right)\nn
&&\qquad\qquad-C_F \Y\cdot [{\cal H}(0)-{\cal H}(\r)]\cdot \Y D_{\rm o}\nn
&&\qquad\qquad+\frac{C_F}{2MT}\left[  \nabla^2 W(0)-\nabla^2 W(\r)-\bmnabla W(\r)\cdot\bmnabla_\r \right](D_{\rm s}-D_{\rm o})\nn
&&\qquad\qquad-\frac{C_F}{2MT}\left(\y\cdot {\cal H}(\r)\cdot\bmnabla_\y \,D_{\rm s}+\y\cdot {\cal H}(0)\cdot\bmnabla_\y \,D_{\rm o} \right)\nn
&&\qquad\qquad-\frac{C_F}{2MT}\Y\cdot [{\cal H}(0)-{\cal H}(\r)]\cdot \bmnabla_\Y D_{\rm o}.
\eeq
One reason why we display this  equation is that it reduces to the QED equation when $D_{\rm s}=D_{\rm o}$. Thus, if color quickly equilibrates, an assumption that we shall exploit in the next section, the dynamics becomes analogous to that of  the abelian case. In this case, color degrees of freedom play a minor role, and the motion of the heavy particles can be described by a Fokker-Planck equation or the associated Langevin equation.

As we have already emphasized, the component $D_0$ corresponds to the maximum (color) entropy state, where all colors are equally probable. This state plays an important role in the calculations to be presented in the next section. Thus, it is useful to write the corresponding equations of motion, or equivalently the operators ${\cal L}$ of the $(D_0,D_8)$ basis, in the semi-classical limit. 
We have:
\beq\label{L00m}
{\cal L}^{00}&=&-C_F\left\{\Y\cdot {\cal H}(0)\cdot \Y+\frac{1}{4} \y\cdot {\cal H}(0)\cdot \y   \right\}\nn
&&-\frac{C_F}{2MT}\left\{\Y\cdot {\cal H}(0)\cdot \bmnabla_\Y+\y\cdot {\cal H}(0)\cdot \bmnabla_\y   \right\},
\eeq
and 
\beq\label{L08m}
{\cal L}^{08}&=&i\frac{C_F}{2N_c} \y\cdot\bmnabla V(\r)\nn
&&-\frac{C_F}{2N_c}\left\{ \frac{1}{4} \y\cdot {\cal H}(\r)\cdot \y-\Y\cdot {\cal H}(\r)\cdot \Y \right\}\nn
&&-\frac{C_F}{2N_c}\frac{1}{2MT}\left\{ \y\cdot {\cal H}(\r)\cdot \bmnabla_\y -\Y\cdot {\cal H}(\r)\cdot \bmnabla_\Y\right\}.\nn
\eeq
We shall also need the operators ${\cal L}^{08}$ and ${\cal L}^{88}$ at leading order in $\y$. These are given by 
\beq\label{L80m}
{\cal L}^{80}=i\ \y\cdot\bmnabla V(\r),\qquad {\cal L}^{88}=-N_c \Gamma(\r).
\eeq

\section{Numerical studies}\label{numerics}

The equations for the time evolution of the reduced density matrix that we have obtained in the previous sections are difficult to solve in their original form, that is, for the operator ${\cal L}$ given in Sect.~\ref{sec:qcd-pair}, or ~\ref{ap:equations}, for a quark-antiquark pair. We shall not attempt to solve them directly in the present paper. In the case of QED, we have seen that an additional approximation, the semi-classical approximation,  allows us to transform these equations into  Fokker-Planck, or equivalently, Langevin equations,  describing the relative and center of mass motions of the heavy particles as simple random walks. In QCD, the presence of transitions between singlet and octet color states complicates the situation, since such transitions are a priori not amenable to a semi-classical description. The purpose of this section is to present numerical studies that illustrate two possible strategies to cope with this problem, namely preserve as much as possible the simplicity of the semi-classical description of the heavy particle motions, while taking into account the effects of color transitions. To simplify the discussion we shall ignore the center of mass motion in most of this section. 

The new feature in QCD, as compared to QED, namely the transitions between distinct color states, is best seen in the infinite mass limit, where the relative motion is entirely frozen.  Then   the only dynamics is that of color: as a result of the collisions with the plasma constituents the colors of the heavy quarks and antiquarks can change in time.  The corresponding equations of motion for the density matrix are easily obtained from the formulae listed in ~\ref{ap:equations}. They read, for a quark-antiquark pair,
\beq\label{eqDsDoMinfty}
&&\frac{\rmd D_{\rm s}}{\rmd t}=-2C_F \Gamma(\r) (D_{\rm s}-D_{\rm o}),\nn
&&\frac{\rmd D_{\rm 0}}{\rmd t}=-\frac{1}{N_c} \Gamma(\r) (D_{\rm o}-D_{\rm s}),
\eeq
where $\r $ is the (fixed) relative coordinate. These equations exhibit the decay widths in the singlet ($2C_F \Gamma(\r)$) and the octet ($(1/N_c)\Gamma(\r)$) channels, respectively. These two coupled equations acquire a more transparent physical interpretation in the $(D_0,D_8)$ basis, where they take a diagonal form
\beq\label{eqD0D8Minfty}
&&\frac{\partial D_0}{\partial t}=0,\nn
&&\frac{\partial D_8}{\partial t}=-  N_c \Gamma(\r) D_8.
\eeq
The first equation merely reflects the conservation of the trace of the density matrix. Recall also that $D_0$ is associated with the maximum color entropy state, where all colors are equally probable (see Eq.~(\ref{Dunpolar})): this component of the density matrix represents an equilibrium state that remains unaffected by the collisions. The second equation  indicates that $D_8\propto D_{\rm s}-D_{\rm o}$ decays on a time scale $(N_c\Gamma(\r))^{-1}$. Thus, at large times only $D_0$ survives, that is, the collisions drive the system to the maximum entropy state.   Note that the distance between the quark and the antiquark enters the rate $\propto \Gamma(\r)$  at which equilibrium is reached. When $|\r|\gtrsim m_D$, $\Gamma(\r)\approx 2\gamma_Q$, where $\gamma_Q $ is the damping factor of one heavy quark (or antiquark): at large separation, the quark and the antiquark equilibrate their color independently (with a rate $N_c \gamma_Q$ -- see ~\ref{ap:singlequark}).  On the other hand, when $|\r|\lesssim m_D$ the two quarks screen their respective colors, hindering the effect of collisions in equilibrating color.\\

The first strategy that we shall explore in order to treat the relative motion of the heavy quarks semi-classically and at the same time take into account the color transitions just discussed, rests on the following observation. There is  one instance where one can recover a situation analogous to that of QED: this is the regime where the color exchanges are fast enough to equilibrate the color on a short time scale (short compared to the typical time scale of the relative motion). We have for instance observed in the previous section that the  component $D_{\rm s}$ of the density matrix,  Eq.~(\ref{eqfinDs}),  obeys the same equation as the QED density matrix when $D_{\rm o}=D_{\rm s}$ (to within the multiplicative color factor $C_F$), which corresponds indeed to the maximum entropy state. In this case, one can explore the dynamics in the vicinity of this particular color state, treating the color transitions in perturbation theory. One can then  derive Langevin equations which contain an additional random force arising from the fluctuations of the color force between the heavy particles. This perturbative approach is easily generalized to the case of a large number of quark-antiquark pairs.

The second strategy  is  based on an analogy between the equations (\ref{eqDsDoMinfty}), and their generalizations to include the semi-classical corrections, and a Boltzmann equation, with the right hand side being viewed as a collision term. That is,  the changes of color that accompany the singlet-octet transitions are then treated as collisions rather than an additional random force in a Langevin equation. This strategy allows us to overcome some of the limitations of the perturbative approach. \\

\subsection{Langevin equation with a random color force: single quark-antiquark pair}
\label{sec:randomc1pair}

We shall now examine the corrections to Eq.~(\ref{eqD0D8Minfty})  that arise in the semi-classical approximation, i.e. taking into account corrections to the infinite mass limit. Note first that the kinetic energy of the heavy quarks leaves ${\cal L}$ as a diagonal operator.  In  the $(D_0,D_8)$ basis, this operator reads
\begin{equation}
 {\cal L}={\cal L}_0+\Gamma({\bf r})\left(\begin{array}{cc}
 0 & 0 \\
 0 & -N_c
  \end{array}\right),\qquad {\cal L}_0= \frac{2i}{M}{\bmnabla}_\r\cdot{\bmnabla}_\y.
 \end{equation} 
 The semi-classical corrections brings ${\cal L}$ to the form 
\beq\label{Lpert}
\partial_t\left(   \begin{array}{c}
D_0\\
D_8
\end{array}
\right)
=\left(
\begin{array}{cc}
{\cal  L}_0+y a^{(1)}_{00}+y^2 a^{(2)}_{00}&y a^{(1)}_{08}+y^2 a^{(2)}_{08}\\
y a^{(1)}_{80}+y^2 a^{(2)}_{80} & {\cal  L}_0+a^{(0)}_{88}+y a^{(1)}_{88}+y^2 a^{(2)}_{88}
\end{array}
\right)
\left(   \begin{array}{c}
D_0\\
D_8
\end{array}
\right),\nn
\eeq
where the various coefficients can be read off the equations recalled in the previous section
 (see Eqs.~(\ref{L00m}) and (\ref{L08m})):
\beq\label{aij}
&&a^{(1)}_{00}=0,\qquad a^{(2)}_{00}=-\frac{C_F}{4}\y\cdot{\cal H}(0)\cdot\y-\frac{C_F}{2MT}\y\cdot{\cal H}(0)\cdot \bmnabla_y,\nn
&&a^{(1)}_{08}=-i\frac{C_F}{2N_c}\y\cdot\F(\r),\qquad a^{(1)}_{80}=-i\y\cdot\F(\r),\qquad a^{(0)}_{88}= -N_c\Gamma(\r),\nn
\eeq
and we have set
\beq
\F(\r)\equiv -\bmnabla V(\r).
\eeq
One can diagonalize this new operator ${\cal L}$, and in particular find the eigenvalue that corresponds to the maximum entropy state in the limit where $y\to 0$. This is given by usual perturbation theory
\beq\label{pertformula}
{\cal L}_0+y a^{(1)}_{00}+y^2\left( a^{(2)}_{00}-\frac{a^{(1)}_{08}a^{(1)}_{80}}{a^{(0)}_{88}}   \right).
\eeq
By using the explicit expressions given above for the various coefficients, one deduces that the corresponding eigenvector, $D_0'$, fulfills the equation
 \begin{eqnarray}
&&\partial_t D_0'=\left(\frac{2i}{M}\bmnabla_\r\cdot \bmnabla_\y-\frac{C_F}{4}\y\cdot{\cal H}(0)\cdot\y-\frac{C_F(\y\cdot \F({\bf r}))^2}{2N_c^2\Gamma({\bf r})}\right.\nonumber\\
&&\qquad\qquad\qquad\left.-\frac{C_F}{2MT}\y\cdot{\cal H}(0)\cdot \bmnabla_y\right)D_0' \equiv {\cal L}'\, D_0'.
\label{eq:D0prime}
\end{eqnarray}
Performing the Wigner transform we obtain
\begin{align}
{\cal L}'=-\frac{2\p\cdot\bmnabla_\r}{M}+\frac{C_F}{4}\bmnabla_\p\cdot {\cal H}(0)\cdot \bmnabla_\p +\frac{C_F (\F({\bf r})\cdot \bmnabla_\p)^2}{2N_c^2\Gamma({\bf r})} +\frac{C_F}{2MT}\bmnabla_\p\cdot {\cal H}(0)\cdot \p.
\end{align}
The comparison  with the Fokker-Planck operators given in Eqs.~(\ref{FPrel0}) to (\ref{FPrel3}), allows us to write the corresponding Langevin equation:
\beq\label{langevinrelqcd}
\v=\dot\r=\frac{2\p}{M},\qquad \frac{M}{2} \ddot \r=-\gamma^{ij} v^j+\xi^i(t)+\Theta^i(t,{\bf r})\,,
\eeq
where
\begin{equation}\label{noisePsi}
\langle \xi^i(t)\xi^j(t')\rangle=\delta(t-t')\eta_{ij},\qquad \eta_{ij}=\frac{C_F}{2}{\cal H}^{ij}(0)=2T \gamma_{ij},
\end{equation}
and
\begin{equation}
\langle \Theta^i(t,{\bf r})\Theta^j(t',{\bf r})\rangle=\delta(t-t')\frac{C_FF^i({\bf r})F^j({\bf r})}{N_c^2\Gamma({\bf r})}\,.
\label{eq:avetheta}
\end{equation}
As compared to the QED case, Eq.~(\ref{langevinrelative}), we note a number of new features. The most noteworthy is the presence of  two random forces. The force $\xi$  is the familiar  stochastic force and is related to the drag force  as indicated in Eq.~(\ref{noisePsi}). The second random force, $\Theta$, has a different nature: it originates from the  fact that the force between a  quark and an antiquark is a function of their color state. Now, $D_0'$ represents a state close to the maximum entropy state, that is, a state  in which the probability to find the pair in an octet is approximately $N_c^2-1$ times bigger than the probability to find it in a singlet. At the same time, the force between heavy quarks in a singlet state is $N_c^2-1$ times bigger than that in an octet state,  and it has opposite sign. The net result is that, on average, the  force between the heavy quark and the heavy antiquark is zero. But this is true only in average. There are fluctuations, and these give rise to $\Theta$. The vanishing of the average force between the quark and the antiquark explains the absence of the force term in the Langevin equation, as compared to the QED case.  Note that this picture is valid as long as transitions between singlet and octet states are fast compared with the rest of the dynamics. This is no longer the case when the size of the pair becomes too small: then, $\Gamma(\r)$ becomes small, reducing the energy denominator in Eq.~(\ref{eq:avetheta}), i.e., amplifying the effect of the random color force. This, as we shall see, can lead to unphysical behavior.

\subsection{Many heavy quarks and antiquarks}
\label{sec:many}
The discussion of the previous subsection can be  generalized to a system containing many heavy quarks and antiquarks. We call $N_Q$ the number of heavy quarks, and for simplicity we assume that it is equal to the number of heavy antiquarks. The density matrix can be written as a product of density matrices of the individual quarks and antiquarks, generalizing the construction of Eq.~(\ref{Dtensorproduct}) for the quark-antiquark density matrix. One can then write
\beq
{\cal D}=D_0\, (\mathbb{I})^{2N_Q}+\cdots
\eeq
where the dots represent all the scalar combinations that can be formed with products of $n\leq 2N_Q$ color matrices $t^a$, with coefficients corresponding to the  components of ${\cal D}$. We shall not need the explicit form of these extra components. As for $D_0$, this is clearly the maximum entropy state, where all colors of individual quarks and antiquarks are equally probable and uncorrelated. It corresponds to the following density matrix (cp. Eq.~(\ref{Dunpolar})):
\beq
{\cal D}=\sum_{\alpha_i, \bar\alpha_i} \ket{\alpha_1,\cdots\bar\alpha_{N_Q}}\bra{\alpha_1,\cdots,\bar\alpha_{N_Q}},
\eeq
where the sum runs over all the colors of the quarks ($\alpha_1,\cdots\alpha_{N_Q}$) and the antiquarks ($\bar\alpha_1,\cdots,\bar\alpha_{N_Q}$).
We want to construct for this system the analog of Eq.~(\ref{eq:D0prime}) which describes how the maximum entropy state is modified by the semiclassical corrections. 

Our starting point is the main equation,  Eq.~(\ref{main}). To proceed, it is useful to have in mind the diagrammatic representation of the density matrix that we have introduced earlier. As compared to the diagrams displayed in Fig.~\ref{fig:densitymatrix2}, in the present case, the diagrams will contain $2N_Q$ lines in the upper part, and $2N_Q$ lines  in the lower part. All the interactions that we are dealing with involve a single gluon exchange, represented by one gluon line joining quark or antiquark lines in various ways. The evolution equation for ${\cal D} $ is still described by an operator ${\cal L}$, which is a matrix in the space of all the independent components. For our perturbative calculation, we need only to consider diagonal (to order $y^2$) and non diagonal (to order $y$) elements of this matrix, the non diagonal elements involving the maximum entropy state as one of their entries.


Consider first the diagonal elements. We have first the kinetic energy, trivially given by
\begin{equation}
\sum_{j\in \{N_Q,N_{\bar{Q}}\}}\left(\frac{2i}{M}\bmnabla_{r_j}\cdot \bmnabla_{p_j}\right)\,.
\end{equation}
The leading order diagonal element that involves the interaction can be obtained in the infinite mass limit. It represents the decay of the components of ${\cal D}$ that are connected to  $D_0$ by one gluon exchange. It is given by
 \begin{equation}
-N_c\Gamma({\r}_{kl})\,.
\end{equation}
where $\r_{kl} =\r_k-\r_l$, $\r_k$ and $\r_l$ denoting the coordinates of the quark or the antiquarks to which the gluon is attached. The factor $N_c$ can be understood from the same argument as that given after Eq.~(\ref{eqD0D8Minfty}): when the separation $\r_{kl}$ is large, the color of the quark (or antiquark) at $\r_k$ and $\r_l$  relax independently at a rate $N_c \gamma_Q$. 
At the order of interest, we need also 
 the diagonal element for the maximum entropy state, including the semi-classical corrections up to order $y^2$ and $\frac{y}{M}$. This is given by 
\begin{equation}
-\frac{C_F}{4} \sum_{j\in \{N_Q,N_{\bar{Q}}\}}\left(\y_j\cdot {\cal H}(0)\cdot \left(\y_j+\frac{2\bmnabla_{\y_j}}{MT}\right)\right)\,.
\end{equation}

We turn now to the non-diagonal elements. To leading order, these involve solely the real part of the potential. Diagrammatically, the corresponding exchanged gluon connects only either the upper or the lower lines among themselves, but does not connect upper with lower lines. 

 Consider first the non-diagonal elements that bring the maximum entropy state to another state. If the pair connected by the exchanged gluon is formed by quark $k$ and antiquark $l$ then the element is (cp. Eq.~(\ref{aij}))
\begin{equation}
-\frac{iC_F}{2N_c}\y_{kl}\cdot F({\bf r}_{kl})\,,
\end{equation}
while if it is formed by quark (antiquark) $l$ and quark (antiquark) $k$ then it is (cp. Eq.~(\ref{D0QQ}))
\begin{equation}
\frac{iC_F}{2N_c}\y_{kl}\cdot F({\bf r}_{kl})\,.
\end{equation}
We also need the non-diagonal elements that bring the system back to the maximum entropy state. If this pair is formed by quark $k$ and antiquark $l$ then the element is (cp. Eq.~(\ref{aij}))
\begin{equation}
-i\,\y_{kl}\cdot F({\bf r}_{kl})\,,
\end{equation}
while if it is formed by quark (antiquark) $l$ and quark (antiquark) $k$ then it is (cp. Eq.~(\ref{D8QQ}))
\begin{equation}
i\y_{kl}\cdot F({\bf r}_{kl})\,.
\end{equation}

With this information, it is straightforward to construct the generalization of Eq.~(\ref{eq:D0prime}) for  an arbitrary number of quark-antiquark pairs. The corresponding equations will be presented later. 

\subsection{Simulation of a Langevin equation with a random color}
\label{sec:randomsimu}
In this subsection we present numerical results obtained by simulating the equations of the previous subsections. We examine successively the evolution of the relative coordinate of a single heavy quark-antiquark pair, and then that of fifty pairs initially produced in a thin layer.
The first case will help us to understand the range of applicability of the perturbative method, while the second will illustrate how the Langevin equation may account for recombination.

In these calculations, we use the standard QCD running coupling constant $\alpha_s$ determined at one loop order for $N_f=3$ massless flavors and  $\Lambda_{QCD}=250$ MeV. The screening Debye mass is given by its HTL approximation, $m_D^2=(2\pi/3)(6+N_f)\alpha_s T^2$, with $\alpha_s$ evaluated at the scale $2\pi T$, with $T$ the temperature. Further details on the parameters of the calculation will be given as we proceed. 

We should emphasize here that the numerical simulations to be presented in this section are meant to illustrate the main physical content, as well as the limitations,  of the equations obtained in  this paper. Although the numbers that go into the calculations are  appropriate for the physics of quarkonia in a quark-gluon plasma, we make no attempt to develop a phenomenological discussion.

\subsubsection{A single heavy quark-antiquark pair}
The Langevin equation for the relative motion is given in Eq.~(\ref{langevinrelqcd}).
The information about the medium is encoded in the functions $ V({\bf r})$ and $W({\bf r})$ which we estimate using the HTL approximation. Note that the resulting value of  $\Delta W({\bf r})$ and $ V({\bf r})$ diverge as $r\to 0$ (for different reasons though, see e.g. \cite{Blaizot:2015hya}).  In \cite{Blaizot:2015hya} the divergence of $\Delta W(\r)$ was regulated with the help of an ultraviolet cut-off. Here, we shall use a simpler procedure and   choose 
\begin{equation}
\gamma=\frac{\Delta W({\bf 0})}{6T}
\end{equation}
as  a free parameter of our simulation, for which we choose the value $\gamma=0.19 T^2$. For the real part of the potential, we proceed as in \cite{Blaizot:2015hya}, that is we define it as the Fourier transform of a Yukawa potential integrated in momentum space up to a cutoff $\Lambda=4$ GeV. The coupling constant that appears in $V(\r)$ is evaluated at a scale corresponding to the inverse Bohr radius of the bottomonium (1348 MeV). The spatial dependence of $W(\r)$ is obtained, as already mentioned, from the HTL approximation, and is of the form $W(\r)=W(0)+\alpha_s T\phi(m_D r)$ \cite{Laine:2006ns}, with $\alpha_s$ evaluated at the scale $2\pi T$, and $\phi(x)$ a monotonously increasing function such that $\phi(x=0)=0$ and $\phi(x\to \infty)=1$. At small separation, i.e., for  $m_D r\ll 1$, $\phi(x)$  can be approximated by
$
 \phi(x)=\frac{x^2}{3}(-\log x+\frac{4}{3}-\gamma_E).
$

\begin{figure}
\begin{center}
\includegraphics[scale=0.5]{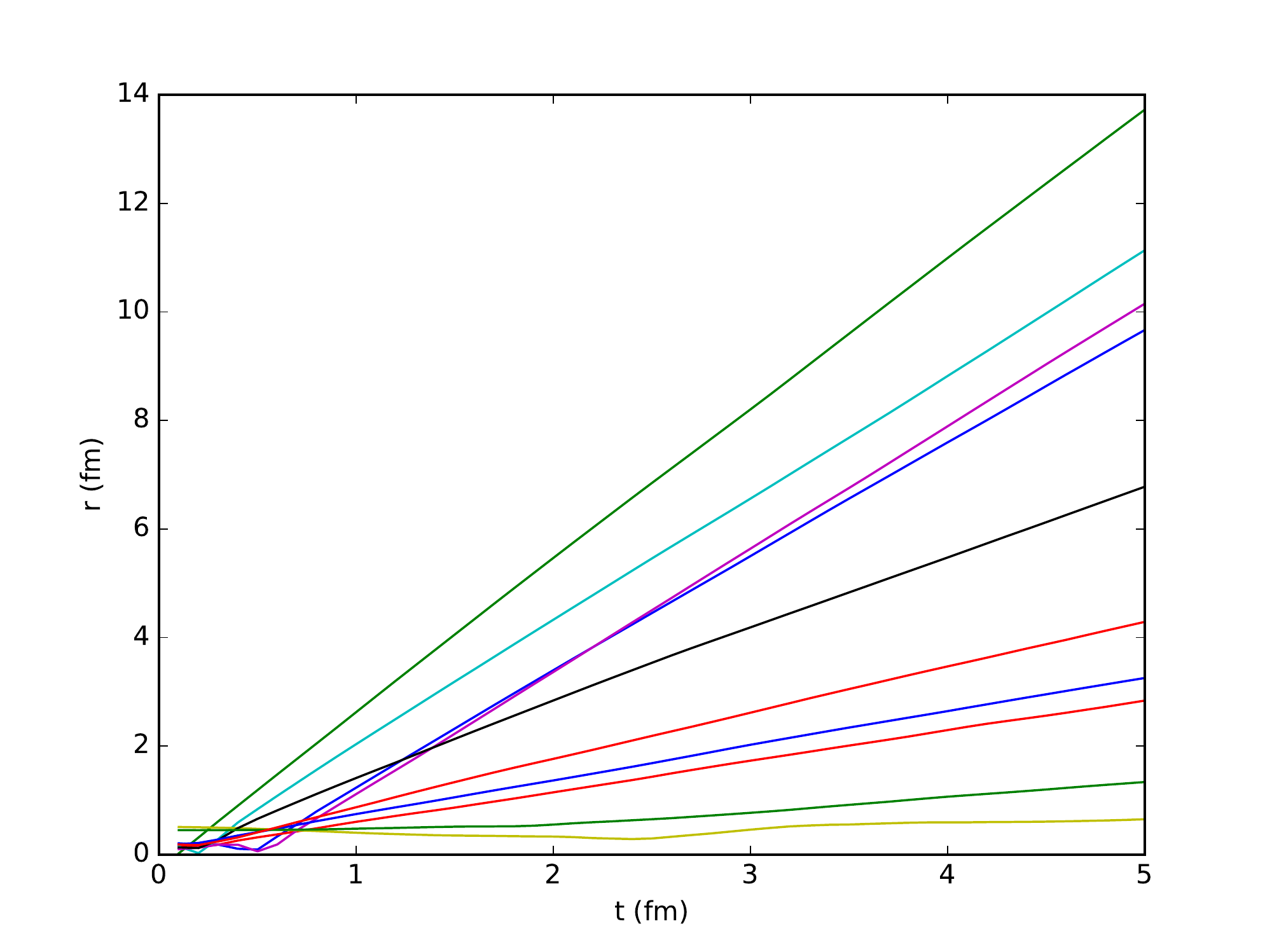}
\caption{Example of ten random trajectories for the relative distance of a bottom quark-antiquark pair prepared in a $1S$ bound state. About half of these trajectories are unphysical, since they correspond to supraluminal velocity, $r(t)\>t$.}
\label{fig:10traj}
\end{center}
\end{figure}

\begin{figure}
\begin{center}
\includegraphics[scale=0.5]{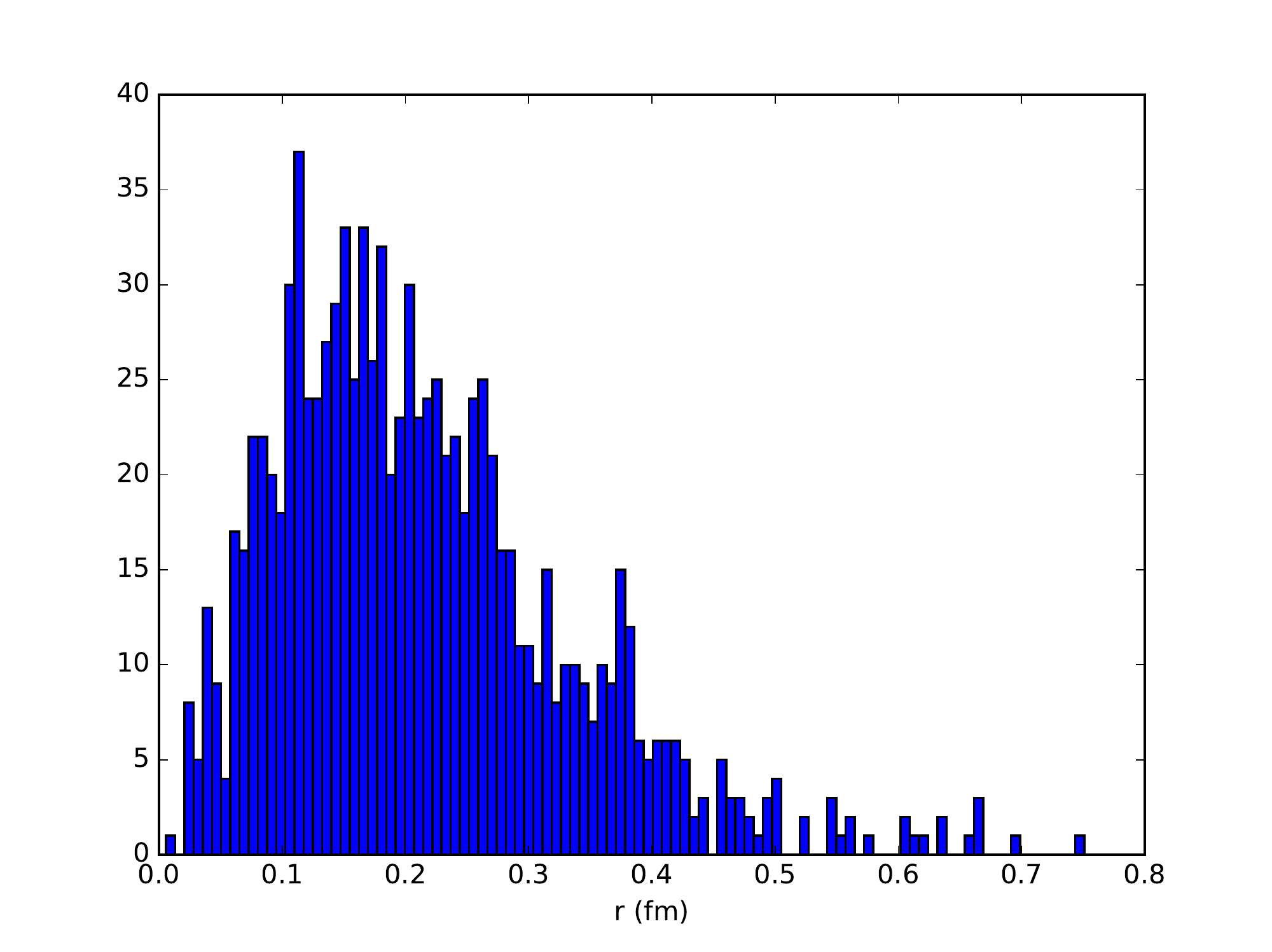}
\caption{Histogram representing the initial distribution of relative distances given by the square of the $1S$ wave-function of the bottomonium.}
\label{fig:histini}
\end{center}
\end{figure}

\begin{figure}
\begin{center}
\includegraphics[scale=0.5]{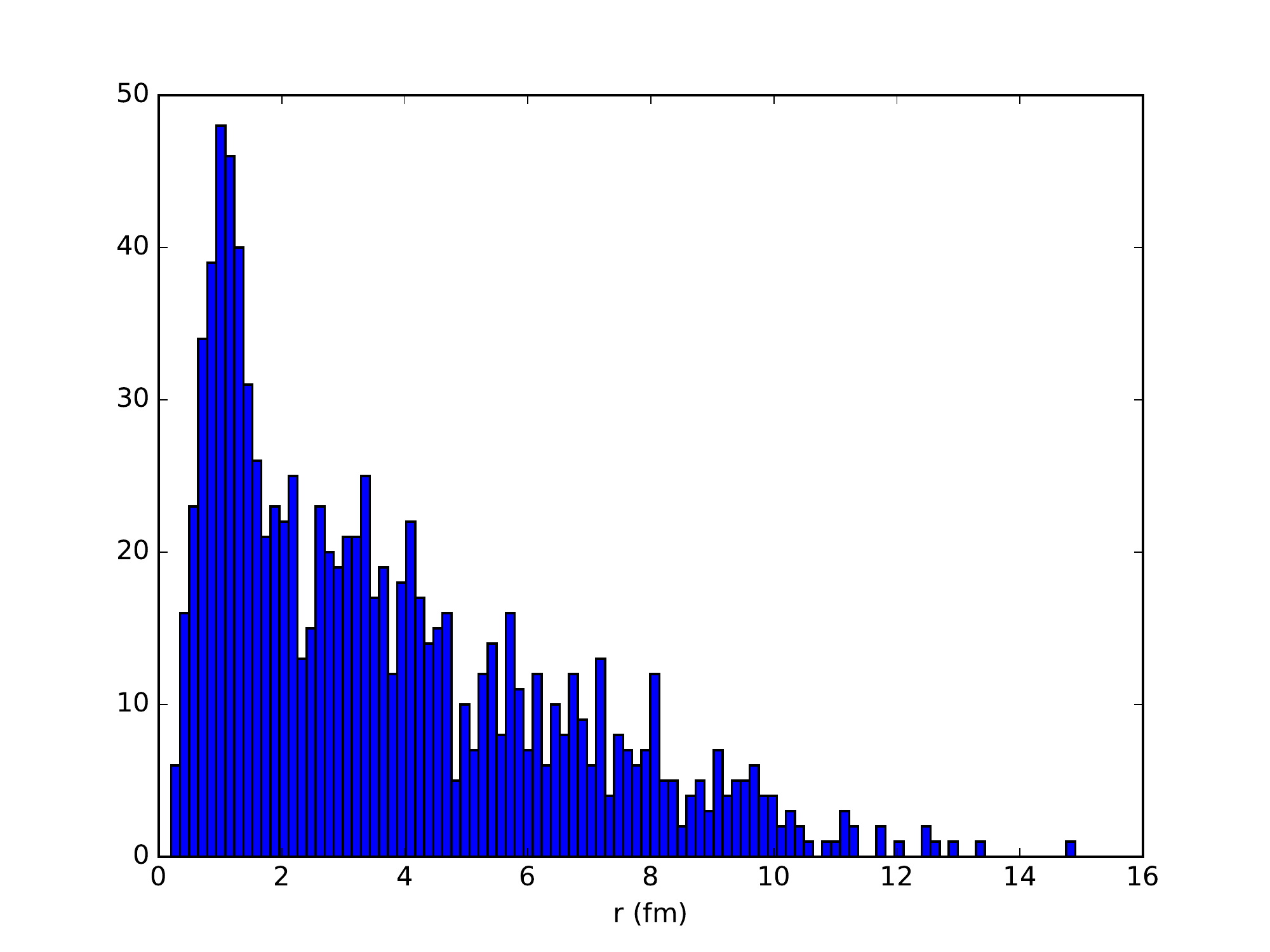}
\caption{Histogram representing the final distribution of relative distances after a time $t=5\,\rm{fm/c}$ assuming the initial distribution of Fig.~\ref{fig:histini}. Note the change in horizontal scale with respect to Fig.~\ref{fig:histini}.}
\label{fig:histfinal}
\end{center}
\end{figure}

In Fig.~\ref{fig:10traj} we show a set of ten random trajectories produced for bottomonium (with mass $M_b=4881\,\text{MeV}$) at a temperature of $T=\,350\, \text{MeV}$.  We take as  initial  distribution of relative distances  that  obtained from the wave function of the $1S$ state (which we plot as an histogram with $1000$ events in Fig.~\ref{fig:histini} for further comparison). We see that many of the trajectories are clearly unphysical since  they involve supraluminal velocities: this is because some of the random kicks can occasionally be very strong due to the amplification produced by the small denominator in Eq.~(\ref{eq:avetheta}) at small $\r$. A more systematic comparison can be done by looking at the distribution of relative distances after some time $t$. This is shown in Fig.~\ref{fig:histfinal} for  $t=5\,\rm{fm}$. The histogram reveals that there remains at this time only a tiny probability to find the pair within a relative distance corresponding to the size of the bound state: the random color force is clearly too efficient in suppressing the bound state!

\subsubsection{Many heavy quark-antiquark pairs}
\label{sec:simumany}

In spite of its shortcomings, the perturbative method remains interesting as it allows us to treat an assembly with an arbitrary number of quark-antiquark pairs, and address in particular the issue of recombination. 
The relevant equations can be constructed along the lines developed in the previous subsection. We need  to take into account not only the relative coordinates but also the center of mass motions. Moreover  the random force does not only act between heavy quarks and antiquarks but also between two heavy quarks and two heavy antiquarks (the equations for the density matrix of a pair of two heavy quark are given in ~\ref{sec:HQpair}). The resulting equations  are given by 
\begin{equation}
M\ddot{\bf r}_a=-C_F\gamma\dot{\bf r}_a+{\bf\Xi}_a(t)+\sum_{b\neq a}^{N_Q}{\bf \Theta}_{ab}(\r_{ab})+\sum_{\hat{b}}^{N_Q}{\bf\Theta}_{a\hat{b}}(\r_{a\hat{b}},t)\,,
\end{equation}
\begin{equation}
M\ddot{{{\bf r}}}_{\hat{a}}=-C_F\gamma\dot{{{\bf r}}}_{\hat{a}}+{\bf \Xi}_{\hat{a}}(t)+\sum_{\hat{b}\neq \hat{a}}^{N_Q}{\bf \Theta}_{\hat{a}\hat{b}}(\r_{\hat{a}\hat{b}},t)+\sum_j^{N_Q}{\bf\Theta}_{\hat{a}b}(\r_{\hat{a}b},t)\,,
\end{equation}
where the noises have vanishing means and correlators given by
\beq
\langle  \Xi_{ia}(t) \Xi_{jb}(t')\rangle=\frac{C_F}{6}\Delta W({\bf 0})\delta_{ab}\delta_{ij}\delta(t-t')\,,
\eeq
\beq
\langle \Theta^{i}_{ab}(t)\Theta^{j}_{cd}(t')\rangle=\frac{C_F}{N_c^2\Gamma_{ab}} F^i_{ab}F^j_{cd}\,\delta_{ac}\delta_{bd}\delta(t-t')\nn
\eeq
\begin{eqnarray}
\langle \Theta^{i}_{a\hat{b}}(t)\Theta^{j}_{c\hat{d}}(t')\rangle=\frac{C_F}{N_c^2\Gamma_{a\hat{b}}} F^i_{a\hat{b}}F^j_{c\hat{d}}\,\delta_{ac}\delta_{\hat{b}\hat{d}}\delta(t-t')\nn
\end{eqnarray}
\begin{eqnarray}
\langle \Theta^{i}_{\hat{a}\hat{b}}(t)\Theta^{j}_{\hat{c}\hat{d}}(t')\rangle=
\frac{C_F}{N_c^2\Gamma_{\hat{a}\hat{b}}}F^i_{\hat{a}\hat{b}}F^j_{\hat{c}\hat{d}}\,\delta_{\hat{a}\hat{c}}\delta_{\hat{b}\hat{d}}\delta(t-t').\nn
\end{eqnarray}  
In these equations $N_Q$ is the number of heavy quarks, equal to the number of heavy antiquarks. The indices $a$ or $b$ are color indices in the $3$ representation while the same with a hat are a color indices in the $\bar{3}$ representation. The nature of the color index specifies whether a given quantity refers to a quark or an antiquark. Thus,  $\r_a$ represents the coordinate of a quark, while $\r_{\hat a}$ represents the coordinate of an antiquark. We use also compact notation, such as  $\r_{ab}=\r_a-\r_b$  to denote the relative distance between a quark of color $a$ and a quark of color $b$, or $\r_{a\hat{b}}=\r_a-\r_{\hat b}$ to denote the relative distance between a quark of color $a$ and an antiquark of color $\hat b$.  Finally, for functions of coordinates, we set $F_{\hat a b}=F(\hat \r_a-\r_b)$, $\Gamma_{ab}=\Gamma(\r_a-\r_b)$, and so on.

\begin{figure}
\begin{center}
\includegraphics[scale=0.6]{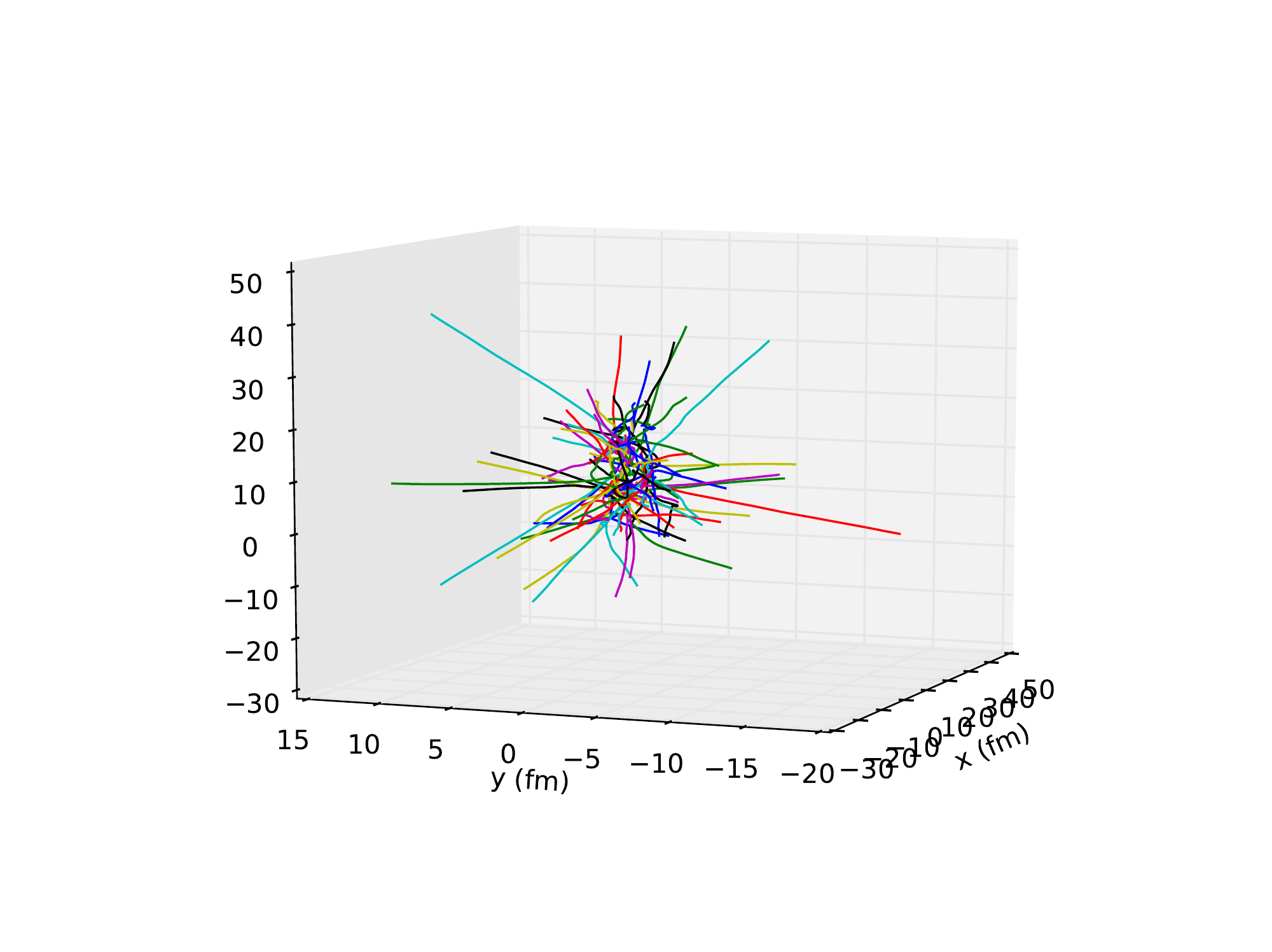}
\caption{Example of  random trajectories of fifty heavy quark-antiquark pairs.  The pairs are prepared as explained in the text, and evolve during a time  $t=5\,\rm{fm/c}$.}
\label{fig:50traj}
\end{center}
\end{figure}
In Fig.~\ref{fig:50traj} we plot some random trajectories of fifty pairs of quarks and antiquarks. The parameters are different from the ones used in the previous section, now the temperature is $T=250\,\text{MeV}$ and the cut-off for $V({\bf r})$ is $\Lambda=1500\,\text{MeV}$. We keep  the same value of $\gamma$. This new choice of parameters makes the problem of the violent hard kicks less severe, at the cost of having a cutoff $\Lambda$ unrealistically small (it is of the order of the inverse of the Bohr radius of the ground state). The system is prepared in the following way: in a square of size $2.5\,\rm {fm}$ we chose fifty random points; in each point we put a quark-antiquark pair  following a probability for the relative coordinate given by
\begin{equation}
\frac{2}{\pi r}\sin(\Lambda r)\,,
\end{equation}
where $\Lambda$ is the same cut-off as for the real part of the potential (this distribution becomes a Dirac delta as $\Lambda\to\infty$). The fifty quark-antiquark pairs then evolve for a time  $t=5\,\text{fm/c}$, according to the stochastic equations displayed above. As can be observed by looking at Fig.~\ref{fig:50traj} some of the trajectories remain close enough to allow for ``recombinations'' into bound state.

\begin{figure}
\begin{center}
\includegraphics[scale=0.5]{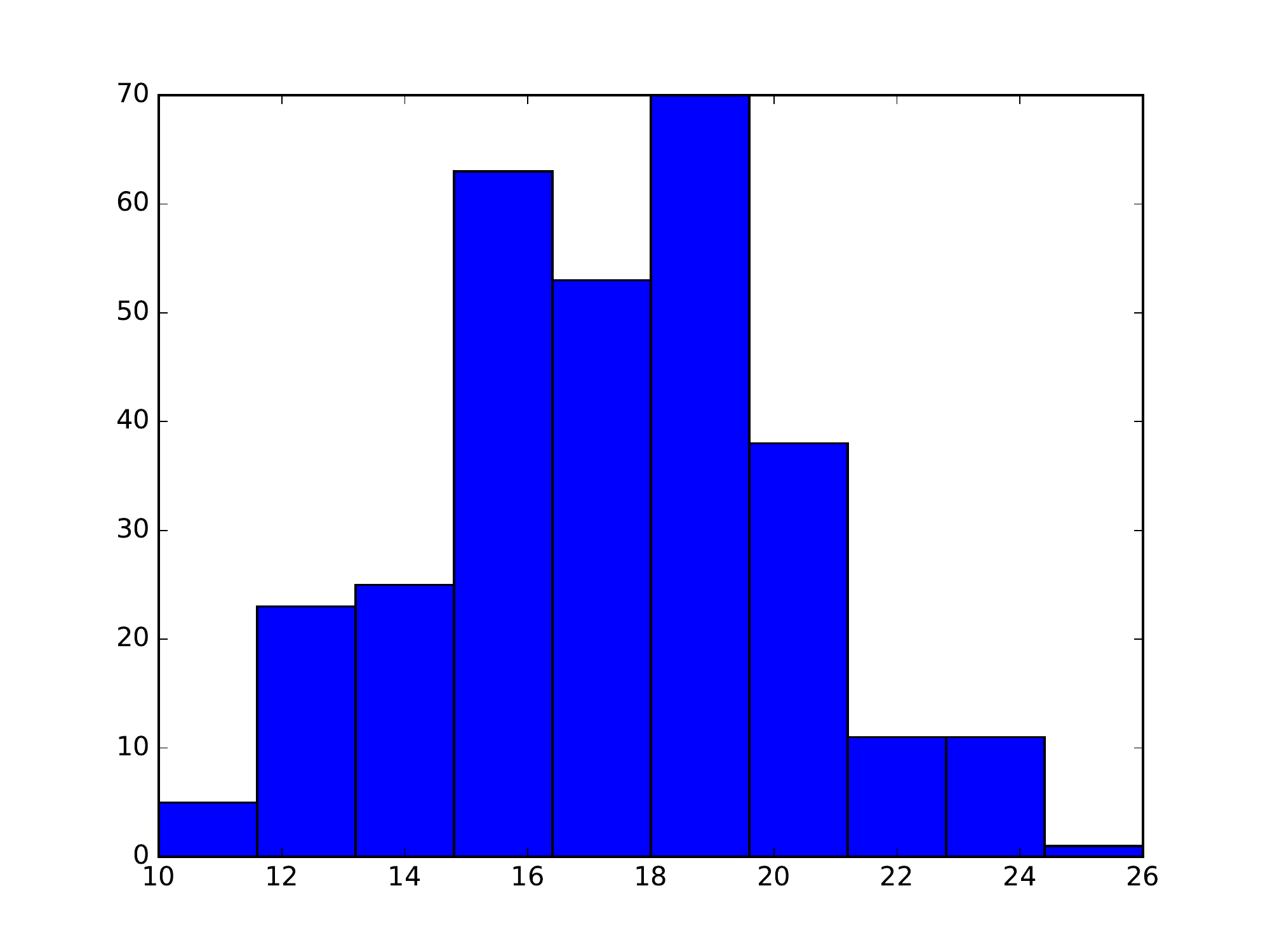}
\caption{Histogram of the number of bound states formed in $300$ simulations with the same initial conditions as in Fig. \ref{fig:50traj}}
\label{fig:histo}
\end{center}
\end{figure}

To quantify the phenomenon,  we perform a statistical analysis of how many bound states are observed at the end of the evolution, starting from the previous initial condition. To define a bound state, we follow the procedure of Ref.~\cite{Blaizot:2015hya}, but with slightly different parameters. We declare a heavy quark-antiquark pair to be bound if the quark and the antiquark  remain at a distance smaller than $0.5\,\rm{fm}$ during a time bigger than $0.1\,\rm{fm/c}$. This procedure can become ambiguous when many quarks and antiquarks are found in a small region, for example in the case in which two quarks and two antiquarks are contained in a sphere of radius smaller than $0.25\,\rm{fm}$. In this situation we count the number of bound states by choosing the combination that yields the maximum number. The results obtained after $300$ simulations are shown in Fig. \ref{fig:histo}. We can  see that from the $50$ initial bound states, on average $17$ remain after a time of $5\,\rm{fm/c}$ spent inside the medium. This is to be contrasted for Fig.~\ref{fig:histfinal} which would suggest that all pairs should become unbound if they were evolving independently of each other. Of course, we should recall that different parameters have been used in the two calculations. However, repeating the simulation of the previous subsection with the present parameters, one finds that about 10 to 15\% of the bound states would survive after a time $t=5\,$fm/c. This is about half of what the present calculation suggests.  We may therefore take this result as  evidence for recombination, in line with what was found in  Ref.~\cite{Blaizot:2015hya}.

\subsection{Langevin equations coupled to random collisions}
\label{sec:mc}
We now turn to the second strategy presented in the introduction of this section. This will allow us, in particular,  to bypass the limitations of the perturbative diagonalization of the matrix in Eq.~(\ref{Lpert}), caused by the vanishing of $\Gamma(\r)$ at small $\r$.  Let us then go back to the small $y$ expansion of Eq.~(\ref{eqfinDs}), and let us temporarily neglect the terms that go like $\frac{y}{M}$ or $y^2$, that is we keep only the kinetic term and the force term. We get 
\begin{equation}\label{DsMC}
\partial_t D_s=\frac{2i}{M}{\bf \nabla_r}\cdot{\bf \nabla_y}D_s-2C_F\Gamma({\bf r})(D_s-D_o)-iC_F{\bf y}\cdot\F(\r)D_s\,,
\end{equation}
\begin{equation}\label{DoMC}
\partial_t D_o=\frac{2i}{M}{\bf \nabla_r}\cdot{\bf \nabla_y}D_o-\frac{1}{N_c}\Gamma({\bf r})(D_o-D_s)+i\frac{1}{2N_c}\y\cdot \F(\r)D_o\,.
\end{equation}
The equation for $D_{\rm o}$ is not given explicitly in Eq.~(\ref{eqfinDs}), but it is easily derived from the material presented in ~\ref{ap:equations}.
We note that only the terms proportional to $\Gamma({\bf r})$ mix singlets and octets, i.e. the terms involving the force preserve the color state of the pair. We now perform a Wigner transform with respect to the variable $\y$,  and define $P_{\rm s}=D_{\rm s}$ and $P_{\rm o}=(N_c^2-1)D_{\rm o}$,  the probabilities for the pair to be in a singlet or octet state, respectively. We get
\begin{equation}
\left[ \partial_t +\frac{2{\bf p}\cdot {\bf \nabla_r}}{M}-C_F \F(\r)\cdot{\bmnabla_\p}\right]P_{\rm s}=-2C_F\Gamma(\r)\left(P_{\rm s}-\frac{P_{\rm o}}{N_c^2-1}\right)\,,
\label{eq:yPs}
\end{equation}
\begin{equation}
\left[  \partial_t +\frac{2{\p}\cdot{\bmnabla_\r}}{M}+\frac{1}{2N_c}\F(\r)\cdot{\bmnabla_p}\right]P_{\rm o}=-\frac{1}{N_c}\Gamma({\bf r})(P_{\rm o}-(N_c^2-1)P_{\rm s})\,.
\label{eq:yPo}
\end{equation}
The right hand sides of these two equations can be interpreted as a ``collision term'' in a Boltzmann equation, with gain and loss terms. Note that these collision terms are opposite in the singlet and octet channels, as expected:
\beq
2C_F \left(P_s-\frac{P_o}{N_c^2-1}\right)=-\frac{1}{N_c}(P_o-(N_c^2-1)P_s).
\eeq
The left hand sides of the  equations (\ref{eq:yPs}) and (\ref{eq:yPo}) describe the relative  motion of the pair under the influence of the color force $\F(\r)$. The corresponding classical equations of motion read
\beq\label{classeqmotion}
\frac{d{\bf r}}{dt}=\frac{2{\bf p}}{M},\qquad \frac{d{\bf p}}{dt}=-C_F{\bf F}({\bf r}),\qquad \frac{d{\bf p}}{dt}=\frac{1}{2N_c}{\bf F}({\bf r}),
\eeq
where the last two equations refer to the singlet and octet channels, respectively.

Thus, instead of treating the singlet-octet transitions as an additional color force in a Langevin equation, we can treat these transitions as ``collisions''. 
In practice we can solve the set of equations (\ref{eq:yPs}) and (\ref{eq:yPo}) using a Monte Carlo method, deciding  at each time step, according to a probability proportional to the respective decay widths,  whether a transition takes place or not, and then evolve the system through the time step according to the classical equations of motion (\ref{classeqmotion}). This is somewhat analogous to the Monte Carlo Wave Function method applied to a 2-level problem in Ref.~\cite{Dalibard:1992zz}. 

The equations (\ref{eq:yPs}) and (\ref{eq:yPo}) capture some of the important physics but they miss the drag forces and the stochastic forces that have to go with them in order to fulfill the fluctuation-dissipation theorem. These come from the semi-classical corrections that we have left out in writing Eqs.~(\ref{DsMC}).  However, if we were to include these corrections as they appear for instance in Eq.~(\ref{eqfinDs}), we would introduce extra couplings between $D_{\rm s }$ and $D_{\rm o}$ that would lead in particular to a collision term involving derivatives of $P_{\rm s,o}$  and we do not know of any efficient numerical tools to solve the resulting equation. However, if the system is not too far from the maximum entropy state,  terms that go like $y^2\left(P_s-\frac{P_o}{N_c^2-1}\right)$ or ${\frac{y}{M}}\left(P_s-\frac{P_o}{N_c^2-1}\right)$ are small and can be safely neglected. We again rely on the assumption that color relaxes faster than the relative motion. 
Under these conditions, and after performing the Wigner transform, we obtain for the singlet 
\beq\label{FPPs}
&&\left\{\partial_t +\v \cdot \bmnabla_\r-C_F\F(\r)\cdot\bmnabla_\p -\frac{1}{2}\bmnabla_\p\cdot \bmeta_{\rm s}(\r) \cdot\left(  \bmnabla_\p+\frac{\v}{T} \right)\right\}P_{\rm s}\nn
&&\qquad=-2C_F\Gamma({\bf r})\left(P_{\rm s}-\frac{P_{\rm o}}{N_c^2-1}\right).
\eeq
with 
\beq
\bmeta_{\rm s}(\r)=\frac{C_F}{2}\left(   {\cal H}(0)+{\cal H}(\r)\right),
\eeq
Comparing the operator in  the first line of this equation with that given in Eqs.~(\ref{FPrel0}) to  (\ref{FPrel3}), we easily derive the following stochastic equations:
\beq
\v=\frac{\rmd{\r}}{\rmd t}=\frac{2{\p}}{M},\qquad
\frac{\rmd\p}{\rmd t}=-C_F\F({\r})-\bmgamma_{\rm s}\cdot \v+\bmxi({\r},t)\,,
\eeq
where 
\beq
\langle \xi^i({\bf r},t)\xi^j({\bf r},t')\rangle=\delta(t-t')\eta^{ij}_{\rm s}(\r)),\qquad \bmgamma_{\rm s}=\frac{1}{2T}\bmeta_{\rm s}.
\eeq
These equations are the analogs of Eq.~(\ref{langevinrelative}) for the QED case.

Performing completely analogous manipulations, we obtain a similar result for the octet. The Fokker-Planck equation reads
\beq\label{FPPo}
&&\left\{\partial_t +\v\cdot \bmnabla_\r+\frac{1}{2N_c}\F(\r)\cdot\bmnabla_\p -\frac{1}{4}\bmnabla\cdot\bmeta_{\rm o}\cdot\left(  \bmnabla_\p+\frac{\v}{T} \right)\right\}P_{\rm o}\nn
&&\qquad=-\frac{1}{N_c}\Gamma({\bf r})\left(P_{\rm o}-(N_c^2-1)P_{\rm s}\right).
\eeq
and the corresponding Langevin equation is
\begin{equation}
\frac{\rmd{\p}}{\rmd t}=\frac{1}{2N_c}\F({\r})-\bmgamma_{\rm o}\cdot \v+\bmxi({\r},t)\,,
\end{equation}
with now
\begin{equation}
\bmeta_{\rm o}(\r)=\frac{1}{2}\left(C_F{\cal H}(0)-\frac{1}{2N_c}{\cal H}(\r)\right),\qquad \bmgamma_{\rm o}=\frac{1}{2T}\bmeta_{\rm o}.
\end{equation}

We shall now present the results of the simulation of these equations, for the case of a single quark-antiquark pair. We consider first the static limit, and then turn to the full equations including the semi-classical corrections. 
 
\subsubsection{The static limit}
\begin{figure}
\begin{center}
\includegraphics[scale=0.5]{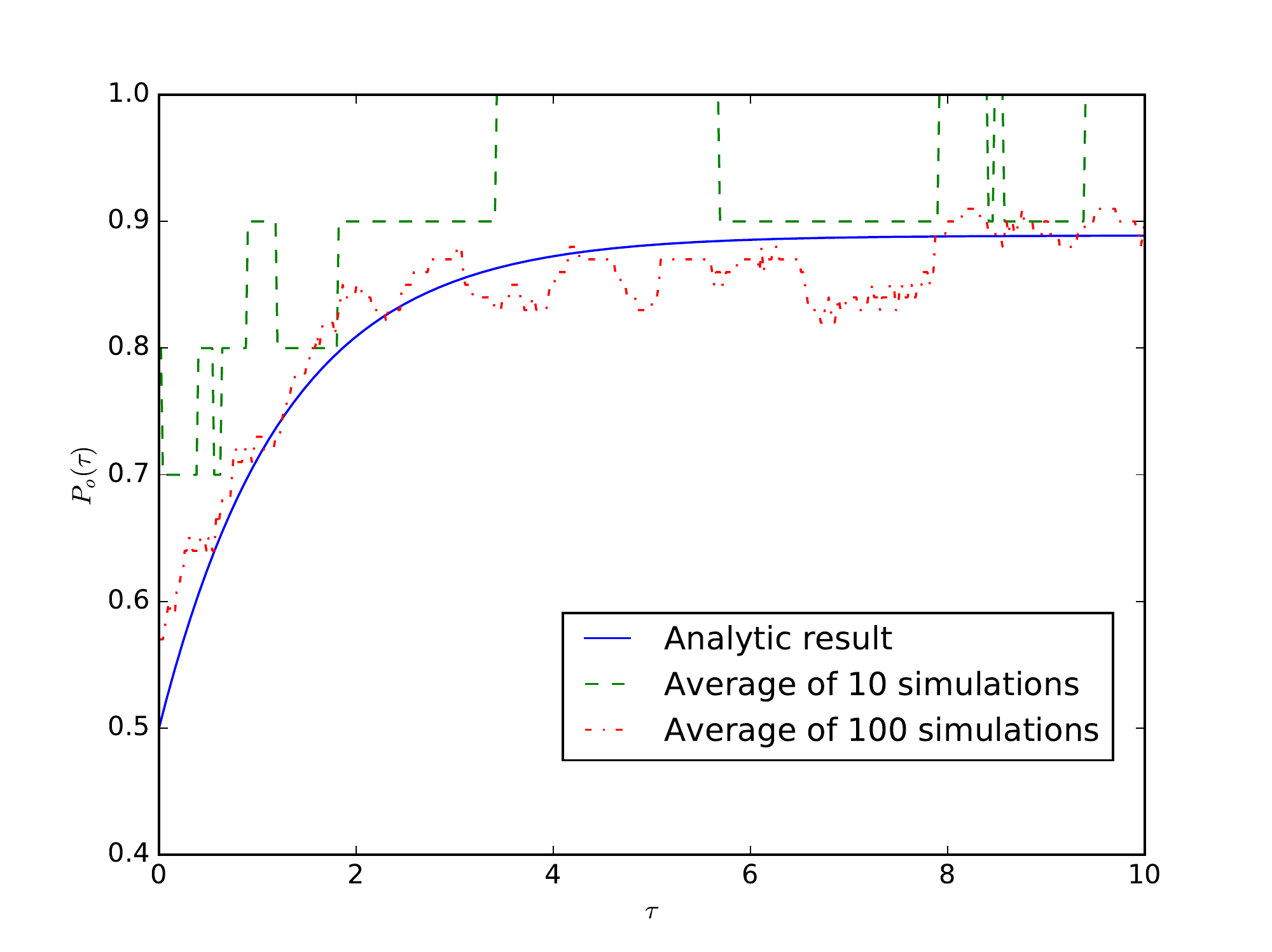}
\caption{Probability of having a static quarkonium in a color state as a function of $\tau={2C_F\alpha_sTt}/{10}$ assuming that the initial probability is ${1}/{2}$ and the temperature $T=250\,\textit{MeV}$. We compare the analytic results with the average of different number of simulations}
\label{fig:colorro}
\end{center}
\end{figure}
The study of the static limit (or infinite mass limit) offers us the possibility to  test the numerical method, since the exact solution can be obtained analytically in this case. In particular, this will give us an idea of the number of iterations that are needed in order to get a good estimate. We consider a heavy quark-antiquark pair, whose relative distance is $r=0.1\,\rm{fm}$, and in a well-defined color state, in a quark-gluon plasma at  temperature  $T=250\,\rm{MeV}$. 

 The equations to be solved are Eqs.~(\ref{eqDsDoMinfty}).
If the initial conditions are such that  singlet or and octet states are equally probable, i.e., $P_{\rm s}(0)=P_{\rm o}(0)$, the probability to be in an octet state at  time $t$ is
\begin{equation}
 P_{\rm o}(t)=\frac{N_c^2-1}{N_c^2}-\frac{N_c^2-2}{2N_c^2}e^{-N_c\Gamma(\r)\,t}\,,
 \end{equation}
and that to be in a singlet state is $P_{\rm s}(t)=1-P_{\rm o}(t)$. 

We can compare this result to that of a simulation using the Monte Carlo  method described above. The results are plotted in Fig. \ref{fig:colorro}, as a function of  $\tau\equiv (2C_F\alpha_sTt)/{10}$, with a time step $\Delta\tau=0.02$. We see that for 100 events the results of the simulation match relatively well the analytic result, although sizeable fluctuations remain. The simulations to be presented next involve 1000 events. 
\subsubsection{Simulation with dynamical quarks}
\begin{figure}
\begin{center}
\includegraphics[scale=0.5]{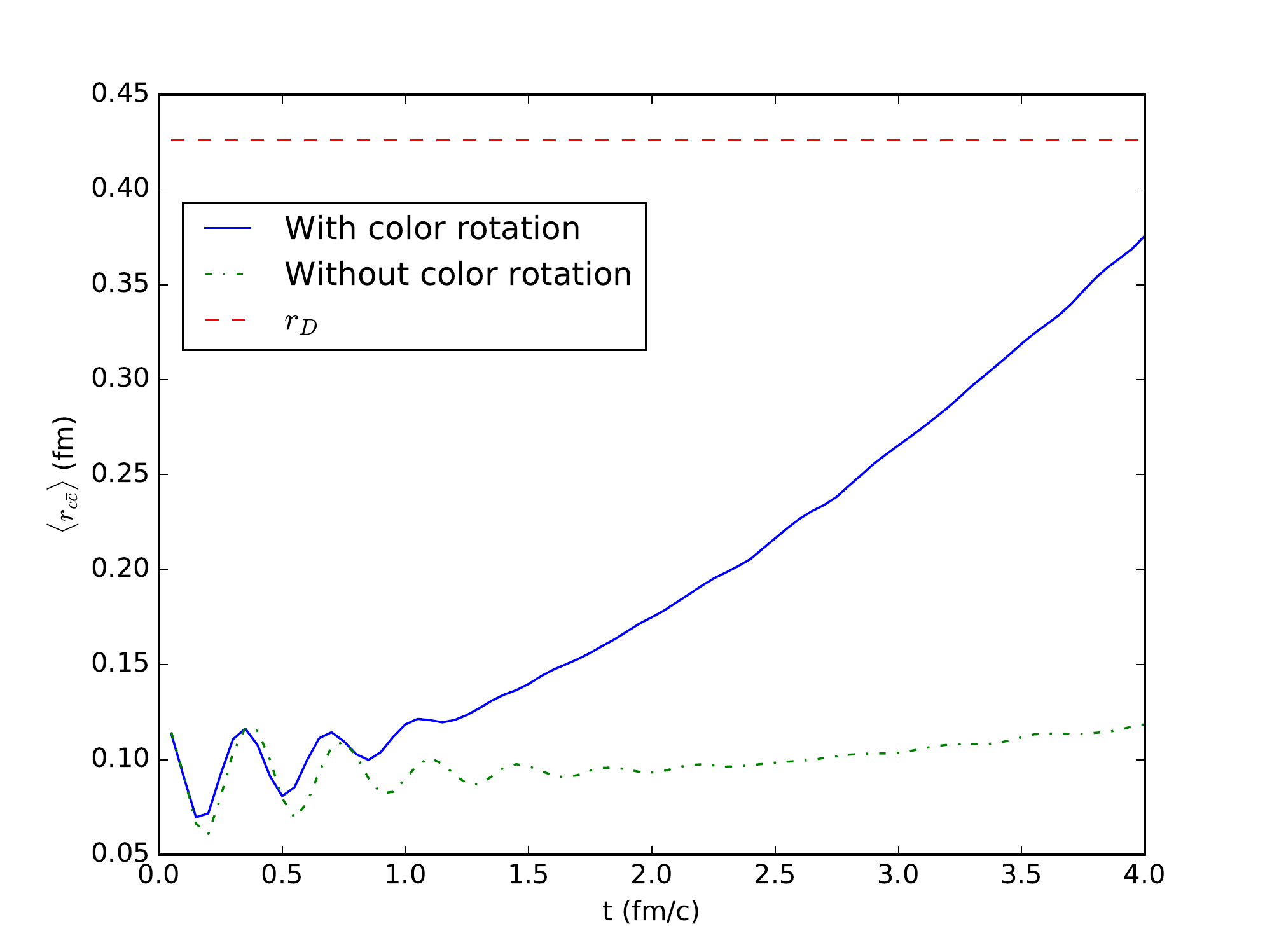}
\caption{Comparison of the evolution of a pair of heavy quarks initially prepared in a $J/\Psi$ state with or without considering the transition into octet states. The screening radius is $r_D=m_D^{-1}$.}
\label{fig:jpsi}
\end{center}
\end{figure}
\begin{figure}
\begin{center}
\includegraphics[scale=0.5]{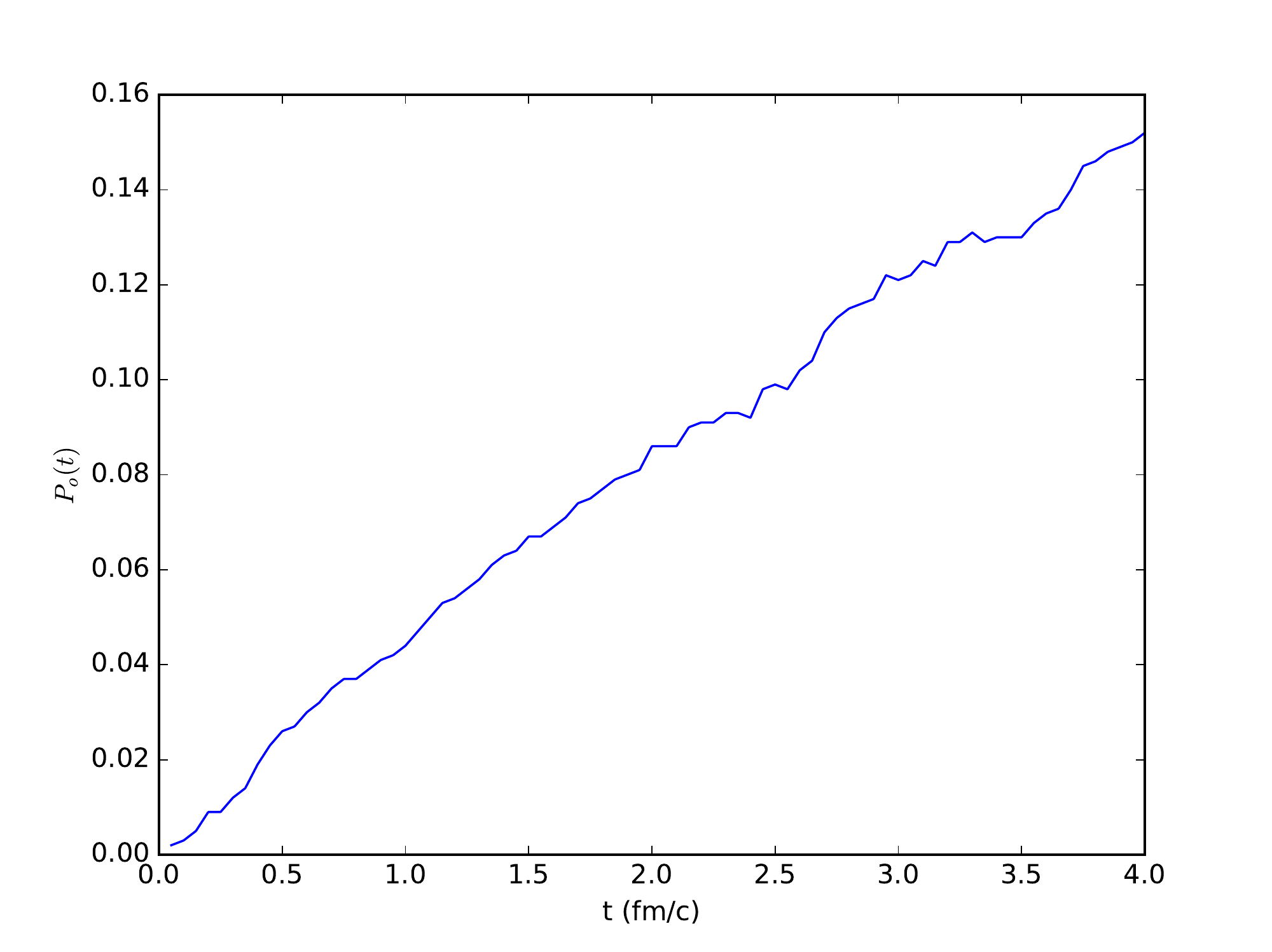}
\caption{Probability to find a heavy quark-antiquark pair in an octet state at time $t$, after it has been prepared at time $t=0$ as a $J/\Psi$ state.}
\label{fig:colora}
\end{center}
\end{figure}

We consider now the full Eqs.~(\ref{FPPs}) and (\ref{FPPo}). In Fig.~\ref{fig:jpsi} we plot the average mean distance $\langle r_{c\bar{c}}\rangle$ of a pair of charm quarks prepared in a $J/\Psi$, according to the prescriptions used in Ref.~\cite{Blaizot:2015hya}. That is, the radius of the pair is chosen randomly between  0 and $1/m_D$, and the relative momentum is chosen according to a Maxwell distribution with most probable velocity given by $v^2=0.3$. Finally one retains only pairs with binding energy  bigger than 500 MeV and  radius bigger than 0.1 fm. The temperature is taken to be $T=160\,\rm{MeV}$, the charm quark mass $M_c=1460 $ MeV, and $\gamma/M_c=0.2$ fm. These conditions  differ slightly from those used earlier: the reason is that we want to check our results against  those of \cite{Blaizot:2015hya} in a domain where they could be compared (see Fig.~\ref{fig:jpsi}). Thus, we also use a different running coupling than above, $\alpha_s=0.5/ (1+0.76 \ln (T/160))$, and $m_D=(16\pi\alpha_s/3)T^2$ (for two massless flavors). The cutoff on $V(\r)$ is 4GeV, while that on $W(\r)$ is  $4.58 \,m_D$. The unit of time in Fig.~\ref{fig:jpsi} is the physical unit fm/c.

We compare the results of the simulation of Eqs.~(\ref{FPPs}) and (\ref{FPPo}) with those obtained by neglecting color rotation, i.e., the singlet-octet transitions. The latter case is equivalent to a  QED simulation, and indeed our result in that case reproduce those obtained in  \cite{Blaizot:2015hya} (cp. the corresponding result in Fig.~\ref{fig:jpsi} with Fig.~5 in \cite{Blaizot:2015hya}). As expected, we see that the bound state tends to remain  bound longer if the transition to octet is not taken into account. The effect of color rotation is clearly to accelerate the melting of the bound state, although, according to the criterion used in \cite{Blaizot:2015hya}, $ \langle r_{c\bar{c}}\rangle\lesssim r_D=m_D^{-1}$, we may consider the system to remain bound at time $t=4$ fm/c.  This is to be contrasted with the result obtained with the Langevin equation with a color random force: in the present case, the disappearance of the bound state is a more gradual phenomenon, not amplified by unphysical violent kicks of a random color force. This gradual transition can be visualized by looking at the evolution of the probability to find the pair in an octet state, which is plotted in Fig.~\ref{fig:colora}. We can see that it takes a non negligible time to lose the information that the system was initially in a singlet state. 
\\
\section{Summary and outlook}

In this paper we have obtained a set of equations for the time  evolution of the reduced density matrix of a collection of quark-antiquark pairs immersed in a quark-gluon plasma in thermal equilibrium. These equations are fairly general (they are valid for an arbitrary number of heavy particles), and rely on two major approximations: weak coupling between the heavy quarks and the quark-gluon plasma, small frequency approximation for the plasma response. In the weak coupling approximation, the plasma sees the heavy quarks as a perturbation, and responds linearly to it. This response is characterized by a set of correlators, expectations values of gauge fields in the equilibrium state of the plasma, and because the heavy quark motion is slow on the typical scale of the plasma dynamics, only static, or nearly static response functions are needed. These functions account for some of the dominant effects of the plasma on the dynamics of the heavy quarks: the screening of the  instantaneous Coulomb interaction between the heavy quarks, and the effect of soft, low momentum transfer, collisions of the heavy quarks with the plasma constituents taken into account by an imaginary potential. 
The main equations that result from these two approximations alone generalize the equations that were obtained for an abelian plasma in the path integral formalism, using the Feynman-Vernon influence functional method \cite{Blaizot:2015hya}. Their structure is close to that of a Lindblad equation, and they are essentially equivalent to the equations obtained for QCD by Akamatsu \cite{Akamatsu:2014qsa}, although the present formulation differs from his in several aspects.  Recently a Lindblad equation was obtained for the evolution of the density matrix of a quark-antiquark pair, using similar approximations, but formulated in the context of a non relativistic effective theory (pNRQCD, \cite{Brambilla:2016wgg,Brambilla:2017zei}). This formalism puts the emphasis on the  singlet-octet transitions, and the validity of the employed effective theory requires specific conditions, viz.  $1/r\gg T\sim \gg E$, with $E$ the typical binding energy.  The corresponding Lindblad equation keeps the quantum features of the problem, however at the price  of a high computational cost.

In the case of abelian plasmas, a further approximation, the semi-classical approximation, leads to a Fokker-Planck equation, and a corresponding Langevin equation, which are relatively easy to solve numerically. When trying to extend this semi-classical approximation to QCD,  we have to face new features related to color dynamics. In the particular case of a quark-antiquark pair, this involves  the transitions between the singlet and the octet color configurations of the pair. Taking these transitions into account yields coupled equations for the two independent components of the density matrix, that are not easily solved, even when the motion of the heavy particles is treated  semi-classically.  

We have then explored numerically two strategies to solve approximately these coupled equations. In the first one, we assume that the color dynamics is fast compared to the motion of the heavy quarks. In this case, the collisions drive the systen quickly to a maximum entropy state where all colors are equally probable and uncorrelated. One can then use the Langevin equations to describe the dynamics in the vicinity of this maximum color entropy state, using a perturbative approach. This is sufficiently simple that it can be generalized to a system of an arbitrary number of quarks and antiquarks. However, the perturbative approach is limited by the fact that the color relaxation is slow  when the size of the quark-antiquark pairs is small, which may lead to unphysical behavior for a physically relevant choice of parameters. To overcome this limitation, we have explored another strategy, which appear more promising. It consists in treating the singlet-octet transitions as collisions, viewing the corresponding equations as Boltzmann equations that we solved using Monte Carlo techniques.  

Although they are fairly general, the equations that we have obtained so far do not yet capture all the relevant physics. For instance, in the particular case of a single quark-antiquark pair, the transitions between singlet and octet color states cause  rapid changes in the heavy quark hamiltonian. These  are not properly handled, and may be in conflict with the assumption that the dynamics of the heavy particles is slow compared to that of the plasma.  In addition we have left aside the possibility of absorption or emission of real gluons from the plasma, that are responsible in particular for gluo-dissociation, known to be an important mechanism in some temperature range. These shortcomings,  and further aspects of the problem, will be addressed in a forthcoming publication. 
\\

\noindent {\bf Acknowledgements}
This work has been supported in part by the European Research Council under the
Advanced Investigator Grant ERC-AD-267258. The work of M.A.E. was supported by the Academy of Finland, project 303756.

\appendix

\section{Correlators}\label{App:correlators}

In this appendix, we recall important properties of the correlators (Eqs.~(\ref{correlators})) which are used in the main text (see also \cite{Beraudo:2007ky}). These correlators depend only on the time difference and on the difference of coordinates, which we shall denote respectively  by $\tau$ and $\x$ in this Appendix. They are invariant under the change $\x\to-\x$.

After Fourier transform with respect to time, the time ordered propagator $\Delta(\tau,\x)$ can be written as  $\Delta(\omega,\x)=\Delta^R(\omega\x)+\iu\Delta^<(\omega,\x)$, where $\Delta^R(\omega,\x)$ is the retarded propagator. The correlator $\Delta^<(\omega,\x)$  is related to  $\Delta^>(\omega,\x)$ by the KMS relation, $\Delta^>(\omega,\x)=\rme^{\beta\omega}\Delta^<(\omega,\x)$, where $\beta=1/T$ is the inverse temperature. The two functions allow us to reconstruct the spectral density $\rho(\omega,\x)=\Delta^>(\omega,\x)-\Delta^<(\omega,\x)$.  From the last two equations, one easily establishes that $\Delta^<(\omega,\x)=N(\omega)\rho(\omega,\x)$, with $N(\omega)=1/(\rme^{\beta\omega}-1)$. From this relation, and using the fact that the spectral function is an odd function of $\omega$, it is easy to show that $\Delta^>(-\omega,\x)=\Delta^<(\omega,\x)$, so that, in particular, $\Delta^<(\omega=0,\x)=\Delta^>(\omega=0,\x)$. It follows then easily that
\beq
\left.\frac{\rmd \Delta^>(\omega,\x)}{\rmd \omega}\right|_{\omega=0}=-\left.\frac{\rmd \Delta^<(\omega,\x)}{\rmd \omega}\right|_{\omega=0}=\frac{\beta}{2} \Delta^<(\omega=0,\x).
\eeq

As argued in the main text,  all we need to describe the effective dynamics of the heavy quarks are the plasma correlators at or near zero frequency. More precisely, we need the following integrals
\beq
&&\int_{-\infty}^\infty \rmd \tau\, \Delta(\tau,\x)=\Delta(\omega=0,\x)=\Delta^R(\omega=0,\x)+i\Delta^<(\omega=0,\x),\nn
&&\int_0^\infty\rmd\tau \tau \,\Delta^>(\tau,\x)=-\frac{i}{2}\Delta'^>(\omega=0,\x),\nn
&&\int_0^\infty \rmd\tau \Delta^>(\tau,\x))=\frac{1}{2}\int_0^\infty \rmd\tau \Delta^>(\tau,\x)+\frac{1}{2}\int_{-\infty}^0 \rmd\tau \Delta^<(\tau,\x)\nn
&&\qquad\qquad\qquad\quad\;\;=-\frac{i}{2}\int_{-\infty}^\infty \rmd\tau \Delta(\tau,\x),\nn
\eeq
where we have used $\Delta^<(\tau,\x)=\Delta^<(\tau,-\x)$ and $\Delta^<(-\tau,\x)=\Delta^>(\tau,\x)$.

It is convenient to relate the zero frequency time-ordered propagator to an effective complex potential, $V(\r)+iW(\r)$ \cite{Laine:2006ns,Beraudo:2007ky}. We set 
  \beq
\Delta^R(0,\x)=-V(\x),\qquad  
 \Delta^<(0,\x)=-W(\r). 
\eeq

\section{Alternative time discretization}\label{App:alternative}

Our starting point is Eq.~(\ref{exprhoSK2b}), with symmetrical time integrations, of which we take the time derivative. We get
\beq\label{exprhoSK2bder}
&& \frac{\rmd }{\rmd t}{\cal D}(t)= \nn
&&\frac{i}{2} \frac{\rmd }{\rmd t}\int_{t_0}^t \rmd t_1 \int_{t_0}^{t} \rmd t'_1\int_{\x\x'}\,{\rm T}[n^a(t_1,\x) n^b(t_1',\x')] {\cal D}(t_0) \Delta(t_1-t_1',\x-\x')] \rangle_0\nn
&&\frac{i}{2} \frac{\rmd }{\rmd t}\int_{t_0}^t \rmd t_2 \int_{t_0}^{t} \rmd t_2'\int_{\x\x'}\, {\cal D}(t_0)\tilde{\rm T}[n^a(t_2,\x) n^b(t'_2,\x')]  \,\langle \tilde\Delta(t_2-t_2',\x-\x')] \rangle_0] \rangle_0\nn
&&+\frac{\rmd }{\rmd t}\int_{t_0}^t \rmd t_1 \int_{t_0}^{t} \rmd t_2\int_{\x\x'}\, [n^a(t_1,\x){\cal D}(t_0) n^b(t_2,\x')] \Delta^>(t_2-t_1,\x'-\x)] \rangle_0,\nn
\eeq
with ${\cal D}$ the heavy quark reduced density matrix in the interaction picture. 
Following the same reasoning as in the main text, we replace ${\cal D}(t_0)$ by ${\cal D}(\bar t)$, where $\bar t$ is an arbitrarily chosen time between $t$ and $t_0$, the error made in this substitution being at least of order $(H_1)^2$. We then exploit the freedom that we have in choosing $\bar t$. Given the symmetry of the expression (\ref{exprhoSK2bder}), it appears natural to  choose\footnote{In the language of stochastic differential equations, this choice corresponds to the Stratonovich choice, while that adopted in the main text, $\bar t=t$, corresponds rather to the It\^ o prescription. }
\beq
\bar t\equiv \frac{t+t'}{2},\qquad \tau\equiv t-t'
\eeq
so that
\beq\label{changetime}
\int_{t_0}^t \rmd t_1 \int_{t_0}^{t} \rmd t{_1'} \longrightarrow \int_{t_0}^t \rmd \bar t \int\rmd\tau,
\eeq
where the bounds on the $\tau$-integrals are $\pm (\bar t-t_0)$ or $\pm (\bar t-t)$ depending on whether $\bar t<t/2$ or $\bar t>t/2$, respectively.
We then exploit the fact that the dynamics of the plasma is fast compared to that of the heavy quark, and expand $n(t)$ and $n(t')$ around $\bar t$, assuming that  $\tau$ remains small. We get
\beq\label{middletermsb}
n(\x,t)=n(\x,\bar t)+\frac{\tau}{2} \frac{\rmd n(\x,\bar t)}{\rmd \bar t},\qquad n(\x',t')=n(\x',\bar t)-\frac{\tau}{2} \frac{\rmd n(\x',\bar t)}{\rmd \bar t}.\nn\eeq
When it is integrated with a symmetric function of $\x-\x'$, which is the case here, we can then write the product $n(\x,t) n(\x',t')$ as
\beq
n(\x,t) n(\x',t')=n(\x,\bar t)n(\x',\bar t)+\frac{\tau}{2}\left[\frac{\rmd n(\x,\bar t)}{\rmd \bar t},n(\x',\bar t) \right].
\eeq
Simlarly, under the same condition, 
\beq
{\rm T}\left[ n(\x,t) n(\x',t') \right]=n(\x,\bar t)n(\x',\bar t)+\frac{1}{2}\left[\frac{\rmd n(\x,\bar t)}{\rmd \bar t},n(\x',\bar t) \right]\, (\tau \theta(\tau)-\tau\theta(-\tau)).\nn
\eeq
Since the $\tau$-integrand is limited to  small $\tau$ (by the correlators $\Delta(\tau)$),  we may extend the boundaries of the $\tau$-integration to $\pm \infty$. The derivative with respect to time in Eq.~(\ref{exprhoSK2bder})  will then force $\bar t=t$ (see Eq.~(\ref{changetime})). 
We then obtain 
\beq\label{eqrhoAt2}
\frac{\rmd {\cal D}}{\rmd t}&=& \frac{i}{2}\int_{\x\x'} n(\x, t)n(\x', t){\cal D}( t) \int_{-\infty}^\infty\rmd \tau \Delta(\tau,\x-\x')\nn
&-&\frac{1}{2}\int_{\x\x'}\left[\dot n(\x,t) ,n(\x', t) \right] {\cal D}( t) \int_{0}^\infty\rmd \tau \tau \Delta^>(\tau,\x-\x')\nn
&+& \frac{i}{2}\int_{\x\x'}{\cal D}( t)n(\x,t) n(\x',t)) \int_{-\infty}^\infty\rmd \tau \tilde \Delta(\tau,\x-\x')\nn
&+&\frac{1}{2}\int_{\x\x'} {\cal D}( t) \left[\dot n(\x,t) ,n(\x', t) \right] \int_{0}^\infty\rmd \tau \tau  \Delta^<(\tau,\x-\x')\nn
&+& \int_{\x\x'}n(\x,t){\cal D}( t) n(\x',t)\int_{-\infty}^\infty\rmd \tau\, \Delta^>(-\tau,\x'-\x)\nn
& +& \frac{1}{2}\int_{\x\x'}\left( \dot n(\x, t) {\cal D}( t)n(\x',t)- n(\x, t) {\cal D}( t)\dot  n(\x't)\right) \int_{-\infty}^\infty\rmd \tau\tau \Delta^>(-\tau,\x'-\x).\nn
\eeq
At this point we use the values of the time integrals of the correlators that are given in  \ref{App:correlators}, and obtain
\beq\label{eqforrhoA3}
\frac{\rmd {\cal D}(t)}{\rmd t}&=&- \frac{i}{2}\int_{\x,\x'} V(\x-\x') \left[ n(\x)n(\x'),{\cal D}\right]\nn
&+&\frac{1}{2}\int_{\x,\x'} W(\x-\x')\left(\left\{n(\x)n(\x'),{\cal D}  \right\} -2 n(\x){\cal D}n(\x')\right) \nn
&-&\frac{i }{4T}\int_{\x,\x'}  W(\x-\x') \left(\dot n(\x){\cal D} n(\x')-n(\x){\cal D} \dot n(\x')) \right)\nn
&-&\frac{i }{8T}\int_{\x,\x'}  W(\x-\x') \left\{ {\cal D},[\dot n(\x),n(\x')]\right\}.
    \eeq
    
    As mentioned in the main text (see the discussion after Eq.~(\ref{main2}) the structure of this equation is close to that of a Lindblad equation  \footnote{Recall that one of the virtues of the Lindblad equation is to maintain the positivity of the density matrix \cite{Lindblad}.}. To make this more obvious, let us  Fourier transform the variables $\x$ and $\x'$. One obtains easily
\beq\label{eqforrhoA3b}
\frac{\rmd {\cal D}(t)}{\rmd t}&=&- \frac{i}{2}\int_{\q} V(\q) \left[ n_\q n_\q^\dagger,{\cal D}\right]\nn
&+&\frac{1}{2}\int_{\q} W(\q)\left(\left\{n_\q^\dagger n_\q,{\cal D}  \right\} -2 n_\q{\cal D}n_\q^\dagger\right) \nn
&-&\frac{i }{4T}\int_{\q}  W(\q) \left(\dot n_\q{\cal D} n_\q^\dagger-n_\q{\cal D} \dot n_\q^\dagger) \right)\nn
&-&\frac{i }{8T}\int_{\q}  W(\q) \left\{ {\cal D},[\dot n_\q,n_\q^\dagger]\right\}, 
    \eeq
    with the shorthand notation $\int_\q=\int\rmd^3\q/(2\pi)^3$, and $\q$ the variable conjugate to $\x$ in the Fourier transform. The second line of this equation has the structure of a Lindblad operator, but this is not so for the last two lines corresponding to the operator ${\cal L}_3$. However, it is easy to see that the substitution $n_\q\to n_\q+(i/4T) \dot n_\q$ in the second line, which obviously preserves its Lindblad structure,  generates all the terms in the last two lines with, in addition, terms that are quadratic in the time derivative, and are therefore suppressed with respect to the other terms by a power $1/MT$. Thus, to within these terms, one may consider Eq.~(\ref{eqforrhoA3b}) as a  Lindblad equation\footnote{The necessity of additonnal terms quadratic in velocities in order to obtain the Lindblad equation is also discussed in \cite{Akamatsu:2014qsa}.}. Note that the substitution mentioned above works only with the present time discretization, and is not immediately applicable to that used in the main text. 

\section{Comparison between the two discretizations}

The two time discretizations differ solely in their respective contributions to ${\cal L}_3$, and only in that part of it that we called ${\cal L}_{3a}$. In this appendix, we examine this difference in the case of QED. The generalization to QCD is straightforward. 
In the main text, we obtained (Eq.~(\ref{QEDanticom}))
\beq
{\cal L}_{3a}&=&-\frac{i}{8T}\int_{\x\x'}\, W(\x-\x')  \left(  2{\cal D}\dot n_{\x'} n_\x   -2n_\x \dot n_{\x'}{\cal D}   \right),
\eeq
while the discretization presented in \ref{App:alternative} yields (Eq.~(\ref{eqforrhoA3}))
\beq
{\cal L}_{3a}'=-\frac{i}{8T}\int_{\x\x'}\, W(\x-\x')  \left(  {\cal D} \dot n_{\x'} n_{\x} - {\cal D} n_{\x} \dot n_{\x'}  +\dot n_{\x'} n_{\x}{\cal D}  - n_{\x} \dot n_{\x'}{\cal D} \right). \nn
\eeq
By taking the difference one obtains
\beq
{\cal L}_{3a}'-{\cal L}_{3a}=\frac{i}{8T}\int_{\x\x'}\, W(\x-\x')  \left[ {\cal D}, \left\{\dot n_{\x'}, n_{\x}\right\}  \right] .
\eeq

It is easy to verify that this difference does not contribute to the matrix elements of the single particle density matrix. 
Let us then consider the two particle density matrix. A straightforward calculation yields
\beq
{\cal L}_{3a}'-{\cal L}_{3a}&=&\frac{1}{4MT}\, \left( \nabla^2 W_c+\bmnabla W_c \cdot \bmnabla_c \right).\nn
\eeq
Changing to the variables of Eq.~(\ref{changeofvar5}), and performing the small $\y$ expansion, one gets
\beq
{\cal L}_{3a}'-{\cal L}_{3a}\approx \frac{1}{2MT}\left(\nabla^2 W(\r)+ \bmnabla W(\r)\cdot \bmnabla_\r+\y\cdot {\cal H}(\r)\cdot  \nabla_\y   \right).
\eeq
Using the expression of ${\cal L}_3$ from Eq.~(\ref{L123QEDa}), and ${\cal L}'_3-{\cal L}_3={\cal L}_{3a}'-{\cal L}_{3a}$, we obtain 
\beq
&&{\cal L}_{3}'= -\frac{1}{2MT}\left\{   \Y\cdot\left({\cal H}(0)-{\cal H}(\r)  \right)\cdot \nabla_\Y
+ \y\cdot {\cal H}(0) \cdot\nabla_\y     \right\}\nn
&&+\frac{1}{2MT}\left(\nabla^2 W(\r)+ \bmnabla W(\r)\cdot \bmnabla_\r  \right).
\eeq
After taking a Wigner transform this yields
\beq\label{calL3p}
&&{\cal L}_{3}'=\frac{1}{2MT}\left[   ({\cal H}_{ij}(\r)-{\cal H}_{ij}(0))\nabla_\P^i P^j+ {\cal H}_{ij}(0) \nabla^i_\p \p^j+\nabla^2 W(\r)+\bmnabla W(\r)\cdot \bmnabla_\r\right] \nn
\eeq
This yields a modified Fokker-Planck equation for the relative coordinate, as compared to that used in the main text, Eq.~(\ref{FPrel3}). However, a simple change of variables allows us to recover the Langevin equation (\ref{langevinrelative}). Let us indeed set
\beq
\p'=\p-\frac{\bmnabla W(\r)}{4T}.
\eeq
Then, to within terms that are suppressed in the semi-classical approximation, we have
\beq
\frac{\rmd \p'}{\rmd t}=\F(\r)-\frac{1}{2MT}({\cal H}(0)+{\cal H}(\r))\cdot \p'+\bmxi(\r,t), \qquad \frac{\rmd \r}{\rmd t}=\frac{2\p'}{M}, 
\eeq
and a simple calculation shows that, in terms of these variables, the Langevin equation corresponding to the operator ${\cal L}'$, with ${\cal L}_{3}'$ given in Eq.~(\ref{calL3p}) and ${\cal L}'_{1,2}={\cal L}_{1,2}$, is identical to Eq.~(\ref{langevinrelative}). 

\section{Color structure of the density matrix}\label{sect:densitymatrix}

The density matrix of a color quark is a $3\times 3$ matrix in color space, which can be written as follows
\beq\label{Dcolor}
{\cal D}=a_0\, \mathbb{I}+ \a\cdot\bmt
\eeq
where $\bmt_i=\lambda_i/2$, with $\lambda_i$  the Gell-Mann matrices. We use the standard normalization
\beq
{\rm Tr}\, \bmt^a\bmt^b=\frac{1}{2}\delta^{ab}.
\eeq
The density matrix (\ref{Dcolor}) depends on 9 real parameters, and contains a scalar as well as a vector (octet) contributions. The density matrix associated to an antiquark may be written as 
\beq
{\cal D}=b_0\, \mathbb{I}- \b\cdot\tilde \bmt.
\eeq

A representation of the density matrix of a quark-antiquark pair may be obtained as the tensor product 
\beq\label{Dtensorproduct}
{\cal D}&=&(a_0 \mathbb{I}+ \a\cdot\bmt)\otimes (b_0 \mathbb{I}- \b\cdot\tilde\bmt)\nn
&=&a_0 b_0\,\mathbb{I}\otimes \mathbb{I}+  b_0 \a\cdot\bmt\otimes \mathbb{I}-a_0\mathbb{I}\otimes \tilde \bmt\cdot \b -a_ib_j\, t_i\otimes \tilde t_j.
\eeq
For a system invariant under color rotations, only the scalar components of ${\cal D}$ survive (e.g. $a_ib_j\propto \delta_{ij}$), and the density matrix takes the simpler form
\beq\label{D0D8def}
{\cal D}=D_0\,\mathbb{I}\otimes \mathbb{I}+D_8\, t_i\otimes \tilde t_i.
\eeq
Taking the matrix element, we get
\beq
\bra{\alpha\beta}{\cal D}\ket{\gamma\delta}&=&D_0\,\delta_{\alpha\gamma}\delta_{\beta\delta}+D_8\bra{\alpha}t^i\ket{\gamma}\bra{\beta}\tilde t^i\ket{\delta}\nn
&=&D_0\,\delta_{\alpha\gamma}\delta_{\beta\delta}+D_8\bra{\alpha}t^i\ket{\gamma}\bra{\delta} t^i\ket{\beta}.
\eeq
Using the identity
\beq
t^i_{\alpha\gamma}t^i_{\delta\beta}=\frac{1}{2}\left( \delta_{\alpha\beta}\delta_{\gamma\delta}-\frac{1}{N_c}  \delta_{\alpha\gamma}\delta _{\beta\delta} \right),
\eeq
one can write ${\cal D}$ as
\beq
\bra{\alpha\beta}{\cal D}\ket{\gamma\delta}=\left(  D_0-\frac{D_8}{2N_c}  \right)\,\delta_{\alpha\gamma}\delta_{\beta\delta}+\frac{D_8}{2}\,\delta_{\alpha\beta}\delta_{\gamma\delta}.
\eeq

Alternatively, one can project the quark-antiquark pairs on singlet or octet configurations:
\beq\label{DsDodef}
{\cal D}=D_{\rm s}\ket{\rm s}\bra{\rm s}+D_{\rm o}\sum_c \ket{{\rm o}^c}\bra{o^c},
\eeq
where $\ket{s}$ denotes a color singlet and $\ket{o^c}$ a color octet, with projection $c$. The states $\ket{\rm s}$ and $\ket{\rm o}$ are normalized to unity $\bra{\rm s}\rm s\rangle=1$, $\bra{\rm o^c}\rm o^d\rangle=\delta^{cd}$.  We have 
\beq
\bra{\alpha\bar \alpha}\rm s\rangle=\delta_{\alpha\bar \alpha}\frac{1}{\sqrt{N_c}},\qquad \bra{\alpha\bar \alpha}\rm o^c\rangle=\sqrt{2} \,t^c_{\alpha\bar \alpha}.
\eeq
Thus,
\beq
\bra{\alpha\beta}{\cal D}\ket{\gamma\delta}&=&\frac{D_s}{N_c}\delta_{\alpha\beta}\delta_{\gamma\delta}+2 D_o t^i_{\alpha\beta}\tilde t^i_{\gamma\delta}\nn
&=&\frac{D_s-D_o}{N_c}\delta_{\alpha\beta}\delta_{\gamma\delta}+D_o\delta_{\alpha\gamma}\delta_{\beta\delta}.
\eeq

The relations between the coefficients in the two basis are easily obtained. They are given by 
\beq\label{relationsD0D8DsDo}
&&D_s=D_0+C_F D_8,\qquad D_o=D_0-\frac{1}{2N_c} D_8,\nn
&&D_0= \frac{1}{N_c^2}[D_s+(N_c^2-1) D_o]  ,\qquad  D_8=   \frac{2}{N_c}(D_s-D_o).
\eeq

Note that the component $D_0$  corresponds to a completely unpolarized system, and can be written as 
\beq\label{Dunpolar}
{\cal D}=\sum_{\alpha\beta}\ket{\alpha\beta}\bra{\alpha\beta}=D_0\,\mathbb{I}\otimes \mathbb{I}.
\eeq
The same density matrix in the singlet-octet basis corresponds to $D_{\rm s}=D_{\rm o}$.

\section{Some useful formulae and matrix elements} \label{Ap:matelem}

In this appendix, we list a number of useful formulae, as well as some matrix elements that facilitate the derivation of the equations presented in the main text. 

We start with relations involving color matrices in the fundamental representation. Using the relations 
\beq
t^a t^b=\frac{1}{2N_c} \delta^{ab}+\frac{1}{2}\left[ if^{abc} +d^{abc}   \right]t^c,\nn
\eeq
and 
\beq
f^{abc} f^{abd}=N_c\delta^{cd}, \qquad d^{abc}d^{abd}=\frac{N_c^2-4}{N_c}\delta^{cd}, \qquad d^{abc}\delta^{ab}=0,
\eeq
 it is easy to establish the following formulae
\beq
&&t^at^b\otimes \tilde t^a\tilde t^b
=\frac{N_c^2-1}{4N_c^2} +\frac{N_c^2-2}{2N_c}\,t^a\otimes\tilde t^{a},\nn
&&t^at^b\otimes \tilde t^b\tilde t^a=\frac{N_c^2-1}{4N_c^2} -\frac{1}{N_c}t^c\otimes\tilde t^{c}.
\eeq
We also need
\beq
t^a t^b t^a=\left( C_F-\frac{N_c}{2} \right) t^b=-\frac{1}{2N_c}t^b.
\eeq

We consider now matrix elements in the singlet-octet basis. 
We have 
\beq
&&\bra{\rm s} t^a \otimes\tilde t^b\ket{\rm s}=\frac{1}{2N_c}\delta^{ab},\qquad \bra{\rm s} t^a \otimes\tilde t^a\ket{\rm s}=C_F\nn
&&\bra{\rm o^c} t^a \otimes\tilde t^a\ket{\rm o^d}=-\frac{1}{2N_c}\delta^{cd},\nn
&&\bra{\rm s} t^a \otimes\mathbb{I}\ket{\rm o^c}=\frac{1}{\sqrt{2N_c}} \delta^{ac},\nn
&&\bra{\rm o^d} t^a \otimes\mathbb{I}\ket{\rm o^c}= \frac{1}{2}\left( d^{dac}+if^{dac}   \right).
\eeq
\\

The following matrix elements of the color charge density, or its time derivative, are also useful. We have singlet-octet matrix elements, 
\beq
\bra{\r_1,\r_2;{\rm s}} \rho^a(\x)\ket{\r_3,\r_4;{\rm o}^c}=\frac{\delta^{ac} }{\sqrt{2N_c}}\,\bra{\r_1,\r_2} n(\x)\ket{\r_3,\r_4}
\eeq
where $n(\x)$ is the QED charge density
\beq
n(\x)=\delta(\x-\hat\r)\otimes\mathbb{I} -\mathbb{I}\otimes \delta(\x-\hat\r).
\eeq
We can write the formula above more simply as
\beq
\bra{{\rm s}} \rho^a(\x)\ket{{\rm o}^c}=\frac{\delta^{ac} }{\sqrt{2N_c}}\,  n(\x).
\eeq
Similarly, we have
 \beq
 \bra{{\rm s}}\dot\rho^a(\x)\ket{{\rm o}^c}=\frac{\delta^{ac}}{\sqrt{2N_c}} \,\dot n(\x).
 \eeq
We have also octet-octet matrix elements, 
\beq
 &&\bra{\r_1,\r_2;{\rm o}^d}\rho^a(\x)\ket{\r_3,\r_4;{\rm o}^c}\nn
 && \qquad= \delta(\r_1-\r_3)\delta(\r_2-\r_4) \bra{{\rm o}^d} \left[ \delta(\x-\r_1)\,t^a\otimes\mathbb{I} -\mathbb{I}\otimes \tilde t^a\,\delta(\x-\r_2)\right] \ket{{\rm o}^c}\nn
 &&\qquad=\frac{1}{2} d^{dac}\bra{\r_1,\r_2}n(\x)\ket{\r_3,\r_4}+\frac{i}{2}f^{dac}  \bra{\r_1,\r_2}m(\x)\ket{\r_3,\r_4},
 \eeq
 or, more simply, 
 \beq
 \bra{{\rm o}^d}\rho^a(\x)\ket{{\rm o}^c}=\frac{1}{2} d^{dac}\,n(\x)+\frac{i}{2}f^{dac}  \,m(\x),
 \eeq
 with
 \beq
 m(\x)=\delta(\x-\hat\r)\otimes\mathbb{I} +\mathbb{I}\otimes \delta(\x-\hat\r).
 \eeq
 Finally
 \beq
  \bra{{\rm o}^d}\dot\rho^a(\x)\ket{{\rm o}^c}= \frac{1}{2} d^{dac}\,\dot n(\x)+\frac{i}{2}f^{dac}  \,\dot m(\x).
 \eeq

\section{The equations of motion for the density matrix of a heavy quark-antiquark pair}\label{ap:equations}

In this Appendix we present the equations of motion for the matrix elements of the reduced density matrix of a heavy quark-antiquark pair. We  consider the two representations of the density matrix that correspond to the $(D_0,D_8)$  and the $(D_{\rm s},D_{\rm o})$ basis. In the main text we have indicated how to obtain the equations for $D_{\rm s}$ and $D_{\rm o}$ from the corresponding equations of the abelian case. Depending on the choice of basis, the color algebra proceeds differently, and the  relevant formulae are listed in \ref{Ap:matelem}. Independently of the way we proceed, the results should eventually lead to formulae for the various components of ${\cal D}_Q$ which satisfy the relations (\ref{relationsD0D8DsDo}). This constitutes a useful check of the results.

The equations to be presented below correspond to the coordinate space matrix element $\bra{\r_1\r_2} {\cal D}\ket{\r_1'\r_2'}$. Thus, for instance, $D_0$ stands for $D_0(\r_1,\r_2;\r_1',\r_2')$, and similarly for $D_8$, or $D_{\rm s}$ and $D_{\rm o}$.  We use the notation $(D_{\rm  s}|{\cal L}|{\cal D})$ for the contribution of the operator ${\cal L}_i$ to the time derivative of $D_{\rm s}$, that is
\beq
\frac{\rmd D_{\rm s}}{\rmd  t}=(D_{\rm  s}|{\cal L}_i|{\cal D}). 
\eeq 
and similarly for the other components of the density matrix. Also we  use the compact notation introduced in the main text, e.g., $W_{12}=W(\r_1-\r_2)$, $\bmnabla_1=\bmnabla_{\r_1}$,  $\bmnabla_{12}=\bmnabla_{\r_1}-\bmnabla_{\r_2}$, and so on, as well as the quantities $W_{a,b,c,\pm}$ defined in the main text (see Eqs.~(\ref{Wabcdef})). 
The various calculations are straightforward, but lengthy and somewhat tedious. They will not be presented here, we just list the final results. Those results, listed in the next two sections for the $(D_0,D_8)$ basis and the $(D_{\rm s},D_{\rm o})$ basis, respectively, are an exact transcription of Eq.~(\ref{main3}) to the density matrix of a quark-antiquark pair, without additional approximation. In the last subsection of this appendix, we list a number of formulae that are useful to implement the semi-classical approximation.

\subsection{ $(D_0,D_8)$ basis}
In the $(D_0,D_8)$ basis the contributions of the operators ${\cal L}_i$ to time derivative of the density matrix are given by:
\beq
&&(D_0|{\cal L}_1 |{\cal D})=i[V_{12}-V_{1'2'}] \frac{N_c^2-1}{4N_c^2}\,  \,D_8, \nn
&&(D_8|{\cal L}_1 |{\cal D})=i [V_{12}-V_{1'2'}] \left(D_0 +\frac{N_c^2-2}{2N_c}D_8\right).
\eeq

 \beq
&&(D_0|{\cal L}_2 |{\cal D})= C_F [2W(0)-W_a] \,D_0 +  \frac{C_F}{2N_c}[W_b-W_c]\,D_8 ,\nn
&&(D_8|{\cal L}_2 |{\cal D})=[W_b-W_c]\,D_0+C_F  [2W(0)-W_c]\,D_8 \nn
&&\qquad\qquad\qquad  +\frac{1}{2N_c} [W_a+W_c-2W_b] \,D_8.
\eeq

\beq
&&(D_0|{\cal L}_3 |{\cal D})=\frac{C_F}{4MT} [2\bmnabla^2 W(0) -\bmnabla^2W_a -\bmnabla W_a\cdot \bmnabla_a] D_0 \nn
&&\qquad-\frac{C_F}{4MT}\left\{ \nabla^2 W_c+ \bmnabla W_c\cdot \bmnabla_c -\bmnabla^2W_b -   \bmnabla W_b\cdot \bmnabla_b\right\}\frac{D_8}{2N_c}. \nn
\eeq
\beq
&&(D_8|{\cal L}_3 |{\cal D})=\frac{1}{4MT}  \left\{\nabla^2 W^b+\bmnabla W^b\cdot \bmnabla^b    -\nabla^2 W_c- \bmnabla W_c\cdot \bmnabla_c  \right\} D_0\nn
&&\qquad+\frac{C_F}{4MT} [2\bmnabla^2 W(0)-\nabla^2 W_c- \bmnabla W_c\cdot \bmnabla_c ] D_8\nn
&&\qquad+\frac{1}{4MT}\left\{ \nabla^2 W^a+\bmnabla W^a\cdot \bmnabla^a +\nabla^2 W_c+ \bmnabla W_c\cdot \bmnabla_c \right.\nn
&&\qquad\qquad\qquad\left.  -2\left( \nabla^2W^b+\bmnabla W^b\cdot \bmnabla^b  \right) \right\} \frac{D_8}{2N_c}.\nn
\eeq

Note that in the infinite mass limit, the contribution $(D_0|{\cal L}_2 |{\cal D})$ vanishes, while the second contribution reduces to $(D_8|{\cal L}_2 |{\cal D})=-N_c \Gamma(\r)D_8$. These results have a simple interpretation discussed in the main text. 

\subsection{ $(D_{\rm s},D_{\rm o})$ basis}
In the $(D_{\rm s},D_{\rm o})$ basis the contributions of the operators ${\cal L}_i$ to time derivative of the density matrix are given by:

\beq\label{DsDoForce}
&&(D_{\rm s}|{\cal L}_1 |{\cal D})=iC_F  [V_{12}-V_{1'2'}]\,D_{\rm s}\nn
&&(D_{\rm o}|{\cal L}_1 |{\cal D})=-\frac{i}{2N_c}[V_{12}-V_{1'2'}]D_{\rm o} .
\eeq

\beq\label{DswooneoverM}
&&(D_{\rm s}|{\cal L}_2 |{\cal D})=C_F  \left[  2W(0)-W_c \right]D_{\rm s}- C_F  W^- \,D_{\rm o} ,
\eeq
\beq\label{DowooneoverM}
(D_{\rm o}|{\cal L}_2 |{\cal D})=-\frac{1}{2N_c} W^- D_{\rm s}+\left[  2C_FW(0) +\frac{1}{2N_c}W_c -\left(\frac{N_c^2-4}{4N_c} W^-+\frac{N_c}{4} W^+\right)\right]D_{\rm o}.\nn
\eeq

\beq\label{DswoneoverM}
&&(D_{\rm s}|{\cal L}_3 |{\cal D})=\frac{C_F}{4MT} \left\{   2\bmnabla^2 W(0) -\nabla^2 W_c-\bmnabla W_c\cdot \bmnabla_c  \right\} D_s \nn
&&\qquad\qquad\qquad-   \frac{C_F}{4MT} \left\{ \bmnabla^2W^- +W^- \bmnabla W^-\cdot \bmnabla^-\right\}D_{\rm o}\nn
\eeq
Note the analogy with the equation (\ref{DswooneoverM}): replace $W(0)\to \nabla^2 W(0)$, $W_c\to \nabla^2 W_c+\bmnabla W_c\cdot W_c$ and $W^-\to \nabla^2 W^-+\bmnabla W^-\cdot \bmnabla^-$ (the color algebra is the same). Also this equation is identical to that in QED when $D_{\rm o}=D_{\rm s}$.

\beq\label{DowoneoverM}
&&(D_{\rm o}|{\cal L}_3 |{\cal D})=\frac{C_F}{2MT} \bmnabla^2 W(0) D_o-\frac{1}{4MT}[\nabla^2 W^-+\bmnabla W^-\cdot \bmnabla^- ]\, \frac{D_{\rm s}}{2N_c}\nn
&&+\frac{1}{4MT}\frac{1}{2N_c}\left(\nabla^2 W_c+ \bmnabla W_c\cdot \bmnabla_c \right)D_{\rm o}\nn
&&-\frac{1}{4MT}\left\{ \frac{N_c^2-2}{2N_c}  \left(  \nabla^2 W_a +\bmnabla W_a\cdot \bmnabla_a\right) +\frac{1}{N_c}\left( \nabla^2 W_b+\bmnabla W_b\cdot \bmnabla_b\right)\right\}\,D_{\rm o}.\nn
\eeq\\

\subsection{Equations in the semi-classical approximation}\label{ap:semiclassical}

To derive the equations in the semi-classical approximation, the following formulae are useful.
\beq
&&W_a=W(\Y+\y/2)+W(\Y-\y/2)\approx 2 W(0)+\Y\cdot {\cal H}(0)\cdot\Y+\frac{1}{4} \y\cdot {\cal H}(0)\cdot\y,\nn
&& W_b=W(\Y+\r)+W(\Y-\r)\approx 2W(\r)+ \Y\cdot{\cal H}(\r)\cdot\Y,\nn
&&W_c=W(\r+\y/2)+W(\r-\y/2)\approx  2W(\r)+\frac{1}{4}\y\cdot{\cal H}(\r)\cdot \y.
\eeq

\beq
&&\nabla^2 W_a=\nabla^2(W_{11'}+W_{22'})=2 \nabla^2 W(0)\nn
&&\nabla^2 W_b=\nabla^2(W_{12'}+W_{21'})=2 \nabla^2 W(\r)\nn
&&\nabla^2 W_c=\nabla^2(W_{12}+W_{1'2'})=2 \nabla^2 W(\r)\nn
\eeq

\beq
&&\bmnabla W_a\cdot \bmnabla_a\equiv\bmnabla  W_{11'} \bmnabla_{11'}+\bmnabla  W_{22'} \bmnabla_{22'}= 2\Y\cdot{\cal H}(0)\cdot\bmnabla_\Y+2\y\cdot{\cal H}(0)\cdot\bmnabla_\y\nn
&&\bmnabla W_b\cdot \bmnabla_b\equiv\bmnabla  W_{12'} \bmnabla_{12'}+\bmnabla  W_{21'} \bmnabla_{21'}=2 \bmnabla W(\r)\cdot\bmnabla_\r+2\Y\cdot {\cal H}(\r)\cdot\bmnabla_\Y,\nn
&&\bmnabla W_c\cdot \bmnabla_c\equiv \bmnabla W_{12}\cdot \bmnabla_{12}+\bmnabla W_{1'2'}\cdot \bmnabla_{1'2'}=2 \bmnabla W(\r)\cdot\bmnabla_\r+2\y\cdot {\cal H}(\r)\cdot\bmnabla_\y,\nn
\eeq

\section{A single heavy quark in the static limit}\label{ap:singlequark}
As an illustration of the color dynamics we study the case of a single heavy quark in the infinite mass limit. The density matrix can be written as
\begin{equation}
{\cal D}=D_0\,\mathbb{I}+D_8^a\, t^a\,,
\end{equation}
and we leave open the possibility of a vector component ($D_8$). In the infinite mass limit, $D_0$ and $D_8$ obey the equations
\begin{align}
&\frac{dD_0}{dt}=0 \,,\nonumber\\
&\frac{dD_8^a}{dt}=\frac{N_cW({\bf 0})}{2}D_8^a=-N_c\gamma_Q D^a_8.
\end{align}
The first equation reflects the conservation of the trace of the density matrix. The second equation indicates that the states that have a preferred direction in color space decay exponentially in time. To illustrate this behaviour imagine that at $t=0$ we have a density matrix corresponding to a heavy quark with a specific color (such that ${\cal D}_{11}=1$ and the rest of the components are $0$). The evolution of this matrix can be written as
\begin{equation}
{\cal D}(t_0)=\frac{1}{3}\mathbb{I}+\left(t^3+\frac{t^8}{\sqrt{3}}\right)e^{-N_c\gamma_Q t},
\end{equation}
which in terms of the different components means that
\begin{align}
&{\cal D}_{11}=\frac{1}{3}\left(1+2e^{-N_c\gamma_Q t}\right)\,,\nonumber \\
&{\cal D}_{22}={\cal D}_{33}=\frac{1}{3}\left(1-e^{-N_c\gamma_Q t}\right)\,,
\end{align}
and the rest of the components are $0$. Thus, in a time of order $(N_c\gamma_Q)^{-1}$, the density matrix becomes diagonal, with all diagonal elements equal: the system is driven to the maximum entropy state, where only the component $D_0$ is non vanishing. 

\section{The two heavy quark case}\label{sec:HQpair}

We consider here  a system formed by two heavy quarks (rather than a heavy quark-antiquark pair). Eq. (\ref{main}) is also fulfilled in this case, but the color structure of the density matrix of a quark pair differs from that of a heavy quark-antiquark pair. Similarly to Eq.~(\ref{D0D8def}) we can write
\beq\label{D0D8QQ}
{\cal D}=D_0 \,\mathbb{I}\otimes \mathbb{I} +D_8 \, t^a\otimes t^a, 
\eeq
while the analog of Eq.~(\ref{DsDodef}) can be written as 
\beq
{\cal D}=D_{\bar 3} P_{\bar{\bf 3}}+D_{6} P_{{\bf 6}},
\eeq
where $P_{\bar{\bf 3}}$ and $P_{{\bf 6}}$ denote respectively the projectors on the representation $\bar{\bf 3}$ and ${\bf 6}$ of SU(3). In writing Eq.~(\ref{D0D8QQ}) we stick to the notation of Eq.~(\ref{D0D8def}), although in the present case there is no octet state involved in $D_8$. As for $D_0$ it conserves the interpretation of the maximal entropy state. 

Calculating, as we did for the quark-antiquark pair, the matrix element $\bra{\r_1\r_2}{\cal D}\ket{\r_1'\r_2'}$ of the two-quark density matrix, we obtain for the first line of Eq. (\ref{main})
\begin{equation}
\frac{\rmd D_0}{\rmd t}=-\frac{iC_F}{2N_c}(V_{12}-V_{1'2'})D_8\,,
\end{equation}
\begin{equation}
\frac{\rmd D_8}{\rmd t}=-i(V_{12}-V_{1'2'})\left(D_0-\frac{D_8}{N_c}\right)\,.
\end{equation}
This set of equations has two eigenvalues with opposite signs ($i (V_{12}-V_{1'2'}) (1\pm N_c)/(2N_c))$. The positive sign corresponds to the color configuration $\bar{\bf 3}$ and represents an attractive interaction, while the minus sign corresponds to the configuration ${\bf 6}$ for which the interaction is repulsive. The equations for $D_{\bf \bar 3}$ and $D_{\bf 6}$ read
\beq
&&\frac{\rmd D_{\bf \bar 3}}{\rmd t}=i(V_{12}-V_{1'2'})\frac{1+N_c}{2N_c}D_{\bf \bar 3},\qquad D_{\bf \bar 3}=D_0-\frac{N_c+1}{2N_c}D_8,\nn
&&\frac{\rmd D_{\bf 6}}{\rmd t}=i(V_{12}-V_{1'2'})\frac{1-N_c}{2N_c}D_{\bf 6},\qquad D_{\bf 6}=D_0+\frac{N_c-1}{2N_c}D_8.
\eeq
Note that the attraction between two quarks in the $\bf \bar 3$ channel is $N_c-1$ times weaker than the attraction of a quark-antiquark pair in the singlet channel. For $N_c=3$ this is only a factor 2, which suggests that the probability to find heavy di-quarks in a plasma may not be too different from that of finding bound heavy quark-antiquark pairs. 

We turn now to the second line of   Eq. (\ref{main}), which yields
\beq
&&\frac{\rmd D_0}{\rmd t}=C_F\left(2 W(0)-W_a\right) D_0+\frac{C_F}{2N_c}(W_c -W_b)D_8,\nn
&&\frac{\rmd D_8}{\rmd t}=(W_c-W_b)D_0 \nn
&&\qquad +\left\{ C_F(2W(0)-W_b)+\frac{1}{2N_c}(W_+-2W_c)\right\} D_8.
\eeq
Finally, for the third line of Eq. (\ref{main}) we obtain
\beq
&&\frac{\rmd D_0}{\rmd t}=\frac{C_F}{4MT}\left[2\nabla^2 W(0)-\nabla^2 W_a-\bmnabla W_a \cdot \bmnabla_a\right]D_0\nonumber\\
&&\qquad+\frac{C_F}{4MT}\left[\nabla^2W_c+\bmnabla W_c \cdot \bmnabla_c-\nabla^2W_b-\bmnabla W_b \cdot \bmnabla_b\right]\frac{D_8}{2N_c},
\eeq
and
\beq
&&\frac{\rmd D_8}{\rmd t}=\frac{1}{4MT}(\bmnabla W_c\cdot \bmnabla_c-\bmnabla W_b\cdot \bmnabla_b)D_0\nn
&&+\frac{C_F}{4MT}\left\{2\bmnabla^2W(0)-\bmnabla^2 W_b-\bmnabla W_b\cdot\bmnabla_b\right\}D_8\nn
&&+\frac{1}{4MT}\frac{1}{2N_c}\left\{2(\bmnabla^2W(0)-\bmnabla^2 W(\r))+\bmnabla W_a\cdot \bmnabla_a+\bmnabla W_b\cdot \bmnabla_b-2 \bmnabla W_c\cdot \bmnabla_c\right\}{D_8}.\nn
\eeq

At this point we may use the formulae listed in ~\ref{ap:equations} in order to perform the small $y$ expansion.  We obtain 
\beq\label{D0QQ}
&&\frac{\partial D_0}{\partial t}=i \left(\frac{\bmnabla_{\cal R}\cdot\bmnabla_\Y}{2M}+\frac{2\nabla_{\r}\cdot\nabla_{\y}}{M}  \right)D_0+i\frac{C_F}{2N_c}\y\cdot \F(\r)D_8\nonumber\\
&&-C_F\left(  \Y\cdot{\cal H}(0)\cdot \Y +\frac{1}{4}\y\cdot{\cal H}(0)\cdot \y     \right)D_0-\frac{C_F}{2N_c}\left(  \Y\cdot{\cal H}(\r)\cdot \Y -\frac{1}{4}\y\cdot{\cal H}(\r)\cdot \y     \right)D_8\nonumber\\
&&-\frac{C_F}{2MT}\left( \Y\cdot{\cal H}(0)\cdot \bmnabla_\Y + \y\cdot{\cal H}(0)\cdot \bmnabla_\y     \right)D_0\nn
&&-\frac{C_F}{2MT}\left[\Y\cdot {\cal H}(\r)\cdot \bmnabla_\Y-\y\cdot {\cal H}(\r)\cdot \bmnabla_\y  \right]\frac{D_8}{2N_c},
\eeq
and 
\beq\label{D8QQ}
&&\frac{\partial D_8}{\partial t}=i \left(\frac{\bmnabla_{\cal R}\cdot\bmnabla_\Y}{2M}+\frac{2\bmnabla_{\r}\cdot\bmnabla_{\y}}{M}  \right)D_8-N_c\Gamma(\r)D_8\nonumber\\
&&+i\y\cdot\F(\r)\left(D_0-\frac{D_8}{N_c}\right)+\frac{1}{2N_c}\left(  \Y\cdot{\cal H}(0)\cdot \Y +\frac{1}{4}\y\cdot{\cal H}(0)\cdot \y     \right)D_8\nonumber\\
&&-\Y\cdot{\cal H}(\r)\cdot \Y\left(\frac{N_c^2-2}{2N_c}D_8+D_0\right)+\frac{1}{4}\y\cdot {\cal H}(\r)\cdot \y\left(D_0-\frac{D_8}{N_c}\right)\nonumber\\
&&-\frac{1}{2MT}\left(\Y\cdot{\cal H}(\r)\cdot\bmnabla_\Y -\y\cdot{\cal H}(\r)\cdot\bmnabla_\y  \right) D_0-\frac{C_F}{2MT}\Y\cdot{\cal H}(\r)\cdot\bmnabla_\Y D_8\nn
&&+\frac{N_c}{4MT}\left[ \bmnabla^2W({\bf 0})-\bmnabla^2W({\bf r})-\bmnabla W({\bf r})\cdot\bmnabla_{\bf r}\right]D_8\nonumber\\
&&+\frac{1}{2MT}\left(  \Y\cdot{\cal H}(0)\cdot \bmnabla_\Y + \y\cdot{\cal H}(0)\cdot \bmnabla_\y     \right)+\Y\cdot {\cal H}(\r)\cdot \bmnabla_\Y-2\y\cdot {\cal H}(\r)\cdot \bmnabla_\y)  \frac{D_8}{2N_c}.\nn
\eeq

This pair of equations forms a system that we can diagonalize perturbatively, following the procedure of Sect.~\ref{sec:randomc1pair}. The relevant coefficients that enter Eq.~(\ref{pertformula}) are easily identified on the equations above. We then find  that the evolution of the component of the density matrix that is close to the maximum entropy configuration is given by 
\beq
&&\partial_t D_0'=\left\{(i \left(\frac{\bmnabla_{\cal R}\cdot\bmnabla_\Y}{2M}+\frac{2\nabla_{\r}\cdot\nabla_{\y}}{M}  \right)-C_F\left(  \Y\cdot{\cal H}(0)\cdot \Y +\frac{1}{4}\y\cdot{\cal H}(0)\cdot \y     \right)\right.\nonumber\\
&&\left.\qquad-\frac{C_F}{2MT}\left(  \Y\cdot{\cal H}(0)\cdot \bmnabla_\Y + \y\cdot{\cal H}(0)\cdot \bmnabla_\y     \right)-\frac{C_F(\y\cdot F(\r))^2}{2N_c^2\Gamma(\r)}\right\}D_0'\,.
\eeq
Apart from keeping the dynamics of the center of mass explicit, this equation is very similar to Eq.~(\ref{eq:D0prime}).

\bibliography{Langepre}
\biboptions{sort&compress}
\bibliographystyle{elsarticle-num.bst}

\end{document}